\documentclass[journal]{IEEEtran}

\usepackage{amssymb,amsmath,amsfonts,rotating}
\usepackage[noadjust]{cite}
\usepackage{color} 
\usepackage{floatfig}

\definecolor{TODO}{rgb}{0.6,0.6,0.6} 

\definecolor{TOCHECK}{rgb}{0.8,0.8,0.8} 

\newtheorem{theorem}{Theorem}
\newcommand{\btheo}{\begin{theorem}}
\newcommand{\etheo}{\end{theorem}}
\newcommand{\bproof}{\begin{proof}}
\newcommand{\eproof}{\end{proof}}
\newtheorem{definition}{Definition}
\newcommand{\bdefi}{\begin{definition}}
\newcommand{\edefi}{\end{definition}}
\newtheorem{fact}{Fact}
\newcommand{\bprop}{\begin{fact}}
\newcommand{\eprop}{\end{fact}}
\newtheorem{corollary}{Corollary}
\newcommand{\bcor}{\begin{corollary}}
\newcommand{\ecor}{\end{corollary}}
\newtheorem{example}{Example}
\newcommand{\bex}{\begin{example}}
\newcommand{\eex}{\end{example}}
\newtheorem{lemma}{Lemma}
\newcommand{\blemma}{\begin{lemma}}
\newcommand{\elemma}{\end{lemma}}
\newtheorem{remark}{Remark}
\newcommand{\bremark}{\begin{remark}}
\newcommand{\eremark}{\end{remark}}
\newtheorem{conjecture}{Conjecture}
\newcommand{\bconj}{\begin{conjecture}}
\newcommand{\econj}{\end{conjecture}}

\newcommand{\naturals}{\ensuremath{\mathbb{N}}}
\newcommand{\expectation}{\ensuremath{\mathbb{E}}}

\newcommand{\defas}{\ensuremath{\stackrel{{\vartriangle}}{=}}} 

\def\0{{\tt 0}} 
\def\1{{\tt 1}} 
\def\?{{\tt *}} 
\def\g{{\tt g}} 

\def\edge{{\tt e}} 
\newcommand{\gp}{\ensuremath{\gamma}} 
\newcommand{\chm}[1]{\ensuremath{{{\mu^{{\tt{\epsilon}}}_{#1}}}}} %
\newcommand{\lrm}[1]{\ensuremath{{{\mu^{{\tt{x}}}_{#1}}}}} 
\newcommand{\rlm}[1]{\ensuremath{{{\mu^{{\tt{y}}}_{#1}}}}} 
\newcommand{\m}{\ensuremath{{{\mu}}}} 
\def\lra{{\tt {x}}} 
\def\rla{{\tt {y}}} 
\newcommand{\graph}{{\ensuremath{\tt G}}}
\newcommand{\ldpc}{{{\ensuremath{\text{LDPC}}}}}
\newcommand{\ddp}{dd~pair~} 
\newcommand{\ddps}{dd~pairs~} 
\newcommand{\maxwell}{M~} 
\newcommand{\n}{\ensuremath{n}} 
\newcommand{\drate}{r} 
\newcommand{\rate}{\ensuremath{r}} 
\newcommand{\ledge}{\ensuremath{\lambda}} 
\newcommand{\redge}{\ensuremath{\rho}} 
\newcommand{\lnode}{\ensuremath{\Lambda}} 
\newcommand{\rnode}{\ensuremath{\Gamma}} 
\newcommand{\ldegree}{{\ensuremath{\ensuremath{\tt l}}}} 
\newcommand{\rdegree}{{\ensuremath{\ensuremath{\tt r}}}} 
\newcommand{\ih}{\ensuremath{\epsilon}} 
\newcommand{\cp}{\epsilon} 
\newcommand{\xh}{\ensuremath{h}} 
\newcommand{\MAP}{\ensuremath{\text{\tiny MAP}}} 
\newcommand{\BP}{\ensuremath{\text{\tiny BP}}} 
\newcommand{\EBP}{\ensuremath{\text{\tiny EBP}}} 
\newcommand{\Sh}{\ensuremath{\text{\tiny Sh}}} 

\newcommand{\xl}{\ensuremath{{\tt{x}}}}
\newcommand{\xr}{\ensuremath{{\tt{y}}}}
\newcommand{\lxl}{\ensuremath{\underline{\xl}}} 
\newcommand{\uxl}{\ensuremath{\overline{\xl}}} 
\newcommand{\exitentropy}{EXIT~} 

\newcommand{\qed}{{\hfill \footnotesize $\blacksquare$}}
\renewcommand{\mid}{\,|\,}
\newcommand{\tablespace}{\vspace{10pt}}

\newcommand{\hb}{\bar{h}}

\newcommand{\de}{\text{d}}
\newcommand{\cX}{{\cal X}}
\newcommand{\nddp}{\Xi}
\newcommand{\zl}{{\tt z}}
\newcommand{\cN}{{\cal N}}
\newcommand{\pr}{{\rm Pr}}
\newcommand{\List}{\Theta}
\newcommand{\ind}{{\mathbb I}}
\newcommand{\cW}{{\cal W}}

\newcommand{\entropy}{\hat{H}} 
\newcommand{\iter}{t} 
\newcommand{\drmax}{{\tt r}_{\text{max}}}

\begin{document}
\initfloatingfigs
{
\title{Maxwell Construction: The Hidden Bridge between 
 Iterative and  Maximum a Posteriori Decoding} %
\author{
Cyril~M\'easson$^\dagger$ %
\thanks{$\dagger$ EPFL, School for Computer and Communication Sciences, CH-1015 Lausanne, Switzerland. E-mail: cyril.measson@epfl.ch}, %
Andrea~Montanari$^*$ %
\thanks{$*$ ENS, Laboratoire de Physique Th\'eorique, F-75231 Paris, France. E-mail: montanar@lpt.ens.fr} %
and R\"udiger Urbanke %
$^\ddagger$\thanks{$\ddagger$ EPFL, School for Computer and Communication Sciences, CH-1015 Lausanne, Switzerland. E-mail: ruediger.urbanke@epfl.ch} %
\thanks{Parts of the material were presented in \cite{MeU03b,Mon01c,MMU04, MMRU04}.}
} %
\maketitle
}

\begin{abstract}
There is a fundamental relationship between belief propagation 
and maximum a posteriori decoding. 
A decoding algorithm, which we call the Maxwell decoder, is introduced and 
provides a constructive description of this relationship.
Both, the algorithm itself and the analysis of the new decoder are reminiscent
of the Maxwell construction in thermodynamics. 
This paper investigates in detail the case of transmission
over the binary erasure channel, while the extension to general binary
memoryless channels is discussed in a companion paper.
\end{abstract}

\begin{keywords}
belief propagation, maximum a posteriori, maximum likelihood, Maxwell construction, 
threshold, phase transition, Area Theorem, \exitentropy curve, entropy
\end{keywords}


\section{Introduction}
\PARstart{I}{t}
is a key result, 
and the starting point of iterative coding, that belief 
propagation (BP) is optimal on trees. See, e.g.,   \cite{Per88,Wib96,AjM00,KFL01}. 
However, trees with bounded state size appear not to be powerful enough
models to allow transmission arbitrarily close to capacity. For instance, it
is known that in the setting of standard binary Tanner graphs the error
probability of codes
defined on trees is lower bounded by a constant
which only depends on the channel and the rate of the code \cite{ETV99,RiU05}.
The general wisdom 
is therefore to apply BP decoding to graphs with loops and to consider this 
type of decoding as a (typically) strictly suboptimal attempt to  perform maximum 
a posteriori (MAP) bit decoding. One would therefore not expect any link 
between the BP and the MAP decoder except for the obvious suboptimality of the 
BP decoder.

This contribution demonstrates that there is a fundamental relationship between 
BP and MAP decoding which appears in the limit of large blocklengths. 
This relationship is furnished by the so-called Maxwell (M) decoder. 
The \maxwell decoder combines the BP decoder with a ``guessing'' device to 
perform MAP decoding. It is possible to analyze the performance of the \maxwell decoder
in terms of the \exitentropy curve introduced in \cite{teB01}.
This analysis leads to a precise characterization of 
how difficult it is to convert the BP decoder into a MAP decoder and
this ``gap'' between the MAP and BP decoder has a pleasing 
graphical interpretation in terms of an 
area under the \exitentropy curve.\footnote{The \exitentropy curve is here the \exitentropy curve associated to the 
iterative coding system and not to its individual component codes. This differs from 
the original \exitentropy chart context presented in \cite{teB01}.} Further, the MAP threshold is determined by 
a balance between two areas representing the number of guesses and the reduction 
in uncertainty, respectively.
The analysis gives also rise to a generalized Area Theorem, see also 
\cite{AKtB02}, and it provides an alternative tool for proving area-like 
results.

The concept of a ``BP decoder with guesses'' itself is not new. In \cite{PiF04}
the authors introduced such a decoder in order to improve the performance
of the BP decoder. Our motivation though is quite different.
Whereas, from a practical point of view, such enhancements work best
for relatively small code lengths, or to clean up error floors, we are interested
in the asymptotic setting in which the unexpected 
relationship between the MAP decoder and the BP decoder emerges.
%
\subsection{Preliminaries}
Assume that transmission takes place over a binary erasure channel with
parameter $\ih$, call it BEC($\ih$). More precisely, the transmitted bit $x_i$ at time $i$, $x_i \in{\cal{X}}\defas\{\0,\1\}$, is erased with probability $\ih$. 
The channel output is the random variable $Y_i$ which takes values in 
${\cal{Y}}\defas\{\0,\?,\1\}$.  To be concrete, we will exemplify all 
statements using Low-Density Parity-Check (LDPC) code ensembles \cite{Gal63}.  
However, the results extend to other ensembles like, e.g., Generalized 
 $\ldpc$ or turbo codes, and we will state the results in a general form.
For an in-depth introduction to the analysis of $\ldpc$ ensembles  see, e.g., 
\cite{LMSS01,LMSS01b,RiU01,RSU01}. For convenience of the reader, and to settle 
notation, let us briefly review some key statements. The degree 
distribution (dd) pair  
$\left(\ledge(\xl),\redge(\xl)\right)=(\sum_j \ledge_j \xl^{j-1},\sum_j \redge_j \xl^{j-1})$
 represents the degree distribution of the graph from the {\em edge} perspective. 
We consider the ensemble $\ldpc(\ledge,\redge,\n)$ 
of such graphs of length $\n$ and we are interested in its 
asymptotic average performance 
(when the blocklength $\n\to \infty$).
This ensemble can equivalently be described by 
$\nddp \defas (\lnode(\xl),\rnode(\xl))  =
(\sum_j \lnode_j \xl^{j}, \sum_j \rnode_j \xl^{j})$,
which is the \ddp from the {\em node} perspective\footnote{The changes of 
representation are obtained via $\lnode(\xl)=(1/\int\ledge){\int_0^\xl \ledge(u)\text{d}u}$, 
$\rnode(\xl)=(1/\int\redge){\int_0^\xl \redge(u)\text{d}u}$,  $\ledge(\xl) = \lnode'(\xl)/\lnode'(1)$ 
and  
$\redge(\xl) = \rnode'(\xl)/\rnode'(1)$.}. 
An important characteristic of the 
ensemble $\ldpc(\ledge,\redge,\n)$ is the {\em design rate}
$\drate \defas 1-\int\redge/\int\ledge = 1-\lnode'(1)/\Gamma'(1)$. We will 
write $\drate=\drate(\ledge,\redge)$ or $\drate = \drate(\lnode,\rnode)$ 
whenever we regard the design rate as a function of the degree distribution pair.

 The BP threshold, call it $\ih^\BP=\ih^\BP(\ledge,\redge)$, is defined in \cite{LMSS01,LMSS01b,RiU01,RSU01} as
$
\ih^\BP \defas \text{sup} \{ \ih\in[0,1]: \ih \ledge(1-\redge(1-\xl))<\xl, \forall \xl\in(0,1]\}.
$
 Operationally, if we transmit at $\ih<\ih^\BP$ and use a BP decoder, then all 
bits except possibly a sub-linear  fraction can be recovered when $\n\to\infty$. 
On the other hand, if $\ih\geq\ih^\BP$, then a fixed fraction of bits remains erased 
after BP decoding when $\n\to\infty$. 
In a similar manner we can define the MAP threshold.   
This threshold was first found 
via the replica method in \cite{Mon01}.
Further, in \cite{Mon01c}  a simple counting argument leading to an upper bound
for this threshold was given. The argument is explained and sharpened 
in Sec.~\ref{sec:Counting}. 
In this paper we develop the  
point of view taken in \cite{MeU03b}. The reference quantity is then the extrinsic\footnote{The term 
extrinsic is used when the observation of the bit itself is ignored, see \cite{BGT93,HOP96}.}  
entropy, in short EXIT.\footnote{The term \exitentropy, introduced in \cite{teB01},
stands for extrinsic (mutual) information 
transfer. Rather than using mutual information we opted to use entropies which
in our setting simply means one minus mutual information. It is natural to
use entropy in the setting of the binary erasure channel since the parameter $\ih$ itself
represents the channel entropy.} 
The \exitentropy curve 
associated to the $i^\text{th}$ variable is a function of the channel entropy 
and it is defined as $H(X_i \mid Y_{[\n]\setminus\{i\}})$.
Hereby, $X_i$ represents the $i^{\text{th}}$ 
input bit and, for $S\subseteq[n]\defas\{1,\dots,\n\}$, $X_S$ represents the $|S|$-tuple of 
all bits indexed by $S$.  For notational simplicity, let us write 
$X_{\sim i}=X_{[\n]\setminus\{i\}}$ when a single bit is omitted and $X=X_{[\n]}$ for the 
entire vector. The uniformly averaged quantity 
 $\frac{1}{\n}\sum_{i=1}^\n H(X_i \mid Y_{\sim i})$ is called the \exitentropy function. 
Recall that if there is a uniform prior on the set of hypotheses 
then the maximum a posteriori 
and the maximum likelihood decoding rule are identical. 
Let $\Phi^\MAP_i=\phi^\MAP_i(Y_{\sim i})$ denote the extrinsic MAP bit 
estimate (sometimes called extrinsic information) associated to the $i^{\text{th}}$ bit. This can be 
any sufficient statistics for $X_i$ given $Y_{\sim i}$. Since we 
deal with binary variables, we can always think
of it as the conditional expectation $\phi^\MAP_i(Y_{\sim i})
\defas \expectation[X_i|Y_{\sim i}]$. 
Observe that $H(X_i \mid Y_{\sim i})=H(X_i \mid \Phi^\MAP_i)$.
%
%
\subsection{Overview of  Results}
\label{OverviewSection}
\begin{figure}[htp]
\centering
\setlength{\unitlength}{0.625bp}%
\begin{picture}(360,190)
\put(0,30)
{
\put(0,0){\includegraphics[scale=0.625]{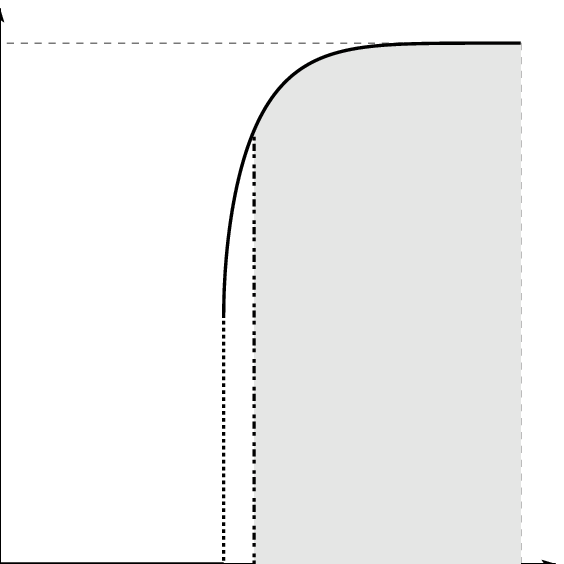}}
\put(150,-5){\makebox(0,0)[t]{\small{$1$}}}
\put(-5,150){\makebox(0,0)[tr]{\small{$1$}}}
\put(73,-5){\makebox(0,0)[tl]{\small{$\ih^\MAP$}}}
\put(64,-5){\makebox(0,0)[rt]{\small{$\ih^\BP$}}}
\put(58,90){\makebox(0,0)[r]{\small{$\xh^\BP(\ih)$}}}
}
\put(200,30)
{
\put(0,0){\includegraphics[scale=0.625]{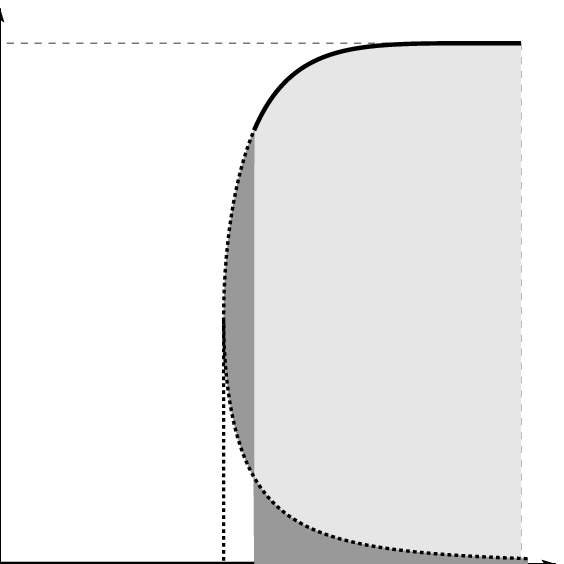}}
\put(150,-5){\makebox(0,0)[t]{\small{$1$}}}
\put(-5,150){\makebox(0,0)[tr]{\small{$1$}}}
\put(73,-5){\makebox(0,0)[lt]{\small{$\ih^\MAP$}}}
\put(79,90){\makebox(0,0)[l]{\small{$\xh^\MAP(\ih)$}}}
\put(64,-5){\makebox(0,0)[rt]{\small{$\ih^\BP$}}}
}
\put(80,0){\makebox(0,0){{\small{(a)}}}}
\put(280,0){\makebox(0,0){{\small{(b)}}}}
\end{picture}
\caption{BP and MAP \exitentropy curves for the \ddp 
$(\ledge(x)=x^2, \redge(x)=x^5)$. 
(a) BP \exitentropy curve $\xh^\BP(\ih)$: its parametric equation is stated 
in (\ref{equ:exitcurve}). It is zero until $\ih^\BP$ 
at which point it jumps. It further 
continues smoothly
until it reaches one at $\ih=1$.
(b) MAP \exitentropy curve $\xh^\MAP(\ih)$.  Note that the figure (b) includes also 
the ``spurious'' branch of Eq. (\ref{equ:exitcurve}).  The spurious branch 
corresponds to unstable fixed points. 
The MAP threshold is determined by the balance of the two dark 
gray areas.
}
\label{fig:exitcurve} 
\end{figure}

Consider a \ddp $(\ledge, \redge)$
and the corresponding sequence of ensembles $\ldpc(\n,\ledge,\redge)$ of increasing length $\n$. 
Fig.~\ref{fig:exitcurve} shows the asymptotic \exitentropy curve for the
regular \ddp $(\ledge(x)=x^2, \redge(x)=x^5)$.

Formally, this \exitentropy curve is  
$\xh^\MAP(\ih)\defas\lim_{\n\rightarrow \infty}\frac{1}{n} \sum_{i=1}^{n} H(X_i \mid Y_{\sim i}(\ih))= \lim_{\n\rightarrow \infty}\frac{1}{n} H(X_i \mid \Phi^\MAP_i)$. 
Its main 
characteristics  
are as follows:
the function is zero below the MAP threshold $\ih^\MAP$,   
it jumps at $\ih^\MAP$ to a non-zero value and
continues then smoothly until it reaches one for $\ih=1$.  
The area under
the \exitentropy curve equals the rate of the code, see \cite{AKtB02}. Compare 
this to the equivalent function of the BP 
decoder which is also shown in Fig.~\ref{fig:exitcurve}. 
The BP \exitentropy curve $\xh^\BP(\ih)\defas \lim_{\n\rightarrow \infty}\frac{1}{n} H(X_i \mid \Phi^{\BP}_i)$ 
corresponds to  running a BP decoder on a very 
large graph until the decoder has reached a fixed point. The extrinsic entropy of the bits 
at this fixed point gives the BP \exitentropy curve.
This curve is given in parametric form by
\begin{align} \label{equ:exitcurve}
\left(
\frac{\xl}{\ledge(1-\redge(1-\xl))}, \lnode(1-\redge(1-\xl))
\right),
\end{align}
where $\xl$ indicates the erasure probability of the variable-to-check messages.  
To see this, note that when transmission takes 
place over BEC($\ih$), then the BP decoder  reaches a fixed point $\xl$ which is given by 
the solution of the density evolution (DE) equation $\ih \ledge(1-\redge(1-\xl))$. 
We can therefore express
$\ih$ as $\ih(\xl)\defas\frac{\xl}{\ledge(1-\redge(1-\xl))}$.
Now the average  extrinsic probability
that a bit is still erased at the fixed point is equal to $\lnode(1-\redge(1-\xl))$. 
Note that the BP EXIT curve is the trace of this parametric equation for $\xl$
starting at
$\xl=1$ until $\xl=\xl^\BP$. This is the critical point  
and $\ih(\xl^\BP)=\ih^\BP$. 
Summarizing, the BP \exitentropy curve is zero up to the BP threshold 
$\ih^\BP$ where it jumps to a non-zero value and then continues 
smoothly until it reaches
one at $\ih=1$. 
\begin{figure}[htp]
\centering
\setlength{\unitlength}{0.625bp}%
\begin{picture}(360,190)
\put(0,30)
{
\put(0,0){\includegraphics[scale=0.625]{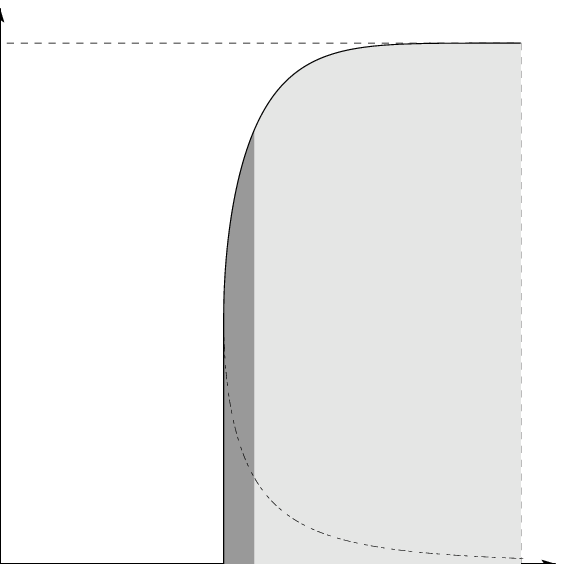}}
\put(150,-5){\makebox(0,0)[t]{\small{$1$}}}
\put(-5,150){\makebox(0,0)[tr]{\small{$1$}}}
\put(64,-5){\makebox(0,0)[tr]{\small{$\ih^\BP$}}}
\put(73,-5){\makebox(0,0)[lt]{\small{$\ih^\MAP$}}}
}
\put(200,30)
{
\put(0,0){\includegraphics[scale=0.625]{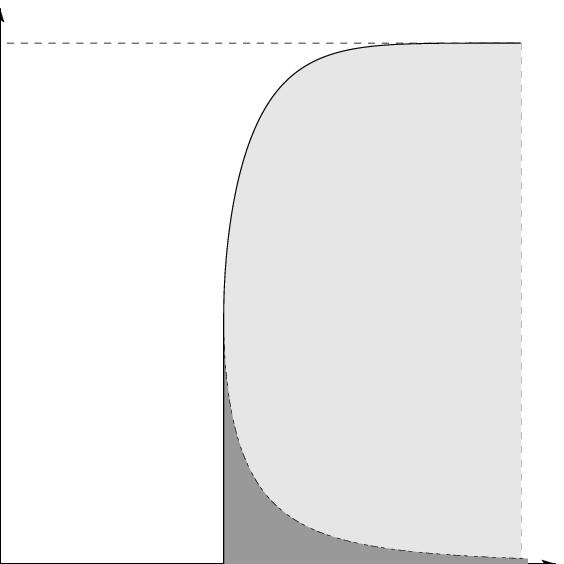}}
\put(150,-5){\makebox(0,0)[t]{\small{$1$}}}
\put(-5,150){\makebox(0,0)[tr]{\small{$1$}}}
\put(64,-5){\makebox(0,0)[rt]{\small{$\ih^\BP$}}}
\put(73,-5){\makebox(0,0)[lt]{\small{$\ih^\MAP$}}}
}
\put(80,0){\makebox(0,0){{\small{(a)}}}}
\put(280,0){\makebox(0,0){{\small{(b)}}}}
\end{picture}
\caption{ 
Balance of areas for the Maxwell decoder
between the number of guesses in (a) and the number of contradictions in (b).
The two dark gray areas are equal at the
MAP threshold. These two areas differ from the areas indicated in
Fig.~\ref{fig:exitcurve} only by a common part. }
\label{fig:exitbalance}
\end{figure}

In \cite{MeU03b} it was pointed out that
for the investigated cases the following two curious relationships between these two
curves hold: 
First, the BP and the MAP curve coincide above $\ih^\MAP$. 
Second, the MAP curve can be constructed from the 
BP curve in the following way.
If we draw the BP curve as parameterized in (\ref{equ:exitcurve}) not only
for $\xl \in [\xl^\BP,1]$ but also 
for $\xl \in (0,\xl^\BP)$ we get the 
curve shown in the right picture of Fig.~\ref{fig:exitcurve}. 
Notice that the branch for $\xl \in (0,\xl^\BP)$ corresponds to unstable
fixed points under BP decoding. Moreover, the fraction of erased 
messages $\xl$ decreases along this branch when the erasure probability 
is increased and it satisfies $\ih(\xl)>\ih$. 
Because of these  peculiar features, it is usually
considered as ``spurious''. 
To determine the MAP threshold take a vertical line at 
$\ih=\ih^\BP$ and shift it  to the right until the area 
which lies to the left of this line and is enclosed by the line
and the BP \exitentropy curve is equal to the area which lies to the right 
of the line and is enclosed by the line and the BP \exitentropy curve 
(these areas are indicated in dark gray in
the picture).
This unique point determines the MAP threshold. 
The MAP \exitentropy curve is now the curve which is zero to the left of the 
threshold and equals the iterative curve to the right
of this threshold. In other words, the MAP threshold is determined by a balance
between two areas.
It turns out that there is an operational meaning to this 
balance condition. We define the so-called Maxwell (M) decoder which performs 
MAP decoding by combining BP decoding with guessing. The dark gray areas 
in in the right picture of 
Fig.~\ref{fig:exitbalance} differ from the ones in Fig.~\ref{fig:exitcurve} 
only by a common part. 
We can show that the gray area 
on the left is connected to  
the number of ``guesses'' the \maxwell 
decoder has to venture, while 
the gray area on the right represents  the 
number of ``confirmations'' regarding these guesses. 
The MAP threshold is determined by the condition that the number of 
confirmations balances the number of guesses (i.e., that each guess
is confirmed), and therefore the two areas
are equal: in other words, at the MAP threshold (and below)
there is just a single codeword compatible 
with the channel received bits. 
\begin{floatingfigure}{152bp}
\vspace{25bp}
\centering
\setlength{\unitlength}{0.5bp}
\begin{picture}(105,153)
\put(-45,0){\includegraphics[scale=0.62]{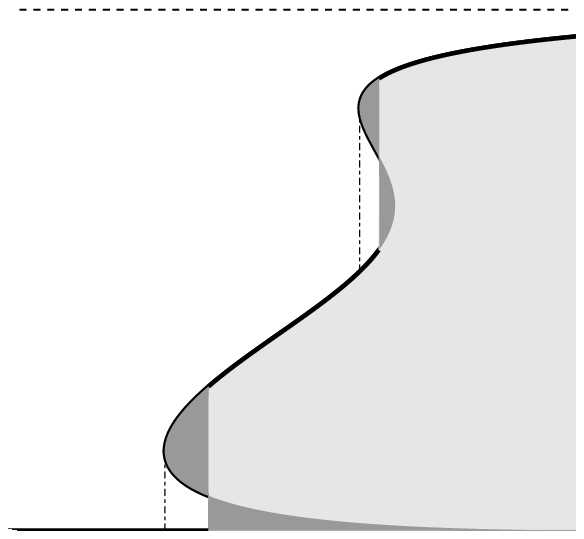}}
\put(36,-7){\makebox(0,0){\small{$\ih^\MAP$}}}
\put(7,-7){\makebox(0,0){\small{$\ih^\BP$}}}
\put(116,88){\makebox(0,0){\small{$\xh^\MAP(\ih)$}}}
\put(46,105){\makebox(0,0){\small{$\xh^\BP(\ih)$}}}
\end{picture}
\caption{\label{fig:multijump} BP (dashed and solid line) and MAP (thick solid line) \exitentropy curves for the
ensemble discussed in Examples \ref{equ:doublejumpexample} and \ref{ex:running2jumps}. 
Both curves have two jumps. The two jumps of the MAP \exitentropy curve are both
determined by a local balance of areas.}
\vspace{5bp}
\end{floatingfigure}
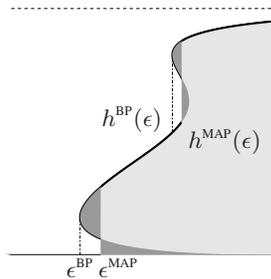

The \exitentropy curves depicted in Fig.~\ref{fig:exitcurve}
are representative for a large family of degree distributions, e.g., those of 
regular LDPC ensembles.
But more complicated scenarios are possible.
Fig.~\ref{fig:multijump} depicts a slightly more general case
in which the BP \exitentropy curve and the MAP \exitentropy curve have two jumps.
As can be seen from this figure, the same kind of balance condition
holds in this case {\em locally} and it 
determines the position of each jump. 

\subsection{Paper Outline}

We start by considering the conditional entropy $H(X \mid Y)$,
where $X$ is the transmitted 
codeword and $Y$ the received sequence, and we derive the so-called
Area Theorem for finite-length codes. 
When applying the Area Theorem to the binary erasure channel, 
the notion of \exitentropy curve enters explicitly. 
Next, we show that when the codes are chosen randomly
from a suitable defined ensemble then the 
individual conditional entropies and \exitentropy curves concentrate
around their ensemble averages. This is the first step towards
the asymptotic analysis.

We continue by defining the three asymptotic 
\exitentropy curves of interest. These are 
the (MAP) \exitentropy curve, the BP \exitentropy curve, and the EBP  
\exitentropy curve (which holds extended BP \exitentropy  and includes the spurious branch). 
We show that the Area Theorem 
remains valid in the asymptotic setting. As an immediate consequence
we will see that for some classes of ensembles (roughly
those for which the stability condition determines the threshold) 
BP decoding coincides with MAP decoding. 

We then present a key point of the paper, which is 
the derivation of an upper-bound for the MAP threshold. Several examples 
illustrate this technique and lead to suggests the tightness of the bound.

The same result is recovered through a counting argument that, 
supplemented by a combinatorial calculation, implies the tightness of 
the bound.

Finally, we introduce the so-called \maxwell decoder which 
provides a unified framework for understanding the
connection between the BP and the MAP decoder. A closer
analysis of the performance of the \maxwell decoder will allow us
to prove a refined upper bound on the MAP threshold
and it will give rise to a pleasing interpretation
of the MAP threshold as that parameter in which two areas under the 
EBP \exitentropy curve are in balance.
   
We conclude the paper by discussing some applications of our method. 
%
%
\section{Finite-Length Codes: Area Theorem and Concentration}

Let $X$ be the transmitted codeword and let $Y$ be the received word.
The conditional entropy $H(X \mid Y)$ is 
 of fundamental importance if we 
consider the question whether reliable communication is possible. Let us 
see how this quantity appears naturally in the context of decoding. 
To this end,  
we first recall the original Area Theorem 
as introduced in \cite{AKtB02}.
\btheo[Area Theorem]
\label{theo:areatheorem0}
Let $X$ be a binary vector of length $n$ chosen with probability $p_X(x)$ from a finite set.
Let $Y$ be the result of passing $X$ through  BEC$(\cp)$.
Let $\Omega$ be a further observation of $X$ so that 
$p_{\Omega \mid X, Y}(\omega \mid x, y) = p_{\Omega \mid X}(\omega \mid x)$.
To emphasize that $Y$ depends on the channel parameter $\cp$ we write $Y(\cp)$. Then
\begin{align}
\label{equ:areatheorem0}
\frac{H(X \mid \Omega)}{\n} =
\int_0^1\frac{1}{\n}\sum_{i\in[\n]} H(X_i \mid Y_{\sim i}(\cp), \Omega)\text{d}\cp.
\end{align}
\etheo

The reader familiar with the original statement in \cite{AKtB02} will
have noticed that we have rephrased the theorem. First, 
we expressed the result in terms of entropy instead of mutual information. 
Second, the observations $Y$ and $\Omega$ represent what in the original theorem
were called the ``extrinsic'' information and the ``channel,'' respectively.
 
In (\ref{equ:areatheorem0}) 
the integration ranges from zero (perfect channel)
to one (no information conveyed). The following is a trivial extension. 
\btheo[Area Theorem]
\label{theo:areatheorem1}  
Let $X$ be a binary vector of length $\n$ chosen with probability $p_X(x)$ from a finite set.
Let $Y$ be the result of passing $X$ through  BEC$(\cp)$.
Let $\Omega$ be a further observation of $X$ so that
$p_{\Omega \mid X, Y}(\omega \mid x, y) = p_{\Omega \mid X}(\omega \mid x)$. Then
\begin{align*}
\frac{H(X \mid Y(\cp^*), \Omega)}{\n} =
\int_0^{\cp^*}\frac{1}{\n}\sum_{i\in[\n]}H(X_i \mid Y_{\sim i}(\cp), \Omega)\text{d}\cp.
\end{align*}
\etheo

{\em Proof of Theorem~\ref{theo:areatheorem1}:}
Let $Y^{(1)}$ be the result of passing $X$ through BEC($\cp$)
and $Y^{(2)}$ be the result of passing $X$ through BEC($\cp^*$).
Let $\Omega$ be the additional observation of $X$. Applying Theorem~\ref{theo:areatheorem0},
with $Y=Y^{(1)}$ and with additional observation $(Y^{(2)}, \Omega)$, we 
 have $p_{\Omega, Y^{(2)} \mid X, Y^{(1)}}(\omega, y^{(2)} \mid x, y^{(1)}) = 
p_{\Omega, Y^{(2)} \mid X}(\omega, y^{(2)} \mid x)$, as required, so that we get  
\begin{align*}
H(X \mid Y^{(2)}(\cp^*), \Omega) =
\int_0^1\sum_{i\in[\n]} 
H(X_i \mid Y^{(1)}_{\sim i}(\cp), Y^{(2)}(\cp^*), \Omega)\text{d}\cp.
\end{align*}
Now note that 
\begin{align*}
H(X_i \mid Y^{(1)}_{\sim i}(\cp), Y^{(2)}(\cp^*), \Omega) & =
\cp^* H(X_i \mid Y_{\sim i}(\cp \cp^*), \Omega).
\end{align*}
This is true since the bits of $Y^{(1)}_{\sim i}(\cp)$ and $Y^{(2)}(\cp^*)$
are erased independently (so that the respective erasure probabilities multiply) and since $Y^{(2)}(\cp^*)$
contains the intrinsic observation of bit $X_i$, which is erased with probability $\cp^*$.
If we now substitute the right hand side of the last expression in our previous integral and
make the change of variables $\cp' = \cp \cdot \cp^*$, 
Theorem \ref{theo:areatheorem1} 
follows.
\qed

Assume that we allow each $X_i$ to be passed through a different channel  BEC$(\cp_i)$.
Rather than phrasing our result specifically for the 
case of the BEC$(\cp_i)$, let us
state the area theorem right away in its general form
as introduced in \cite{MMRU04}. 
In this paper we will only be interesting
in the consequences as they pertain to transmission over the BEC$(\cp)$.
The investigation of the general case is relegated to the companion paper 
\cite{MMRU05}.

In order to state this and subsequent results in a more compact form we introduce the
following definition. 
\bdefi[Channel Smoothness]
Consider a family of memoryless channels with input and output alphabets ${\cal X}$ 
and ${\cal Y}$, respectively, and characterized by their transition 
probability distribution functions (pdf's) 
$p_{Y \mid X}(y \mid x)$. If ${\cal Y}$ is discrete, we interpret
$p_{Y \mid X}(\cdot \mid x)$ as a pdf with respect to the counting measure. 
If ${\cal Y}$ is continuous, $p_{Y \mid X}(y \mid x)$ is a density with respect
to Lebesgue measure.
Assume that the family is parameterized by $\cp$, 
where $\cp$ takes values in some interval $I \subseteq {\mathbb R}$.
The channel is said to be {\em smooth} 
with respect to the parameter $\cp$  
if the pdf's $\{p_{Y \mid X}(y \mid x): x \in {\cal X}, \, y \in{\cal Y}\}$
are differentiable functions of $\cp \in I$.
\edefi
Notice that, if a channel family is smooth, then several basic properties 
of the channel are likely to be differentiable with respect to the channel
parameter. A basic (but important) example is the channel conditional
entropy $H(Y|X) = \expectation[-\log \{p_{Y \mid X}(Y \mid X)\}]$
given a reference measure $p_{X}(x)$ on ${\cal X}$. Suppose that 
${\cal Y}$ is finite, and that, for any $\cp\in I$, $p_{Y|X}(y|x)>0$
for any $x \in {\cal X}, \, y \in{\cal Y}$. Then 
\begin{align*}
\frac{\de H(X|Y)}{\de \cp} &= \sum_{x,y}p_X(x)\log \left(
\frac{1}{p_{Y|X}(y|x)}\right)\,
\frac{\de p_{Y|X} }{\de \cp}(y|x)\, .
\end{align*}
In other words, differentiability of $H(Y|X)$ follows from differentiability
of $p_{Y|X}(y|x)$ and of $-x\log x$. 
In this paper we consider families of binary erasure channels which are trivially 
smooth with respect to the parameter $\cp$. 
 
\btheo[General Area Theorem-\cite{MMRU04}] 
\label{theo:generalareatheorem}
Let $X$ be a binary vector of length $n$ chosen with probability $p_X(x)$ from a finite set.
Let the channel from $X$ to $Y$ be memoryless, where 
$Y_i$ is the result of passing $X_i$ through a smooth channel with parameter $\cp_i$, $\cp_i \in I_i$.
Let $\Omega$ be a further observation of $X$ so that
$p_{\Omega \mid X, Y}(\omega \mid x, y) = p_{\Omega \mid X}(\omega \mid x)$.
Then
\begin{align} 
\label{eq:generalareatheorem}
\text{d} H(X \mid Y,\Omega) & = 
\sum_{i=1}^\n \frac{\partial H(X_i \mid Y,\Omega ) }{\partial \cp_i} \text{d}\cp_i.
\end{align}
\etheo
\bproof 
For $i \in [\n]$, the entropy rule gives $H(X \mid Y,\Omega)=H(X_i \mid Y,\Omega)+H(X_{\sim i} \mid X_i,Y,\Omega)$.
We have 
$p_{X_{\sim i} \mid X_i,Y,\Omega} = p_{X_{\sim i} \mid X_i,Y_{\sim i},\Omega}$ 
 since the channel is memoryless and $p_{\Omega \mid X, Y} = p_{\Omega \mid X}$. 
Therefore, $H(X_{\sim i} \mid X_i,Y,\Omega)=H(X_{\sim i} \mid X_i,Y_{\sim i},\Omega)$ and 
$\frac{\partial H(X \mid Y,\Omega ) }{\partial \cp_i} = \frac{\partial H(X_i \mid Y,\Omega) }{\partial \cp_i}$.
From this the total derivate as stated in  
(\ref{eq:generalareatheorem}) follows
immediately.
\eproof

{\em{\hspace{16pt} Alternative proof of Theorem~\ref{theo:areatheorem1}:}}  
Keeping in mind that transmission takes place over a binary erasure channel, we write
\begin{align*}
H(X_i \mid Y, \Omega)&= \sum_{y_i\in\{\0,\?,\1\}}p_{Y_i}(y_i)
H(X_i\mid  Y_i=y_i,Y_{\sim i}, \Omega)\, .
\end{align*}
The terms corresponding to $y_i\in\{\0,\1\}$ vanish because $X_i$ is then completely 
determined by the channel output. The remaining term yields
$H(X_i \mid Y, \Omega) = \epsilon_i H(X_i\mid  Y_{\sim i}, \Omega)$,
%
because $p_{Y_i}(\?) =\epsilon_i$, and the occurrence at the channel output of an erasure 
at position $i$ is independent from $X$, $Y_{\sim i}$ and $\Omega$.
 We can then write 
\begin{align*}
\text{d} H(X \mid Y (\cp),\Omega) & = 
\sum_{i \in [\n]} \frac{\partial H(X_i \mid Y,\Omega ) }{\partial \cp_i} \text{d} \cp_i \\
& = \sum_{i \in [\n]}  H(X_i \mid  Y_{\sim i},\Omega) \text{d} \cp_i,  
\end{align*}
 which, when we assume that $\cp_i=\cp$ for all $i \in [n]$, gives Theorem~\ref{theo:areatheorem1}.\qed
 
A few remarks are in order.  
First, the additional degree of freedom afforded by allowing
an extra observation $\Omega$ is useful when studying the 
dynamical behavior of certain iterative coding schemes via 
\exitentropy chart arguments.  
(For example, in a parallel concatenation, 
$Y$ typically represents the observation of the systematic bits 
and $\Omega$ represents the fixed channel observation of the 
parity bits.)  
For the purpose of this paper however, the additional observation
$\Omega$ is not needed since we are not concerned by componentwise  
\exitentropy charts. We will therefore skip $\Omega$ in the sequel. 
Second, as emphasized in the last step in the previous proof, we 
 can assume at this point, more generally, that the individual channel
parameters $\cp_i$ are not the same but that the individual 
channels  are all 
 parametrized by a common parameter $\cp$.
For instance one may think of a families  $\{$BEC($\ih_i$)$\}$
where $\ih_i(\ih)$ are smooth functions of $\ih\in[0,1]$.
In the simplest case some parameter might be chosen to be constant.
This degree of freedom allows for an elegant proof of
Theorem \ref{theo:ebpareatheorem}. 

One of the main aims of this paper is to investigate the MAP performance
of sparse graph codes in the limit of large blocklengths.  Our task is
made much easier by realizing that we can restrict our study to the
{\em average} such performance. More precisely, let $\graph$ be chosen
uniformly at random from $\ldpc(\ledge,\redge,\n)$ and let $H_\graph(X
\mid Y)$ denote the conditional entropy for the code $\graph$.
We state the following theorems right away 
for general binary memoryless symmetric (BMS) channels.
\btheo[Concentration of Conditional Entropy] 
\label{theo:concentrationml} 
Let $\graph$ be chosen uniformly at random from $\ldpc(\n,\ledge,\redge)$.
Assume that $\graph$ is used to transmit over a BMS channel.
By some abuse of notation, let
$H_{\graph(n)} =H_\graph(X \mid Y)$ be the associated conditional entropy.
Then for any $\xi > 0$
\begin{align*}
\Pr \left\{|H_{\graph(n)}-\expectation\left[H_{\graph(n)}\right]|>\n \xi\right\} &\leq 2\, 
\text{e}^{-\n B \xi^2},\\
\end{align*}
where $B= 1/(2 (\drmax+1)^2(1-r))$ and where $\drmax$ is the maximal check-node degree.
\etheo
\bproof 
The proof uses the standard technique of first constructing a Doob's martingale
with bounded differences and then applying the Hoeffding-Azuma inequality. 
The complete proof can be found in 
\cite{Mon04} and it is reported in an adapted and streamlined form in
Appendix~\ref{app:ConcentrationProofs}.
\eproof
Let us now consider the concentration of the MAP \exitentropy curve.
For the BEC this curve is given equivalently by 
$\frac{1}{\n} \sum_{i=1}^\n H_{\graph(n)}(X_i \mid Y_{\sim i}(\ih))$ or
by $\frac{1}{\n} H_{\graph(n)}'(X \mid Y(\ih))$. We 
choose the second representation and phrase the statement in terms 
of the derivative of the conditional entropy with respect 
to the channel parameter $\ih$.
\btheo[Concentration of MAP \exitentropy Curve] 
\label{theo:concentrationexitml} 
Let $\graph$ be chosen uniformly at random from $\ldpc(\n,\ledge,\redge)$
and let $\{\text{BMS}(\cp)\}_{\cp \in I}$ denote a family of BMS channels ordered by physical degradation (with $\text{BMS}(\cp')$ physically degraded 
with respect to $\text{BMS}(\cp)$ whenever $\cp'>\cp$)
and smooth with respect to $\cp$.
Assume that $\graph$ is used to transmit over the BMS$(\cp)$ channel.
Let $H_{\graph(n)} =H_\graph(X \mid Y)$ be the associated conditional entropy.
Denote by $H'_{\graph(n)}$ the derivative of 
$H_{\graph(n)}$ with respect to $\cp$ (such a derivative exists because of the explicit 
calculation presented in Theorem \ref{theo:generalareatheorem}) and 
let $J \subseteq I$ be an interval on which 
$\lim_{n\to\infty} \frac{1}{n}\expectation \left[H_{\graph(n)}\right]$ 
exists and is differentiable with respect 
to $\cp$. Then, for any $\epsilon\in J$ and $\xi>0$ there exist an 
$\alpha_{\xi}>0$ such that, for $n$ large enough
\begin{align*}
\Pr \left\{|H'_{\graph(n)}-\expectation [H'_{\graph(n)} ]|>\n \xi\right\} &\leq 
 \text{e}^{-\n \alpha_{\xi}}.
\end{align*}
Furthermore, if 
$\lim_{n\to\infty} \frac{1}{n}\expectation \left[H_{\graph(n)}\right]$
is twice differentiable with respect 
to $\cp\in J$, there exists a strictly positive constant $A$ such that $\alpha_\xi>A\xi^4$.
\etheo
The proof is deferred once more to Appendix~\ref{app:concentrationproofs}. 

Notice the two extra hypothesis with respect to
Theorem~\ref{theo:concentrationml}. First, we assumed that the channel
family $\{\text{BMS}(\cp)\}_{\cp \in I}$ is ordered by physical
degradation. This ensures that $H'_n$ is non-negative.  This condition
is trivially satisfied for the family $\{\text{BEC}(\cp)\}_{\cp \in
[0, 1]}$. More generally, we can let $\cp$ be any function of the
erasure probability differentiable and increasing
from zero to one.  The second condition, namely the
existence and differentiability of the expected entropy per bit in the 
limit, is instead crucial. As discussed in the previous section (see, e.g., 
Fig.~\ref{fig:exitcurve}), the asymptotic \exitentropy curve may have jumps. By
Theorem~\ref{theo:areatheorem1} these jumps correspond to discontinuities
in the derivative of the conditional entropy. At a jump $\cp_*$, the
value of the \exitentropy
curve may vary dramatically when passing from one element of the ensemble
to the other. Some (a finite fraction) of the codes will perform well,
and have an \exitentropy curve close to the asymptotic value at $\cp_*-\delta$,
while others (a finite fraction) may have an \exitentropy function close to the
asymptotic value at $\cp_*+\delta$ ($\delta$ is here a generic small
positive number).

\btheo[Concentration of BP \exitentropy Curve] 
\label{theo:concentrationexitbp} 
Let $\graph$ be chosen uniformly at random from $\ldpc(\n,\ledge,\redge)$.
Assume that $\graph$ is used to transmit over a BMS channel and let 
$\Phi_i^{\BP, \iter}= \phi_i^{\BP, \iter}(Y_{\sim i})$ denote the 
extrinsic estimate (conditional mean)
of $X_i$ produced by the BP decoder after $\iter$ iterations.
Denote by $H_{\graph,i}^{\BP, \iter} = 
H_\graph \bigl(X_i \mid \Phi_i^{\BP, \iter} \bigr)$ the 
resulting (extrinsic) entropy of the binary variable $X_i$.
Then, for all  $\xi>0$, there exists $\alpha_\xi>0$, such that
\begin{multline}
\Pr\Bigl\{\Bigl|\sum_{i=1}^\n \Bigl(H_{\graph,i}^{\BP, \iter} - 
\expectation_\graph \Bigl[ H_{\graph,i}^{\BP, \iter} \Bigr] \Bigr)\Bigr|> n \xi\Bigr\} 
\leq \text{e}^{-\alpha_\xi \n}.
\end{multline}
\etheo

\bproof 
The proof is virtually identical to the ones given in \cite{LMSS01,RiU01} where the
probability of decoding error is considered. 
\eproof
%
%

\section{Asymptotic Setting}
 
\subsection{(MAP) \exitentropy}

The next definition and theorem define our main object of study. 
\bdefi \label{def:asymptoticexit}
Let ${\cal C}(n)$ be a sequence of code ensembles of diverging blocklength $\n$
and let $\graph(n)$ be chosen uniformly at random from ${\cal C}(n)$.
Assume that $\lim_{n \rightarrow \infty} \expectation_{\graph}\left[\frac{1}{\n}
\sum_{i=1}^\n H_{\graph(n)}'(X \mid Y(\ih))\right]$ exists.
Then this limit is called the asymptotic \exitentropy function 
of the family of ensembles and
we denote it by $\xh^\MAP(\ih)$.  We define the MAP threshold
$\cp^\MAP$ to be the supremum of all values $\cp$ such that $\xh^\MAP(\cp) = 0$.
\edefi
Given a \ddp $(\ledge,\redge)$, consider the sequence of 
ensembles $\{\ldpc(\ledge,\redge,\n)\}_{\n}$. 
It is natural to conjecture that the associated asymptotic 
\exitentropy function exists.
Note that from Theorem \ref{theo:concentrationexitml} we know 
that if this limit exists, then individual code
instances are closely concentrated around the ensemble average.
It is therefore meaningful to define in such a setting the
MAP threshold in terms of the ensemble average. 

Unfortunately, no {\em general} proof of the existence of the 
MAP \exitentropy curve is known.
But we will show how one 
can in {\em most cases} compute the asymptotic \exitentropy function explicitly for a given ensemble,
thus proving existence of the limit in such cases. 
See also \cite{ZC95} for a discussion on asymptotic thresholds. 

It is worth pointing out that we defined the MAP threshold to be
the channel parameter at which the conditional entropy becomes sublinear.
At this point the average conditional bit entropy converges to zero, so
that this point is the bit MAP threshold. We note that for some ensembles the
block MAP threshold is strictly smaller than the bit MAP threshold.

\btheo[Asymptotic Area Theorem] 
\label{theo:asymptoticareatheorem} 
Consider a \ddp $(\ledge,\redge)$. Assume that the associated asymptotic \exitentropy function 
as defined in  Definition~\ref{def:asymptoticexit} exists for all 
$\cp\in[0,1]$. 
Assume further that the limit  
$\drate_{\text{as}}   = 
\lim_{\n \rightarrow \infty}\expectation_{\graph} \Bigl[\frac{H(X)}{\n} \Bigr]$
exists.
Then 
\begin{align*}
\drate_{\text{as}} & =
\int_0^{1} \xh^\MAP(\ih)\mbox{d} \ih.
\end{align*}
\etheo
\bproof 
Let $\xh^\MAP_{\graph(\n)}(\ih)$ denote the \exitentropy function associated to a particular   
$\graph \in\ldpc(\ledge,\redge,\n)$ with rate $r_{\graph(\n)}$. 
We have
\begin{align}
\notag \int_0^1 \expectation_{\graph}\left[\xh^\MAP_{\graph(\n)}(\ih)\right]   \text{d}\ih & =\expectation_{\graph}\left[\int_0^1 \xh^\MAP_{\graph(\n)}(\ih)  \text{d}\ih \right] \\
\label{eq:fubbini} & = \expectation_{\graph}\Bigl[\frac{H(X)}{\n}\Bigr] ~~{\underset{\n \to \infty }{\longrightarrow}}~~ \drate_{\text{as}}
\end{align}  
The first equality is obtained by noticing that
the function $\xh^\MAP_{\graph(\n)}(\ih)$ is  non-negative. We are therefore
justified by Fubini theorem to switch the order of integration.
The second step follows from the Area Theorem (the rate being equal to $\frac{H(X)}{\n}$).

On the other hand, the Dominated Convergence Theorem can be applied to the sequence 
 $\left\{\expectation_{\graph}\left[\xh^\MAP_{\graph(\n)}(\ih)\right]\right\}_{\n}$ since it
converges (as assumed in the hypothesis) to $\xh^\MAP(\ih)$ 
and is trivially upper-bounded by 1. We therefore get
\begin{align}
\notag \lim_{\n \to \infty} \int_0^1 \expectation_{\graph}\left[\xh^\MAP_{\graph(\n)}(\ih)\right]   
\text{d}\ih &= \int_0^1 \lim_{\n \to \infty}  
\expectation_{\graph}\left[\xh^\MAP_{\graph(\n)}(\ih)\right]   \text{d}\ih\\
\notag & = \int_0^1 \xh^\MAP(\ih) \text{d} \ih.
\end{align}
which, combined with (\ref{eq:fubbini}), concludes the proof. 
\eproof

Lemma \ref{WELemma} gives a sufficient condition for the limit 
$\drate_{\text{as}}$ to exists. Note that under this condition
the asymptotic rate $\drate_{\text{as}}$ is equal to 
to the design rate $\drate(\ledge,\redge)$.
Most \ddps $(\ledge,\redge)$ encountered in practice fulfill this condition.
This condition is therefore not
very restrictive.

%
%
\subsection{BP \exitentropy} \label{sec:bpexitfunction}
Recall that the MAP \exitentropy curve can be expressed as 
$H(X_i \mid \Phi^\MAP_i)$ where $\Phi^\MAP_i = \phi_i^\MAP(Y_{\sim i})$
is the posterior estimate (conditional mean) of $X_i$ given $Y_{\sim i}$.
Unfortunately this quantity is not easy to evaluate.
In fact, the main aim of this paper is to accomplish this task.

\begin{floatingfigure}{115bp}
\centering
\setlength{\unitlength}{0.5bp}
\begin{picture}(210,230) 
\put(19,10)
{
\put(0,0){\includegraphics[scale=0.62]{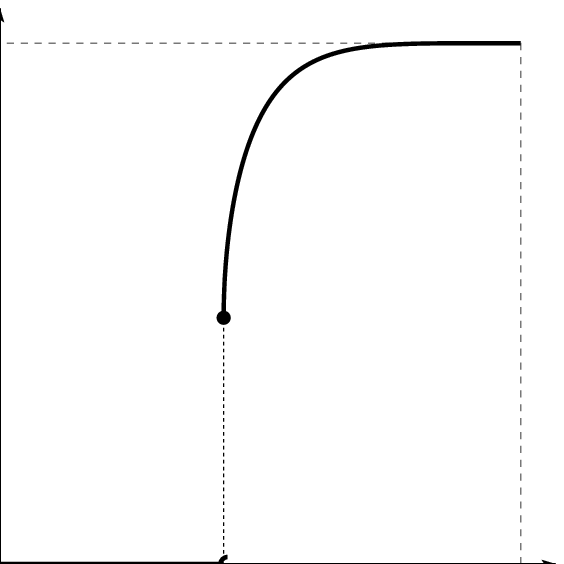}}
\setlength{\unitlength}{0.62bp}
\put(7,177){\makebox(0,0){\small{$\xh^\BP$}}}
\put(170,0){\makebox(0,0){\small{$\ih$}}}
\put(-8,-8){\makebox(0,0){\small{$0$}}}
\put(153,-8){\makebox(0,0){\small{$1$}}}
\put(-8,154){\makebox(0,0){\small{$1$}}}
}
\end{picture}
\caption{\label{fig:bpexit} BP \exitentropy function $\ih \mapsto \xh^\BP(\ih)$.}
\vspace{2bp}
\end{floatingfigure}
A related quantity which is much easier to compute is 
the BP \exitentropy curve shown in Fig.~\ref{fig:bpexit} 
for the \ddp $(\xl^2,\xl^5)$. The BP \exitentropy corresponds to
$H(X_i \mid \Phi_i^\BP)$, where $\Phi_i^\BP = \phi_i^\BP(Y_{\sim i})$
is the extrinsic estimate of $X_i$ delivered by the BP decoder
Here a fixed number of iterations, let us say $\iter$, is understood.
Asymptotically, we consider $\iter\to\infty$ {\em after} $\n\to\infty$.
An exact expression for the average asymptotic BP \exitentropy curve for LDPC ensembles 
is easily computed via the DE method  \cite{LMSS01,LMSS01b,RiU01,RSU01}. 

\begin{figure}[htp]
\centering
\setlength{\unitlength}{1bp}
\begin{picture}(240,140)
\put(5,28)
{
\put(5,5){\includegraphics[width=100bp]{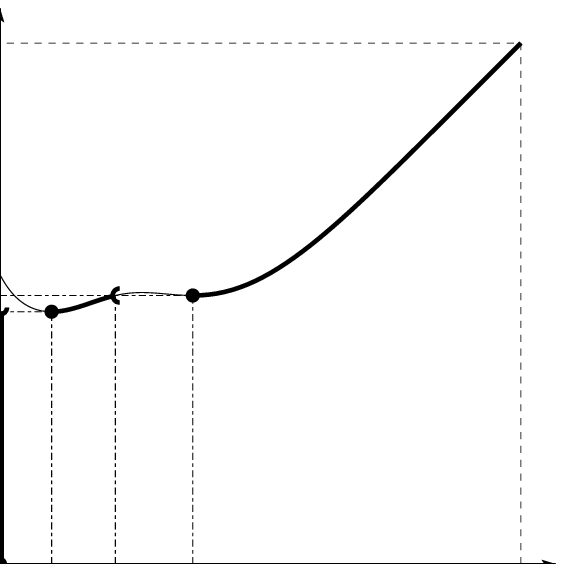}}
\put(135,5){\includegraphics[width=100bp]{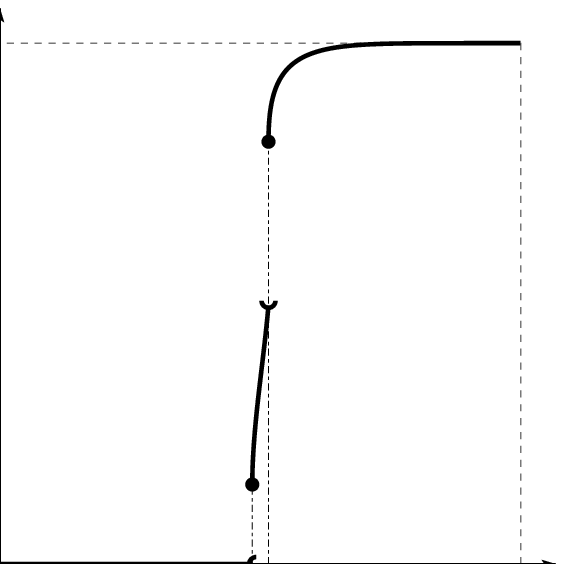}}
\put(0,0){\makebox(0,0){\small{$0$}}}
\put(100,0){\makebox(0,0){\small{$1$}}}
\put(0,100){\makebox(0,0){\small{$1$}}}
\put(130,0){\makebox(0,0){\small{$0$}}}
\put(230,0){\makebox(0,0){\small{$1$}}}
\put(130,100){\makebox(0,0){\small{$1$}}}
\put(7,-6){\makebox(0,0){\small{$\lxl^0$}}}
\put(16,-6){\makebox(0,0){\small{$\lxl^1$}}}
\put(28,-5){\makebox(0,0){\small{$\uxl^1$}}}
\put(42,-6){\makebox(0,0){\small{$\lxl^2$}}}
\put(101,-6){\makebox(0,0){\small{$\uxl^2$}}}
\put(0,48){\makebox(0,0){\small{$\ih^1$}}}
\put(0,57){\makebox(0,0){\small{$\ih^2$}}}
\put(60,80){\makebox(0,0){\small{$\ih(\xl)$}}}
\put(178,-5){\makebox(0,0){\small{$\ih^\BP$}}}
\put(177,13){\makebox(0,0){\small{$\ih^1$}}}
\put(180,68){\makebox(0,0){\small{$\ih^2$}}}
\put(200,80){\makebox(0,0){\small{$\xh^\BP(\ih)$}}}
}
\put(60,7){\makebox(0,0){{\small{(a)}}}}
\put(190,7){\makebox(0,0){{\small{(b)}}}}
\end{picture}
\caption{
BP \exitentropy curve with two discontinuities ($J$=2): 
(a) Channel entropy function $\xl\mapsto\ih(\xl)$ (b) BP \exitentropy function $\ih\mapsto\xh^\BP(\ih)$. 
This example corresponds to the \ddp $(\ledge,\redge)=(0.3x+0.3x^2+0.4x^{13},x^6)$, which has 
design rate $\drate\approx0.48718$. 
The BP threshold is $\ih^\BP\approx0.48437$ at $\xl^\BP\approx0.09904$. 
This is also the first discontinuity, i.e.,  $\ih^1=\ih^\BP$, $\lxl^1=\xl^\BP$ and $\uxl^1\approx0.22156$. 
The second discontinuity occurs for $\ih=\ih^2\approx0.51553$ at  $\lxl^2\approx0.37016$ ($\uxl^1=1$).
}
\label{fig:epsilon}
\end{figure}

Consider the fixed-point condition for the density evolution equations,
\begin{align*}
\ih \ledge(1-\redge(1-\xl)) & =  \xl.
\end{align*}
Solving for $\ih$, we get $\ih(\xl)\defas\frac{\xl}{\ledge(1-\redge(1-\xl))},~\xl\in(0,1]$.
In words, for each non-zero fixed-point $\xl$ of density evolution, there is a unique channel
parameter $\ih$.  At this fixed-point the asymptotic average BP \exitentropy function equals 
$\lnode(1-\redge(1-\xl))$.
If $\cp(\xl)$ is monotonically increasing in $\xl$ over the whole range $[0, 1]$, 
then the BP \exitentropy curve is given in parametric form by
\begin{equation}
\label{equ:bpexitcurve}
\bigl(
\cp(\xl), \lnode(1-\redge(1-\xl)) \bigr).
\end{equation}
For some ensembles (e.g., regular cycle-code ensembles) $\cp(\xl)$ is indeed 
monotone increasing over the whole range $[0, 1]$, but for most ensembles this
is not true.  In this case we have to restrict the above 
parameterization to the unique union of intervals
\[
{\cal{I}}\defas\bigcup_{i\in[J]}[\lxl^i,\uxl^i)\cup\{1\},
\]
which has the property that $\cp(\xl)$ is continuously and monotonically
increasing from $\cp^\BP$ to one as $\xl$ takes on increasing values in ${\cal I}$
and for all $i \in [J]$, $\lxl^i=0$ or $\cp'(\lxl^i)=0$. An example of such a partition is
shown in Fig.~\ref{fig:epsilon}. That such a partition exists and is unique
follows from the fact that $\cp(\xl)$ is a differentiable function for $\xl \in [0, 1]$
as can be verified by direct computation.
Set $\uxl^J=1$ and note that $\cp(1)=1 \geq 0$.
Define $\lxl^J$ as the largest nonnegative value of $\xl \leq \uxl^J$ for which $\cp'(\xl)=0$.
If no such value exists then $\cp(\xl)$ is monotonically increasing over the whole
range $[0, 1]$. In this case $J=1$ and we set $\lxl^J=0$.
Now proceed recursively. Assume that the intervals $[\lxl^{i+1}, \uxl^{i+1})$
have been defined and that $\lxl^{i+1}>0$. 
Define $\uxl^i$ as the largest nonnegative value of $\xl < \lxl^{i+1}$
such that $\cp(\xl)=\cp(\lxl^{i+1})$. Note that if such a value exists then
we must have $\cp'(\xl) \geq 0$. If no such value exists then we have already found the sought
after partition and we stop. Otherwise define 
$\lxl^i$ as the largest nonnegative value of $\xl \leq \uxl^i$ for which $\cp'(\xl)=0$. 
As before, if no such value exists then set $\lxl^i=0$ and stop. 
Without loss we can eliminate from the resulting partition any interval of zero length.
Let $J$ denote the number of remaining intervals of nonzero length. Note, if the BP threshold 
happens at a discontinuous phase transition (jump), then $\xl^\BP=\lxl^1$
and $\cp^\BP=\cp(\lxl^1)$, otherwise, if the BP threshold is given by the stability condition, 
then  $\xl^\BP=\lxl^0=0$
and $\ih^\BP=\ih(\lxl^0)$. See also Fig. \ref{fig:epsilonwith1JandStabCond}.
\bcor 
\label{cor:asymptoticbpexit} 
Assume we are given a \ddp $(\ledge,\redge)$ and that transmission takes place over the BEC. 
Let ${\cal{I}}\defas\bigcup_{i\in[J]}[\lxl^i,\uxl^i)\cup\{1\}$ be the partition associated to 
$(\ledge,\redge)$. Define $\cp^\BP=\cp(\lxl^1)$. 
Then the BP \exitentropy function $\xh^\BP(\ih)$ is equal to zero for $0 \leq \cp < \cp^\BP$ and
for $\cp > \cp^\BP$ it has the parametric characterization
\begin{align*}
(\cp(\xl),\lnode(1-\redge(1-\xl))),
\end{align*}
where $\xl$ takes on all values in ${\cal I}$.
\ecor

\bprop[Regular LDPC Ensembles  ``Jump'' at Most Once]
\label{fact:regular} Consider the 
regular \ddp $(\ledge(x),\redge(x))=(x^{\ldegree-1},x^{\rdegree-1})$. Then the function 
$\ih(\xl)\defas\frac{\xl}{\ledge(1-\redge(1-\xl))}$ has a unique minimum in the range $[0, 1]$.
Let  $\xl^\BP$ denote the location of this minimum. Then $\ih(\xl)$ is
strictly decreasing on  $(0,\xl^\BP)$ 
 and strictly increasing on $(\xl^\BP,1)$. 
Moreover, $\xl^\BP=0$ if and only if $\ldegree=2$.
\eprop
\bproof
Note that $\cp(1)=1$ and by direct calculation we see that $\cp'(1)=1$. 
Therefore, either $\cp(\xl)$ takes on its minimum value within the interval $[0, 1]$
for $\xl=0$ or its minimum value is in the interior of the region $[0, 1]$.
Computing explicitly the derivative of $\cp(\xl)$, 
we see that the location of the minima of $\cp(\xl)$ must be a root of
$W(\xl)\defas 1-(1-\xl)^{\rdegree-1}-(\ldegree-1)(\rdegree-1)(1-\xl)^{\rdegree-2}\xl$. Furthermore 
 $W'(\xl) = -(\rdegree-1)(1-\xl)^{\rdegree-3}\{(\ldegree-2)-
[(\ldegree-1)(\rdegree-1)-1]\xl\}$.
Notice that $W(0) = 0$, $W'(0) = -(\rdegree-1)(\ldegree-2)<0$ and $W(1) = 1$.
By the Intermediate Value Theorem, $W(\xl)$ vanishes at least once in $(0,1)$.
Suppose now that $W(\xl)$ vanishes more than once in $(0,1)$, and consider
the first two such zeros $\xl_1$, $\xl_2$. It follows that 
$W'(\xl)$ must vanish at least twice: once in $(0,\xl_1)$ and once 
in $(\xl_1,\xl_2)$. On the other end, the above explicit expression
implies that $W'(\xl)$ vanishes just once in $(0,1)$, at 
$\xl =   (\ldegree-2)/[(\ldegree-1)(\rdegree-1)-1]$.
Therefore $W(\xl)$ has exactly one root in $(0,1)$. See also \cite{BRU04}. 
\eproof
A dynamic interpretation of the convergence of the BP decoding when the number of 
iterations $\iter\to\infty$ 
 is shown in Appendix \ref{app:gapinterpretation} using component EXIT curves. It is 
further shown in Appendix \ref{app:areass} and  Theorem \ref{theo:bpintegrationfact}
how to compute the area under the BP EXIT curve. 
The calculations show that this area is always 
larger or equal the design rate. Moreover, some calculus 
reveals that, whenever the BP EXIT function has discontinuities, then the 
area is strictly larger than the design rate $\drate$.

\subsection{Extended BP \exitentropy Curve}
%

\begin{figure}[htp]
\centering
\setlength{\unitlength}{0.5bp}
\begin{picture}(210,230) 
\put(19,10)
{
\put(0,0){\includegraphics[scale=0.62]{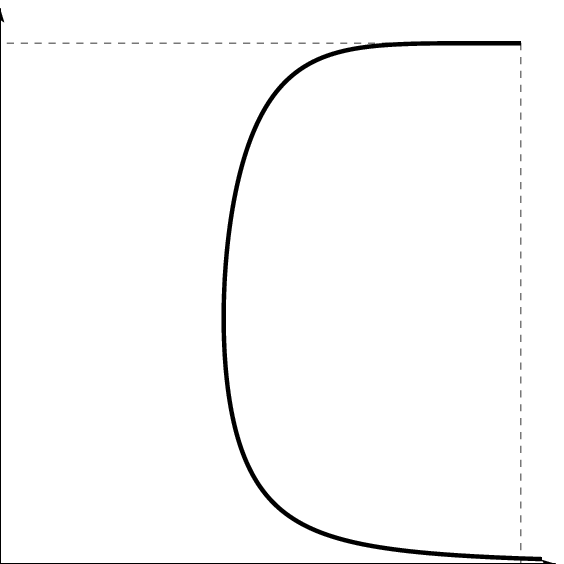}}
\setlength{\unitlength}{0.62bp}
\put(7,177){\makebox(0,0){\small{$\xh^\EBP$}}}
\put(170,0){\makebox(0,0){\small{$\ih$}}}
\put(-8,-8){\makebox(0,0){\small{$0$}}}
\put(153,-8){\makebox(0,0){\small{$1$}}}
\put(-8,154){\makebox(0,0){\small{$1$}}}
}
\end{picture}
\caption{EBP \exitentropy function $\left\{(\ih(\xl),\lnode(\xr(\xl)))\right\}_\xl$.}
\label{fig:ebpexit}
\vspace{2bp}
\end{figure}

Surprisingly, we can apply the Generalized Area Theorem also to BP decoding if 
we  consider the {\em Extended} BP \exitentropy (EBP) curve. 
Fig.~\ref{fig:ebpexit} shows this
EBP \exitentropy curve for the running example, i.e.,  for the \ddp $(\xl^2,\xr^5)$. 
We will see shortly that this
EBP \exitentropy curve plays a central role in our investigation. First, let us give its formal 
definition. 

\bdefi 
Assume we are given a \ddp $(\ledge,\redge)$. 
The EBP \exitentropy curve, denote it by $\xh^{\EBP}$, is given in parametric form by 
\begin{align*}
(\ih,\xh^{\EBP}) & =  \left( \ih(\xl), \lnode(1-\redge(1-\xl)) \right),
\end{align*}
where $\ih(\xl) = \frac{\xl}{\ledge(1-\redge(1-\xl))}$ and $\xl \in [0, 1]$.
\edefi
\btheo[Area Theorem for EBP Decoding] \label{theo:ebpareatheorem} 
Assume we are given a \ddp $(\ledge,\redge)$ of design rate $\drate$. 
Then the EBP \exitentropy curve satisfies 
\begin{align*}
\int_{0}^1 \xh^{\EBP}(\xl) \text{d} \ih(\xl) = \drate.
\end{align*}
\etheo
\bproof
We will give two proofs of this fact. 
\par
\begin{figure}[hbt]
\centering
\setlength{\unitlength}{0.5bp}
\begin{picture}(160,160) 
\put(20,0){\includegraphics[scale=0.5]{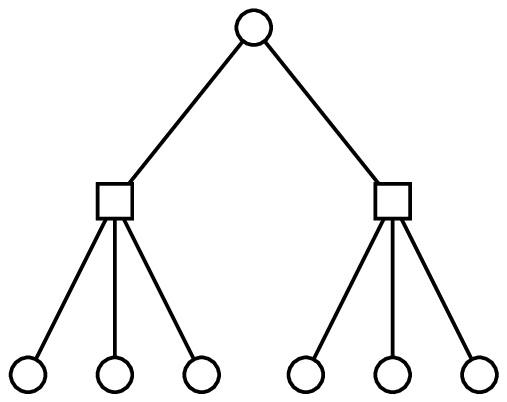}}
\put(100,10){\makebox(0,0){\small{leaves}}}
\put(100,150){\makebox(0,0){\small{root}}}
\end{picture}
\caption{Graph of a small tree code: 
computation tree of depth one for the regular (2,4) 
LDPC ensemble.}
\vspace{7bp}
\label{fig:smallgraph} 
\end{figure}
$(i)$ The first proof applies only if $\ih(\xl)\leq1$ for $\xl\in(0,1]$. This in turn happens only if $\ledge'(0) > 0$, i.e., if the ensemble has a non-trivial
stability condition.
We  use the (General) Area Theorem for transmission over binary 
erasure channels where we allow the parameter of the channel to vary as a function
of the bit position.
First, let us assume that the ensemble is $(\ldegree, \rdegree)$-regular. Consider a 
variable node and the corresponding computation tree of depth one as shown in Fig.~\ref{fig:smallgraph}. 
Let us further define two channel families. The first is the family $\{\text{BEC}(\xl) \}_{\xl=0}^{1}$.
The second one is the family\footnote{Recall 
that $0 \leq \ih(\xl)\leq 1$ for all $\xl\in[0,1]$ by 
assumption.}
$\{\text{BEC}(\ih(\xl)) \}_{\xl=0}^{1}$ where 
 $\ih(\xl) \defas \frac{\xl}{\ledge(1-\redge(1-\xl))}$). The two families are parametrized 
by a common parameter $\xl$ which is the 
fixed-point of density evolution: they are smooth since  $\ih(\xl)$ is differentiable with
respect to $\xl$. Let us now assume that the bit associated to the 
root node is passed through a channel BEC$(\cp(\xl))$, while the ones 
associated to the leaf nodes are  passed through a channel BEC$(\xl)$. 
 We can apply the General Area Theorem: let $X=(X_1,\dots,X_{1+\ldegree\times(\rdegree-1)})$ 
be the transmitted codeword  chosen uniformly at random from the tree code and $Y(\xl)$ be the result of 
passing $X$ through the respective erasure channels parameterized by the common 
 parameter $\xl$. The General Area Theorem  states that
$H(X \mid Y(\xl=1))-H(X \mid Y(\xl=0)) =H(X)$ is equal to the sum of the integrals
of the individual \exitentropy curves, where the integral extends from $\xl=0$ to $\xl=1$. 
 There are two types of individual \exitentropy curves, namely the one
 associated to the root node, call it $\xh_{\text{root}}(\xl)$ 
and the $\ldegree (\rdegree-1)$ ones associated to the leaf nodes, 
call them $\xh_{\text{leaf}}(\xl)$. To summarize, the General Area Theorem 
states
\begin{align*}
H(X) & = \int_0^1\xh_{\text{root}}(\xl)\, \text{d}\cp(\xl)
+\ldegree (\rdegree-1) \int_0^1\xh_{\text{leaf}}(\xl)\text{d}\xl.
\end{align*}
Note that $H(X) = 1+\ldegree (\rdegree-1)-\ldegree=1-\ldegree (\rdegree-2)$ since the computation tree
contains $1+\ldegree(\rdegree-1)$ variable nodes and $\ldegree$ check nodes. Moreover, $\int_0^1\xh_{\text{leaf}}(\xl)\text{d}\xl=\int_0^1 1-\redge(1-\xl)\text{d}\xl=({\rdegree-1})/{\rdegree}$ since the message flowing from the root node to the check nodes is erased with probability $\xl$ (Recall that $\xl=\ih(\xl)\ledge(1-\redge(1-\xl))$, where $(\ledge(\xl),\redge(\xl))=(\xl^{\ldegree-1},\xl^{\rdegree-1})$. Moreover, observe that the result could also be obtained by applying the Area Theorem locally to the Single-Parity-Check code).  
Collecting these observations and solving for $\int_{0}^{1} \xh_{\text{root}}(\xl) d \cp(\xl)$, we get 
$$
\int_{0}^{1} \xh_{\text{root}}(\xl) d \cp(\xl) = 1- \ldegree/\rdegree=\drate,
$$
as claimed since $\xh_{\text{root}}=\xh^\EBP$.
The irregular case follows in the same manner: we consider the ensemble of computation
trees of depth one where the degree of the root note is chosen according to the
node degree distribution $\lnode(\xl)$ and each edge emanating from this root node
is connected to a check node whose degree is chosen according to the edge degree distribution
$\redge(x)$. As before, leaf nodes experience the channel BEC$(\xl)$, whereas the root node
experiences the channel BEC$(\cp(\xl))$. We apply the General Area Theorem to each such choice and
average with the respective probabilities. 

$(ii)$ The second proof applies in all cases. 
Applying integration by parts twice we can write 
\begin{align*}
\int_{0}^1 \xh^{\EBP}(\xl) \text{d} \ih(\xl) 
&= \xh^{\EBP}(\xl)  \ih(\xl) \bigr|_{\xl=0}^{1} - \int_{0}^1 \frac{\text{d}\xh^{\EBP}(\xl)}{\text{d}\xl}  \ih(\xl) \text{d} \xl\\
&\overset{(a)}{=} 1- \lnode'(1) \int_{0}^1 x \redge'(1-x) \text{d} \xl\\
&= 1- \frac{\Bigl(x \redge(1-x) \bigr|_{\xl=0}^{1} + \int_{0}^1 \redge(1-x) \text{d} \xl \Bigr)}{\int_0^1 \ledge(\xl) \text{d} \xl}\\
&= 1- \lnode'(1)/\rnode'(1) = \drate,
\end{align*} 
where $(a)$ follows since $\xh^{\EBP}(\xl)=\lnode'(1) \int_0^{1-\redge(1-\xl)}\ledge(\xl)\text{d}\xl$ and  $\lnode'(1)=1/\int_0^1 \ledge$. Similar computations will be performed several 
times throughout this paper. In this respect it is handy to be able to refer to two
basic facts related to this integration which are summarized as 
Lemma \ref{lemma:ibp1}  and Lemma \ref{lemma:ibp2} 
in Appendix \ref{sec:tricks}. 
\eproof
%
%
\section{An Upper-Bound for the Maximum A Posteriori Threshold}

\label{UpperSection}

Assume that transmission takes places over BEC$(\ih)$. Given a \ddp 
$(\ledge,\redge)$, we trivially have the relations 
\begin{align}
\label{equ:trivialrelations}
\ih^\BP\leq \ih^\MAP \leq \min\{\ih^\text{Sh},\ih^\text{Stab}\},
\end{align}
where $\cp^\text{Sh}$ and $\cp^\text{Stab}$ denote, respectively,
the Shannon and stability threshold.
As we have discussed, it is straightforward to compute $\ih^\BP$ by means of
DE and $\ih^\BP\leq \ih^\MAP $ follows from the sub-optimality
of BP decoding.
The inequality  $\ih^\MAP\leq \ih^\text{Sh}=1 -\drate$ is a rephrasing of 
the Channel Coding Theorem.
Finally $\ih^{\MAP}\le \ih^{\text{Stab}}=1/(\ledge'(0)\redge'(1))$
can be proved through the following graph-theoretic argument.
Assume, by contradiction that $\ih^{\MAP} > \ih^{\text{Stab}}$
and let $\ih$ be such that $\ih^{\text{Stab}}<\ih <\ih^{\MAP}$.
Notice that $\ih^{\text{Stab}}<\ih $ is equivalent to
$\ih \ledge'(0)\redge'(1)>1$. Consider now the residual Tanner graph once the 
received variable nodes have been pruned, and focus on the subgraph of 
degree 2 variable nodes. Such a  Tanner  graph can be identified with an 
ordinary graph by mapping the check nodes to vertices and the variable nodes 
to edges. The average degree of such a graph is $\ih \ledge'(0)\redge'(1)>1$
and therefore a finite fraction of its vertices belong to loops
\cite{Bol01}. If a bit belongs to such a loop, it is not determined by the 
received message: in particular $\expectation[X_i|Y] = 1/2$.
In fact, there exist a codeword such that $x_i=1$: just set $x_j=1$ if 
$j$ belongs to some fixed loop through $i$ and $0$ otherwise. Since there is
a finite fraction of such vertices $h(\cp)>0$ (if the limit exist)
and therefore $\cp>\cp^{\MAP}$. We reached a contradiction, therefore
$\ih^{\MAP} \le \ih^{\text{Stab}}$ as claimed.

While $\ih^\text{Stab}$ and $\ih^\text{Sh}$ are simple quantities, 
the threshold $\ih^\MAP$ is not as easy to compute. 
In this section we will prove an {\em upper-bound} on $\ih^\MAP$
in terms of the (extended) BP \exitentropy curve. In the next sections,
we will see that in fact this bound
is tight for a large class of ensembles. The key to this
bound is to associate the Area Theorem with the following intuitive inequality.
\blemma \label{lemma:orderingviaphysicaldegradation} Consider a \ddp $(\ledge, \redge)$ and the associated EXIT functions $\xh^\BP$ 
and $\xh^\MAP$. Then  
$\xh^\MAP\leq \xh^\BP$.
\elemma
\begin{proof}
Note that Lemma \ref{lemma:orderingviaphysicaldegradation} expresses the natural
statement that BP processing is in general suboptimal.
For a given length $\n$, pick a code at random from $\ldpc(\ledge,\redge,\n)$. 
Call $\Phi^{\BP}_i$ the extrinsic BP
estimate of bit $i$  and note that $\Phi^{\BP}_i=\Phi^{\BP}_i(Y_{\sim i})$,
i.e., the extrinsic BP estimate is a well defined function of $Y_{\sim i}$.
The Data Processing Theorem asserts that
$H(X_i|Y_{\sim i})\leq H(X_i|\Phi^{\BP}_i(Y_{\sim i}))$. This is true for 
all codes in  $\ldpc(\ledge,\redge,\n)$. Therefore taking first the 
average over the ensemble and second the limit  
when the blocklength $\n\to \infty$ (assuming the limit of the
MAP EXIT function exists), we get $\xh^\MAP(\ih)\leq \xh^\BP(\ih)$.
\end{proof}

Because of Lemma \ref{lemma:orderingviaphysicaldegradation}, it is 
 of 
course not surprising that the integral under $\xh^\BP$
is larger or equal than the asymptotic rate of the code $r_{\text{as}}$ 
as pointed out in Section \ref{sec:bpexitfunction}. In most of 
the cases encountered in practice, $\drate=r_{\text{as}}$, (see Section \ref{sec:Counting}), 
the area under the MAP EXIT curve is therefore $\drate$ and the area 
under the BP EXIT curve is strictly larger than $\drate$ if and only if 
the curve exhibits discontinuities (in the absence of discontinuities, the
two curves coincide and the MAP/BP threshold is given by the stability condition).

\begin{figure}[htp]
\centering
\setlength{\unitlength}{1bp}
\begin{picture}(240,140)
\put(5,28)
{
\put(5,5){\includegraphics[width=100bp]{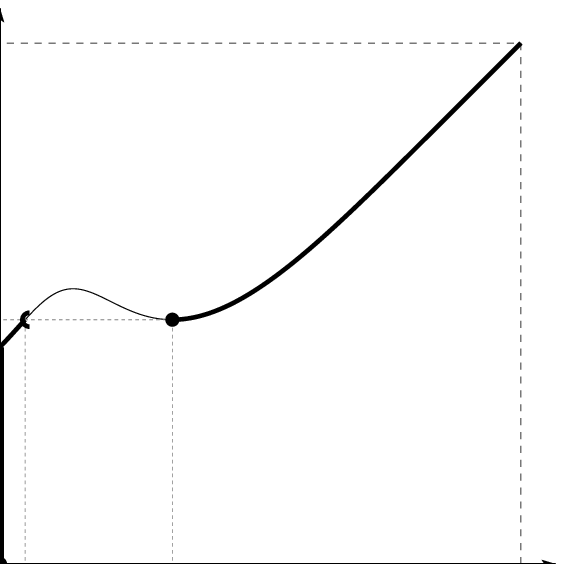}}
\put(135,5){\includegraphics[width=100bp]{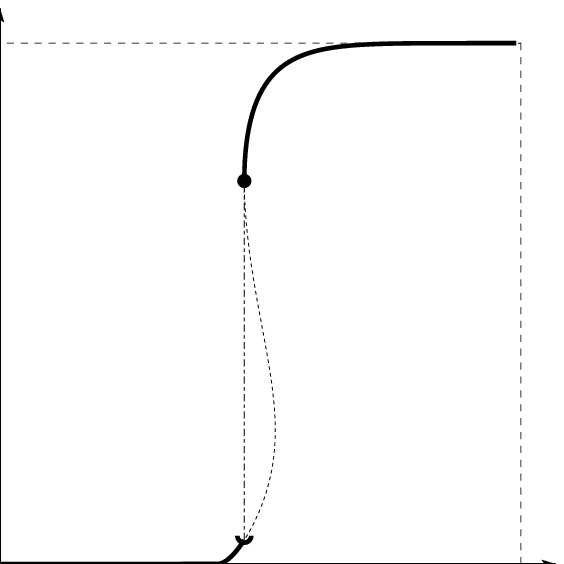}}
\put(0,0){\makebox(0,0){\small{$0$}}}
\put(100,0){\makebox(0,0){\small{$1$}}}
\put(0,100){\makebox(0,0){\small{$1$}}}
\put(130,0){\makebox(0,0){\small{$0$}}}
\put(230,0){\makebox(0,0){\small{$1$}}}
\put(130,100){\makebox(0,0){\small{$1$}}}
\put(5,-6){\makebox(0,0){\small{$\lxl^0$}}}
\put(-1,43){\makebox(0,0){\small{$\ih^\BP$}}}
\put(-1,53){\makebox(0,0){\small{$\ih^1$}}}
\put(60,80){\makebox(0,0){\small{$\ih(\xl)$}}}
\put(185,17){\makebox(0,0){\small{$\ih^1$}}}
\put(200,80){\makebox(0,0){\small{$\xh^\BP(\ih)$}}}
}
\put(60,7){\makebox(0,0){{\small{(a)}}}}
\put(190,7){\makebox(0,0){{\small{(b)}}}}
\end{picture}
\caption{
BP \exitentropy entropy curve with 1 discontinuity ($J$=1) for which the BP threshold $\ih^\BP=\ih^\MAP$ is given by the stability condition: 
(a) Channel entropy function $\xl\mapsto\ih(\xl)$ (b) BP \exitentropy function $\ih\mapsto\xh^\BP(\ih)$. 
}
\label{fig:epsilonwith1JandStabCond}
\end{figure}

Example \ref{ex:stco} refines and illustrates this observation by showing that the 
BP and MAP threshold might be equal even if their respective \exitentropy functions  
are not pointwise equal. 
\bex\label{ex:stco} 
Consider the \ddp $(\ledge,\redge)=(0.4 x + 0.6 x^6,x^6)$ and the  
corresponding $\ldpc$ ensemble with design rate $\drate=0.5$. Using a weight enumerator
 function, see, e.g., Section \ref{sec:Counting}, one can show that  
$\drate=r_{\text{as}}=\int \xh^\MAP$. A quick look 
shows that the BP threshold is given by the stability condition, i.e., 
it is $\ih^\BP\approx0.4167$ obtained for $\xl\approx\lxl^0=0$.  When the 
parameter is $\uxl^0\approx0.04828$, i.e., at $\ih^1\approx0.4691$, a discontinuity 
of the BP \exitentropy curve appears and the edge erasure 
probability $\xl$ ``jumps'' to $\lxl^1\approx0.3309$. This situation is 
shown in Fig.~\ref{fig:epsilonwith1JandStabCond}. 
Since the BP threshold is determined by the 
stability condition,
as explained 
previously 
we have $\ih^\BP=\ih^\MAP$. 
This is true despite the fact that the integral 
under the BP \exitentropy is larger than $\drate=r_{\text{as}}$! 
\eex

Recall that the Area Theorem asserts that 
$\int_0^1\xh^\MAP(\cp)\text{d}\cp=\drate_{\text{as}}$, where
$\drate_{\text{as}}$ is the asymptotic rate of the ensemble defined in 
Theorem \ref{theo:asymptoticareatheorem}. By definition $\xh^\MAP(\cp)=0$ for
$\cp \leq \cp^\MAP$. Therefore we have in fact 
$\int_{\cp^\MAP}^1\xh^\MAP(\cp)\text{d}\cp=\drate_{\text{as}}$.
Now note that the BP decoder is in general suboptimal so that 
$\xh^\MAP(\cp) \leq \xh^\BP(\cp)$. Further, in general $\drate_{\text{as}} \geq \drate(\ledge, \redge)$.
Combining these statements we see that
if $\overline{\ih}^\MAP$ is a real number in $[\cp^\BP, 1]$ such that
$\int_{\overline{\cp}^\MAP}^1 \xh^\BP(\cp) \text{d}\cp=\drate(\ledge, \redge)$ then
$\int_{\overline{\cp}^\MAP}^1 \xh^\MAP(\cp) \text{d}\cp \leq \drate_{\text{as}}$.
We conclude that for such a $\overline{\ih}^\MAP$, 
$\cp^\MAP \leq \overline{\cp}^\MAP$. Let us summarize a slightly strengthened version
of this observation as a lemma. 
\blemma[First Upper Bound on $\cp^\MAP$]
\label{First}
Assume we are given a \ddp $(\ledge, \redge)$. Let $\xh^\BP(\cp)$
denote the associated BP \exitentropy function and let 
$\overline{\ih}^\MAP$ be the unique real number in $[\cp^\BP, 1]$ such that
$\int_{\overline{\cp}^\MAP}^1 \xh^\BP(\cp) \text{d}\cp=\drate(\ledge, \redge)$.
Then $\cp^\MAP \leq \overline{\cp}^\MAP$. If in addition 
$\overline{\cp}^\MAP=\cp^\BP$ then $\cp^\MAP = \cp^\BP$,
and in fact $\xh^\MAP(\cp)=\xh^\BP(\cp)$ for all $\cp \in [0, 1]$.
\elemma
\bproof
We have already discussed the first part of the lemma. To see the second part,
if $\overline{\cp}^\MAP=\cp^\BP$ then by (\ref{equ:trivialrelations})
we have a lower and an upper bound that match and therefore we have equality.
This can only happen if the two \exitentropy functions are in fact identical
(and if $\drate_{\text{as}}=\drate(\ledge, \redge)$).
\eproof
\bex
For the \ddp $(\ledge(x),\redge(x))=(x,x^{3})$, 
 we obtain $\overline{\ih}^\MAP = 1/3 = \ih^\BP$. 
Therefore, for this case the MAP \exitentropy function is equal to the BP \exitentropy function
and in particular both decoders have equal thresholds.
\eex
\bex 
For the \ddp $(\ledge(x),\redge(x))=(x^{2},x^{3})$,  
we obtain $\overline{\ih}^\MAP = \frac{102-7 \sqrt{21}}{108} \approx 0.647426$.
Note that this \ddp has rate $1/4$ so that this upper bound on the 
threshold should be compared to the Shannon limit $3/4=0.75$.
\eex
\bex
\label{exa:36upperbound}
For the \ddp $(\ledge(x),\redge(x))=(x^2, x^5)$ of our running example, we get 
\centerline{
$\overline{\ih}^\MAP  =  \frac{7 - {\sqrt{-1 - a +  b}} - {\sqrt{-2  + a - b + 
\frac{4}{{\sqrt{-1 - a + b}}}}}}{ 6 {\left( -1 +   {\left( - \frac{1}{6}   
+ \frac{{\sqrt{-1 - a +      b  }}}{6} + \frac{{\sqrt{- 2   + a -  b +  
\frac{4}{{\sqrt{-1-a+b}}}}}}{6} \right) }^5 \right) }^2},$} 
with $a \defas \frac{7\cdot 5^{\frac{2}{3}}}{{\left( 11 + 
6\,{\sqrt{51}} \right) }^{\frac{1}{3}}} $ and
 $b \defas {\left( 55 + 30\,{\sqrt{51}} \right) }^{\frac{1}{3}}$.
Numerically, $\overline{\ih}^\MAP \approx 0.4881508841915644$.
The Shannon threshold for this ensemble is $0.5$.
\eex
For a \ddp which exhibits a single jump the computation of this upper bound
is made somewhat easier by the following lemma. Note that by Fact \ref{fact:regular}
this lemma is applicable to regular ensembles.
\blemma \label{theo:exactmlthreshold}
Assume we are given a \ddp $(\ledge,\redge)$.
Define the polynomial $\xr(\xl)\defas1-\redge(1-\xl)$ and, for $\xl\in(0,1]$ the
function $\ih(\xl)\defas\frac{\xl}{\ledge(\xr(\xl))}$. Assume that $\ih(\xl)$ is
increasing over $[\xl^\BP,1]$. Let $\xl^*$ be the unique 
root of the polynomial 
$$
P(\xl) \defas  \lnode'(1)\xl(1-\xr(\xl)) -\frac{\lnode'(1)}{\Gamma'(1)}
[1-\Gamma(1-\xl)]+\epsilon(\xl)\lnode(\xr(\xl))\, ,\label{eq:Pdef}
$$
in the interval $[\xl^\BP,1]$.
Then $\overline{\ih}^\MAP = \ih(\xl^*)$.
\elemma
\bproof
Recall that if $\ih(\xl)$ is increasing over $[\xl^\BP,1]$ then 
we have the parametric representation of $\xh^\BP(\cp)$ as given
in (\ref{equ:bpexitcurve}). Using Lemmas \ref{lemma:ibp1} and 
\ref{lemma:ibp2} we can express the integral 
$\int_{\overline{\cp}^\MAP}^1 \xh^\BP(\cp) \text{d}\cp$  
as a function of $\overline{\cp}^\MAP$. More precisely, we 
parametrize $\overline{\cp}^\MAP$ by $\xl$ and express the integral
as a function of $\xl$.
Equating the result to $\drate(\ledge, \redge)=1-\lnode'(1)/\rnode'(1)$
and solving for $\xl$ leads to the polynomial 
condition $P(\xl)=0$ stated above.
\eproof
\bex 
The following table compares the thresholds and bounds for various ensembles. 
Hereby 
$\ledge^{(1)}(x)=x$, 
$\ledge^{(2)}(x)=\frac{7  x^2 + 2 x^3 + 1  x^4}{10}$, 
$\ledge^{(3)}(x)=\frac{2857 x +3061.47 x^2+ 4081.53 x^9}{10000}$, 
$\ledge^{(4)}(x)=\frac{7.71429 x^2 + 2.28571 x^7}{10}$, and 
$\ledge^{(5)}(x)=\frac{9 x^2 + 1 x^7}{10}$.
The threshold of the first ensemble is given by the stability condition. Its
exact value is $7/28\approx0.1786$.
\tablespace
\small{
\begin{tabular}{cccccc}
$\ledge(x)$  & $\redge(x)$ & $\ih^\BP$ & $\overline{\ih}^\MAP$  & $\cp^\MAP$ & $\ih^\Sh$  \\ \hline
$\ledge^{(1)}(x)$     &   $\frac{2 x^5 + 3 x^6}{5} $   &  0.1786    &   0.1786    & 0.1786 &  0.3048  \\ 
$\ledge^{(2)}(x)$     & $\frac{2 x^5 + 3 x^6}{5} $  &  0.4236     &   0.4948    & 0.4948 &  0.5024  \\
$\ledge^{(3)}(x)$ &  $x^6$   &  0.4804   &   0.4935    & 0.4935 &  0.5000 \\
$\ledge^{(4)}(x)$     &  $x^4$   &  0.5955   &   0.6979    & 0.6979 &  0.7000 \\ 
$\ledge^{(5)}(x)$     &  $x^7$   &  0.3440   &   0.3899    & 0.3899 &  0.4000  \\ 
\end{tabular}
}
\tablespace
\eex
The polynomial $P(\xl)$ provides in fact a fundamental characterization
of the MAP threshold and has some important properties.
These are more conveniently stated in terms of a slightly more general concept.
\bdefi\label{TrialDef}
The {\em trial entropy} for the channel BEC$(\cp)$
associated to the 
\ddp $(\ledge,\redge)$ is the bi-variate polynomial
$$
P_{\cp}(\xl,\xr) \defas  \lnode'(1)\xl(1-\xr) -\frac{\lnode'(1)}{\rnode'(1)}
[1-\rnode(1-\xl)]+\cp\lnode(\xr)\, .
$$
\edefi
A few properties of the trial entropy are listed in the following.
\blemma\label{TrialLemma}
Let $(\ledge,\redge)$ be a \ddp and $P_{\cp}(\xl,\xr)$ the corresponding 
trial entropy. Consider furthermore the DE equations
for the ensemble $\xl_{\iter+1} = \cp\ledge(\xr_{\iter})$,
$\xr_{\iter+1} = 1-\redge(1-\xl_\iter)$, $\iter$ being the iteration number. Then
(in what follows we always consider $\xl,\xr\in[0,1]$)
\begin{enumerate}
\item The fixed points of density evolution are stationary points of the trial
entropy. Vice versa, any stationary point of the trial entropy 
is a fixed point of density evolution.\label{FixedPoints}
\item $P(\xl) = P_{\cp(\xl)}(\xl,\xr(\xl))$.\label{Polynomial}
\item $P(\xl = 1) = P_{\cp=1}(\xl=1,\xr=1)=\drate(\ledge, \redge)$.\label{PolRate} 
\item \label{Integral}
Let $a\defas (\cp_a=\cp(\xl_a),h^{\EBP}(\xl_a))$ and 
$b\equiv (\cp_b=\cp(\xl_b),h^{\EBP}(\xl_b))$ be two points
on the EBP \exitentropy curve (with $\xl_{a/b}\in (0,1]$)
and define $\xr_{a/b} = 1-\redge(1-\xl_{a/b})$. Then
\[
\int_{a}^{b}h^\EBP(\cp(\xl))\, d\cp(\xl) = P_{\cp_{b}}(\xl_b,\xr_b)-
P_{\cp_{a}}(\xl_a,\xr_a)\, .
\]
\end{enumerate} 
\elemma
\bproof
(\ref{FixedPoints}) is proved by explicitly computing the partial derivatives
of $P_{\cp}(\xl,\xr)$ with respect to $\xl$ and $\xr$:
$\partial_{\xl} P_{\cp}(\xl,\xr) = \lnode'(1)[1-\xr-\redge(1-\xl)]$,
$\partial_{\xr} P_{\cp}(\xl,\xr) = \lnode'(1)[-\xl+\cp\ledge(\xr)]$.
Since $\lnode'(1)>0$, the stationarity conditions
$\partial_{\xl} P_{\cp}(\xl,\xr) = 0$ and 
$\partial_{\xr} P_{\cp}(\xl,\xr) = 0$ are equivalent to the
fixed point conditions for DE.
(\ref{Polynomial}) and (\ref{PolRate}) are elementary algebra.
In order to prove (\ref{Integral}), notice that we have
$\partial_{\xl} P_{\cp}(\xl,\xr) = \partial_{\xr} P_{\cp}(\xl,\xr) = 0$ 
at any point $(\xl,\xr(\xl),\cp(\xl))$ along the EBP \exitentropy curve.
This follows from the fact that points on the EBP \exitentropy curve are
fixed points of density evolution. Therefore
\begin{align*}
\frac{\de \phantom{\xl}}{\de \xl} P_{\cp(\xl)}(\xl,\xr(\xl)) & = 
\lnode(\xr(\xl))\,  \frac{\de\cp}{\de \xl} (\xl) = h^\EBP(\cp(\xl))
\,  \frac{\de\cp}{\de \xl} (\xl)\, .
\end{align*}
The thesis follows by integrating over $\xl$. Equivalently, we could 
have used again Lemmas \ref{lemma:ibp1} and \ref{lemma:ibp2}.
\eproof

Unfortunately, the upper-bound stated in Lemma \ref{First} is not always 
tight. In particular, this can happen if the EBP \exitentropy curve exhibits 
multiple jumps (i.e., if $\cp(\xl)$ has more than one local maximum 
in the interval $(0,1]$). We will state a precise sufficient condition for tightness
in the next section.
An improved upper bound is obtained as follows.
\btheo[Improved Upper-Bound on $\cp^\MAP$]\label{ImprovedUB}
Assume we are given a \ddp $(\ledge, \redge)$. Let $\xh^\EBP(\cp)$
denote the associated EBP \exitentropy function and let
$(\overline{\ih}^\MAP=\cp(\xl^*),\xh^{\EBP}(\xl^*))$ 
be a point on this curve. Assume that 
$\int_{\xl^*}^1 \xh^\EBP(\xl) \text{d}\cp(\xl)=\drate(\ledge, \redge)$
and that
there exist 
no  $\xl'\in (\xl^*, 1]$ such that $\cp(\xl') = \cp(\xl^*)$.
	Then $\cp^\MAP \leq \overline{\cp}^\MAP$. 
\etheo
The proof of this theorem will be given in Section \ref{MaxwellSection}
using the so-called Maxwell construction. Notice that in general there can be more than one
value of $\epsilon$ satisfying the theorem hypotheses. We shall always
use the symbol $\overline{\cp}^\MAP$ to refer to the smallest such value. On the other
hand, it is a consequence of the proof of theorem that there always exists
at least one such value.

As before, the following lemma simplifies the computation of the 
upper bound by stating the following more explicit characterization.
\blemma
\label{lem:exactmlthreshold2}
Consider a \ddp $(\ledge,\redge)$.
Let $\xl^*\in(0,1]$ be a root of the polynomial
$P(\xl)$ defined in (\ref{eq:Pdef}),  such that  there exist 
no  $\xl'\in (\xl^*, 1]$  with $\cp(\xl') = \cp(\xl^*)$.
Then $\ih^\MAP \le \ih(\xl^*)$, and $\overline{\epsilon}^\MAP$
is the smallest among such upper bounds.
\elemma
\begin{proof}
Let $\xl^*$ be defined as in the statement. Then, by Lemma
\ref{TrialLemma}, points (\ref{Polynomial}), (\ref{PolRate}) and 
(\ref{Integral}):
\[
\int_{\xl^*}^1h^\EBP(\xl)\,\de\cp(\xl) = P(1)-P(\xl^*) = \drate(\ledge, \redge)-P(\xl^*)\, .
\]
Therefore, $\int_{\xl^*}^1h^\EBP(\xl)\,\de\cp(\xl) = \drate(\ledge, \redge)$ if and
only if $P(\xl^*) = 0$.
\end{proof}

For a large family of \ddps the upper bound stated
in Theorem \ref{ImprovedUB}
is indeed tight. Nevertheless, it is possible to construct
examples where we can not evaluate the bound at all roots
$\xl^*$ of $P(\xl)$ since for some of those roots there exists a point 
$\xl'\in (\xl^*, 1]$  with $\cp(\xl') = \cp(\xl^*)$.
In these cases we expect the bound not to be tight.
Indeed, we conjecture that the extra condition on the roots
of $P(\xl)$ are not necessary and that the MAP threshold is
in general given by the following statement.  
%
\bconj
\label{conj:mainconjecture}
Consider a degree distribution pair $(\ledge,\redge)$ and the associated 
polynomial $P(\xl)$ defined as in (\ref{eq:Pdef}).
Let $\cX\subset (0,1]$ be the set of positive roots of 
$P(\xl)$ in the interval $(0,1]$ (since $P(\xl)$ is a polynomial,
$\cX$ is finite). Equivalently, $\cX$ is the set of
$\xl_*\in (0,1]$ such that $\int_{\xl^*}^1h^\EBP(\xl)\,\de\cp(\xl) = \drate(\ledge, \redge)$.
Then  $\cp^{\MAP}=\min\{\cp(\xl^*);\, \xl\in\cX\}$.   
\econj

%
%
\section{Counting Argument}
\label{sec:Counting}

We will now describe a counting argument which yields
an alternative proof of Lemma \ref{First}. More interestingly,
the argument can be strengthened to obtain an easy-to-evaluate
sufficient condition for tightness of the upper-bound. 

The basic idea is quite simple. Recall that we define the
MAP threshold as the maximum of all channel parameters for which the normalized
conditional entropy converges to zero as the block length tends to infinity. 
For the binary erasure channel, the conditional entropy is equal to the logarithm of the number of codewords
which are compatible with the received word.
Therefore, a first naive way of upper bounding the MAP threshold consists in lower bounding 
the expected number of codewords in the residual graph, after 
eliminating the received variables. If, for a given channel parameter, this
lower bound is exponential with a strictly positive exponent,
then the corresponding conditional entropy is strictly positive and
we are operating above the threshold. It turns out that a much better
result is obtained by considering the residual graph after iterative decoding 
has been applied. In fact, this simple modification allows one to obtain matching upper 
and lower bounds in a large number of cases.

Let $\graph$ be chosen uniformly at random from the ensemble characterized by 
$\nddp \defas (\lnode,\rnode)$. 
Assume further that transmission takes place over BEC$(\cp)$ and 
that a BP decoder is applied to the received sequence. 
Denote by $\graph(\ih)$
the residual graph after decoding has halted, and by
$\nddp_{\graph(\cp)} = (\lnode_{\graph(\cp)},\rnode_{\graph(\cp)})$ 
its degree profile (i.e., the fraction of nodes of any given degree).
We adopt here the convention of normalizing the \ddp of $\graph(\cp)$
with respect to the number of variable nodes and check nodes in the
{\em original} graph. Therefore, $\lnode_{\graph(\cp)}(1)\le 1$
is the number of variable nodes in $\graph(\cp)$
divided by $\n$. Analogously, $\rnode_{\graph(\cp)}(1)\le 1$ is the number
of check nodes in $\graph(\cp)$ divided by $n\lnode'(1)/\rnode'(1)$.

It is shown in \cite{LMSS01b} that, conditioned on the degree profile
of the residual graph,  $\graph(\cp)$ is uniformly distributed.
The \ddp $\nddp_{\graph(\cp)}$ itself is of course a random quantity because
of the channel randomness. 
However, it is  is sharply concentrated around its expected value. For
increasing blocklengths this expected value converges to 
$\nddp_{\cp}=(\lnode_{\cp},\rnode_{\cp})$, which is given by\footnote{The standard \ddp from
the node 
perspective of the residual graph when transmission takes place over BEC($\ih$) 
is then simply given by $\left(\frac{\lnode_{\cp}(\xl)}{\lnode_{\cp}(1)},\frac{\rnode_{\cp}(\xl)}{\rnode_{\cp}(1)}\right)$. }
\begin{align}
\lnode_{\cp}(\zl) & \defas \cp\lnode(\zl \xr)\, ,
\label{LambdaRes}\\
\Gamma_{\cp}(\zl) &\defas \Gamma(1-\xl+\zl\xl)-\Gamma(1-\xl)-
\zl\xl\Gamma'(1-\xl)\, .
\label{GammaRes}
\end{align}
Here, $\xl$ and $\xr$ denote the fraction of erased messages at the fixed point 
of the BP decoder. More precisely, $\xl\in [0,1]$ 
is the largest solution
of $\xl=\epsilon\ledge(1-\redge(1-\xl))$ and $\xr =  1-\redge(1-\xl)$.
The precise concentration statement follows.
\blemma\label{ConcentrationDegreeLemma}
Let  $\ih\in (0,1]$ be a continuity point of  
$\xl(\ih)$ (we shall call such an $\ih$ {\em non-exceptional}).
Then, for any $\xi > 0$, 
\begin{align}
\lim_{\n\to\infty}\Pr\{d(\nddp_{\graph(\ih)},\nddp_\ih)\ge \xi\} &=0\, .
\end{align}
Here, $d(\cdot,\cdot)$ denotes the $L_1$ distance
\begin{align}
d(\nddp,\tilde{\nddp}) & \defas \sum_{\ldegree}
|\lnode_{\ldegree}-\tilde{\lnode}_{\ldegree}|+\sum_{\rdegree}
|\rnode_{\rdegree}-\tilde{\rnode}_{\rdegree}|\, .
\end{align}
\elemma

The proof is deferred to Appendix \ref{ProofCounting}.

Under the zero-codeword assumption, the set of codewords compatible with 
the received bits coincides with the set of codewords of the residual graph.
Their expected number can be computed through standard combinatorial tools.
The key idea here is that,
under suitable conditions on the \ddp (of the residual graph),
the actual rate of codes from the (residual) ensemble is close to the design rate.
We state here a slightly strengthened version of this result from \cite{DMU04}.
\blemma
\label{WELemma}
Let $\graph$ be chosen uniformly at random from the ensemble  
LDPC$(\n,\nddp) = $LDPC$(\n,\lnode,\rnode)$, let $\rate_\graph$ be its rate and
$\drate \defas 1-\lnode'(1)/\rnode'(1)$ be the design rate. 
Consider the function $\Psi_{\nddp}(u)$,
\begin{align}
\Psi_{\nddp}(u) =& -\lnode'(1)\log_2\left[\frac{(1+uv)}{(1+u)(1+v)}\right]
\nonumber \\&+\sum_\ldegree\lnode_\ldegree\log_2\left[
\frac{1+u^\ldegree}{2 (1+u)^\ldegree}\right]
\nonumber\\
&+\frac{\lnode'(1)}{\Gamma'(1)}\sum_{\rdegree}\Gamma_\rdegree
\log_2\left[1+\left(\frac{1-v}{1+v}\right)^\rdegree\right]\, ,\label{Psidef} \\
v  = & \left(\sum_{\ldegree}\frac{\ledge_\ldegree}{1+u^\ldegree}
\right)^{-1}\left(\sum_{\ldegree}\frac{\ledge_{\ldegree}u^{\ldegree-1}}
{1+u^\ldegree}\right)\, .\label{xtoy}
\end{align}
Assume that $\Psi_{\nddp}(u)$ takes on its global maximum
in the range $u\in[0,\infty)$ at $u=1$. Then there exists $B>0$ such that, 
for any  $\xi>0$, and $\n>\n_0(\xi,\Xi)$, 
\begin{align*}
\Pr\{|\rate_\graph- \drate(\lnode, \rnode)|>\xi\}&\le e^{-B n\xi}\, .
\end{align*}
Moreover, there exist $C>0$ such that, for $n>n_0(\xi,\Xi)$, 
\begin{align*}
\expectation[ |\rate_\graph- \drate(\lnode, \rnode)|] &\le C\frac{\log n}{n}\, .
\end{align*}
\elemma
\bproof
The idea of the proof is the following. For any parity-check ensemble 
we have $\rate_{\graph}\ge \drate(\lnode, \rnode)$. If it is true
that the expected value of the rate (more precisely, the logarithm of the 
expected number of codewords divided by the length) is close
to the design rate, then we can use the Markov inequality to
show that most codes have rate close to the design rate.

Let us start by computing the exponent of the expected number
of codewords. We know from
\cite{Gal62,MB04,LS02,LS03,DRU01,DRU04,Di04,DMU04,SeF96,FlS94} 
that the expected number
of codewords involving $E$ edges is given by
\begin{align*}
\expectation[N_{\graph}(E)] & = 
\frac{
\text{coef} 
\left\{ 
\prod_\ldegree (1+u^\ldegree)^{n \lnode_\ldegree} 
\prod_\rdegree q_\rdegree(v)^{n \frac{\lnode'(1)}{\rnode'(1)} \rnode_\rdegree}, u^E v^E \right\} }{
\binom{n \lnode'(1)}{E}}, 
\end{align*}
where $q_\rdegree(v) = ((1+v)^\rdegree+(1-v)^\rdegree)/2$. Let $n$ tend to infinity and define $e=E/(n \lnode'(1)$.
From standard arguments presented in the cited papers it is known that, 
for a fixed $e$, the exponent 
$\lim_{n \rightarrow \infty} \frac{1}{n}\log_2\bigl(\expectation[
N_{\graph}(e n \lnode'(1))]\bigr)$
is given by the infimum with respect to $u, v > 0$ of
\begin{multline}
\sum_{\ldegree}\lnode_{\ldegree}\log_2(1+u^{\ldegree})-\lnode'(1)
e\log_2 u  
+\frac{\lnode'(1)}{\rnode'(1)}\sum_{\rdegree}
\Gamma_{\rdegree}\log_2 q_{\rdegree}(v) \\
-\lnode'(1) e\log_2 v-\lnode'(1)h(e).
\label{equ:exponent}
\end{multline}
We want to determine the exponent corresponding to the expected number
of codewords, i.e.,
$\lim_{n \rightarrow \infty} \frac{1}{n}\log_2\bigl(\expectation[
N_{\graph}]\bigr)$, where $N_{\graph}=\sum_{E} N_{\graph}(E)$.
Since there is only a polynomial number of ``types'' 
(numbers $E$) this exponent is
equal to the supremum of (\ref{equ:exponent}) over all $0 \leq e \leq 1$.
In summary, the sought after exponent is given by a stationary point of the function stated in
(\ref{equ:exponent}) with respect to 
$u$, $v$ and $e$.

Take the derivative with respect to $e$. 
This gives $e=uv/(1+uv)$. If we substitute this expression for $e$
into (\ref{equ:exponent}), subtract the design rate 
$\drate(\lnode, \rnode)$, 
and rearrange the terms somewhat we get (\ref{Psidef}).
Next, if we take the derivative with respect to
$u$ and solve for $v$ we get get (\ref{xtoy}).
In summary, $\Psi_{\nddp}(u)$ is a function so that
\begin{align*}
\log_2 \expectation [N_{\graph}] &=
n\{\drate(\lnode, \rnode)+\sup_{u\in[0,\infty)}\Psi_{\nddp}(u) +\omega_n\}\, ,
\end{align*}
where $\omega_n = o(1)$. In particular, by explicit computation we see that
$\Psi_{\nddp}(u=1) = 0$. A closer look shows that
$u=1$ corresponds to the exponent of codewords of weight $n/2$.
Therefore, the condition that the global maximum
of $\Psi_{\nddp}(u)$ is achieved at $u=1$ is equivalent to the condition
that the expected weight enumerator is dominated by codewords of weight
(close to) $n/2$. Therefore, 
\begin{align*}
\Pr\{\rate_{\graph}\ge \drate(\lnode, \rnode)+\xi\} &= 
\Pr\left\{N_{\graph}\ge 2^{n(\xi-\omega_n)} \expectation[N_{\graph}]
\right\} \\
& \le e^{-Bn\xi}\, ,
\end{align*}
where the step follows from the Markov inequality if 
$B=(\log 2)/2$ and $\omega_n\le\xi/2$ for any $n\ge n_0$. 

Finally, we observe that, since $\rate_{\graph}\le 1$
\begin{align*}
\expectation [|\rate_\graph- \drate(\lnode, \rnode)|] & \le \xi + e^{-Bn\xi}\, ,
\end{align*}
and the second claim follows by choosing $\xi = \log n/Bn$.
\eproof

We would like to apply this result to the residual graph $\graph(\cp)$.
Since the degree profile of $\graph(\cp)$ is a random variable, we need 
a preliminary observation on the ``robustness'' of the hypotheses 
in the Lemma \ref{WELemma}.
\blemma\label{PsiProp}
Let $\Psi_{\nddp}(\cdot)$ be defined as in Lemma \ref{WELemma}.
Then $\Psi_{\nddp}(u)$ achieves its maximum over $u\in[0,+\infty)$
in  $[0,1]$.

Moreover, 
there exists a constant $A>0$ such that, for any two 
degree distribution pairs
$\nddp =(\lnode,\rnode)$ and
$\tilde{\nddp}=(\tilde{\lnode},\tilde{\rnode})$, and
any $u\in [0,1]$,
\begin{align}
|\Psi_{\nddp}(u)-\Psi_{\tilde{\nddp}}(u)| &
\le A\, d(\nddp,\tilde{\nddp})\, (1-u)^2\, .
\label{PsiLip}
\end{align}
\elemma
For the proof we refer to Appendix \ref{ProofCounting}.

We turn now to the main result of this section.
\btheo\label{CountingTheorem}
Let $\graph$ be a code picked uniformly at random from the ensemble LDPC$(\n,\lnode,\Gamma)$ 
and let $H_{\graph}(X|Y)$ be the conditional entropy of the transmitted
message when the code is used for communicating over BEC$(\cp)$.
Denote by $P_{\epsilon}(\xl,\xr)$ the corresponding trial entropy.
Let $\nddp_{\cp}=(\lnode_{\cp},\Gamma_{\cp})$ be the typical degree 
distribution pair of the residual graph, see 
Eqs.~(\ref{LambdaRes}), (\ref{GammaRes}),  and $\Psi_{\nddp_{\cp}}(x)$  
be defined as in Lemma \ref{WELemma}, Eq. (\ref{Psidef}).

Assume that $\Psi_{\nddp_{\epsilon}}(u)$ achieves its global maximum as 
a function of $u\in[0,\infty)$ at $u = 1$, with 
$\Psi''_{\nddp_{\epsilon}}(1)<0$, and that $\epsilon$  is non-exceptional. 
Then
\begin{align}
\label{equ:conditonalentropy}
\lim_{n\to\infty}\frac{1}{n}\expectation [H_{\graph}(X|Y)] & = P_{\epsilon}(\xl,\xr)
\, ,
\end{align}
where $\xl\in [0,1]$ is the largest solution
of $\xl=\epsilon\ledge(1-\redge(1-\xl))$ and $\xr =  1-\redge(1-\xl)$.
\etheo

\bproof
As above, we denote by $\graph(\epsilon)$ the residual graph after 
BP decoding and by $\rate_{\graph(\epsilon)}$ its rate normalized 
to the original blocklength $n$.
Notice that $H_{\graph}(X|Y) = n\rate_{\graph(\epsilon)}$: iterative decoding
does not exclude any codeword compatible with the received bits.
Furthermore, the design rate (always normalized to $n$) for the 
\ddp of the residual graph is 
\begin{align*}
\drate(\nddp_{\graph(\cp)}) &= \lnode_{\graph(\cp)}(1)-
\frac{\lnode'(1)}{\Gamma'(1)}\, \Gamma_{\graph(\cp)}(1)\, .
\end{align*}
We further introduce the notation $r_{\cp}$ for the design rate of the typical
\ddp of the residual graph. Using Eqs.~(\ref{LambdaRes}) and
(\ref{GammaRes}), we can find 
\begin{align*}
\drate_{\cp} &= \lnode'(1)\redge(1-\xl)\xl-\frac{\lnode'(1)}{\Gamma'(1)}
[1-\Gamma(1-\xl)]+\cp\lnode(\xr) \\
& = P_{\cp}(\xl,\xr),
\end{align*}
where the last step follows from the fixed-point condition
$\xr=1-\redge(1-\xl)$. 

Since by assumption $\Psi_{\nddp_{\cp}}(u)$ achieves its global maximum at 
$u = 1$, with $\Psi''_{\nddp_{\cp}}(1)<0$, and $\Psi_{\nddp_{\cp}}(1)=0$,
there exists a positive constant $\delta$ such that 
$\Psi_{\nddp_{\cp}}(u)\le -\delta (1-u)^2$ for any $u\in[0,1]$.
As a consequence of Lemma \ref{PsiProp}, there exist a $\xi>0$
such that, for any \ddp $\nddp$, with $d(\nddp,\nddp_{\cp})\le \xi$,
$\Psi_{\nddp}(u)\le -\delta (1-u)^2/2$ for $u\in[0,1]$.

Let $\pr_{\cp}(\tilde{\nddp})$ be the probability that  
the degree distribution pair of the residual graph 
$\graph(\cp)$ is $\tilde{\nddp} = (\tilde{\lnode},\tilde{\Gamma})$. 
Denote by $\tilde{\expectation}$ expectation with respect to 
a uniformly random code in the $(\tilde{n},\tilde{\lnode},\tilde{\Gamma})$ ensemble (here $\tilde{n}\defas n\tilde{\lnode}(1)$). 
Denote by $\cN(\xi)$ the set of \ddps $\tilde{\nddp}$, such that  
$d(\tilde{\nddp},\nddp_{\cp})\le \xi$. 
The above remarks imply that we can apply Lemma \ref{WELemma} to any 
ensemble in $\cN(\xi)$.
Then
\begin{align*}
\frac{1}{n}\expectation [H_{\graph}(X|Y)]  & =  \sum_{\tilde{\nddp}}
\pr_{\cp}(\tilde{\nddp})\, 
\tilde{\expectation}[ \rate_{\graph(\epsilon)}]  \\
& =  \!\!\!\sum_{\tilde{\nddp}\in {\cal N}(\xi)}\!\!\!
\pr_{\epsilon}(\tilde{\nddp})\, 
\tilde{\expectation}[ \rate_{\graph(\epsilon)} ]  + \omega(n,\xi).
\end{align*}
The remainder can be estimated by noticing that 
$\rate_{\graph(\epsilon)}\le 1$ while the probability of
$\tilde{\nddp}\not\in {\cal N}(\epsilon)$ is bounded by Lemma
\ref{ConcentrationDegreeLemma}. Therefore
\begin{align*}
\lim_{n\to\infty}\omega(n,\xi) & = 0\, .
\end{align*}

Now we can apply Lemma \ref{WELemma} to get
\begin{align*}
\left|\frac{1}{n}\expectation [ H_{\graph}(X|Y)] - 
\drate_{\cp}\right| \le &~~ \!\!\!\!\!\!
\sum_{\tilde{\nddp}\in {\cal N}(\xi)}\!\!\!\!
\pr_{\epsilon}(\tilde{\nddp})\, |\tilde{\expectation}[ \rate_{\graph(\epsilon)}]-
\drate(\tilde{\nddp})|  \\
&~+\!\!\!\!\!\!\sum_{\tilde{\nddp}\in {\cal N}(\xi)}\!\!\!\!
\pr_{\epsilon}(\tilde{\nddp})\, |\drate(\tilde{\nddp})-\drate_{\cp}| +
\omega(n,\xi) \\
\le&~ \!\!\!\!\!\!\sum_{\tilde{\nddp}\in {\cal N}(\xi)}\!\!\!\!
\pr_{\epsilon}(\tilde{\nddp})\, |\drate(\tilde{\nddp})-\drate_{\cp}| +
\omega'(n,\xi)\, ,
\end{align*}
where $\omega'(n,\xi) = \omega(n,\xi)+C\log n/n$.
Notice that there exist $B>0$ such that for any pair $\nddp_1$, $\nddp_2$
\begin{align*}
|\drate(\nddp_1)-\drate(\nddp_2)| & \le
B\, d(\nddp_1;\nddp_2)\, .
\end{align*}
Therefore,
\begin{align*}
\lim_{n\to\infty}\left|\frac{1}{n}\expectation [H_{\graph}(X|Y)] - 
\drate_{\cp}\right|&\le
B \xi\, .
\end{align*}
The claim follows by noticing that $\xi$ can be chosen arbitrarily small.
\eproof

Theorem \ref{CountingTheorem} allows to compute the exact
MAP threshold whenever the required conditions are verified.
An explicit characterization is given below.
\bcor
Consider transmission over BEC$(\cp)$ using  
elements picked uniformly at random from the  ensemble $(\lnode,\Gamma)$. Let $\xl^*, \xr^*>0$
be the DE fixed-point achieved by the BP
decoder at a non-exceptional 
erasure probability $\cp^*$ (i.e., $\xl^*\in (0,1]$ is the largest 
solution of $\xl^* = \cp^*\ledge(1-\redge(1-\xl^*))$).
Assume that $P_{\epsilon^*}(\xl^*,\xr^*)=0$ and that 
$\Psi_{\nddp_{\epsilon^*}}(u)\le 0$ for $u\in[0,+\infty)$ together with
$\Psi''_{\nddp_{\epsilon^*}}(1)<0$. Let $\cW\subseteq[0,+\infty)$ be
the set of points $u\neq 1$ such that $\Psi_{\nddp_{\epsilon^*}}(u)= 0$.
If, for any $u\in\cW$, $\partial_{\cp}\Psi_{\nddp_{\epsilon^*}}(u)<
\partial_{\cp}\Psi_{\nddp_{\epsilon^*}}(1)$, then $\cp^\MAP = \cp^*$.
\ecor

\bproof
We claim that there exist a $\delta>0$ such that
the hypothesis of Theorem \ref{CountingTheorem} are verified for any 
$\cp\in (\cp^*,\cp^*+\delta)$. Before proving this claim,
let us show that it implies the thesis. Consider any $\cp\in (\cp^*,\cp^*+\delta)$
and let $\xl,\xr$ be the corresponding density evolution fixed point.
Then 
\begin{align*}
\lim_{n\to\infty}\frac{1}{n}\expectation[H(X|Y)] &= 
P_{\cp}(\xl(\cp),\xr(\cp))\;\;\;\;\;\;\; \forall  
\cp\in (\cp^*,\cp^*+\delta)\, .
\end{align*}
Moreover $P_{\cp^*}(\xl(\cp^*),\xr(\cp^*))=0$ by hypothesis
and
\begin{align*}
\frac{\de\phantom{\cp}}{\de\cp}P_{\cp}(\xl(\cp),\xr(\cp)) & = \lnode(\xr(\cp))>0
\, .
\end{align*}
Therefore $P_{\cp}(\xl(\cp),\xr(\cp))>0$ for any $\cp>\cp^*$. This implies
$\cp^\MAP\le\cp^*$. On the other hand $\expectation[H(X|Y)]$ is strictly
increasing with $\cp$. This implies
\begin{align*}
\lim_{n\to\infty}\frac{1}{n}\expectation[H(X|Y)] &= 0,
\;\;\;\;\;\;\; \forall  \cp\in [0,\cp^*] ,
\end{align*}
which in turn implies $\cp^\MAP\ge \cp^*$ and, therefore, $\cp^\MAP = \cp^*$.

Let us now prove the claim. By assumption $\cp^*$ is non-exceptional
and therefore the residual \ddp $\nddp_{\cp}$ is  continuous at
$\cp^*$. This implies, via Lemma \ref{PsiProp}
that, for any $\xi>0$, there exist $\delta$ such that
for $\cp\in[\cp^*,\cp^*+\delta)$ and any $u\in[0,1]$,
\begin{align*}
|\Psi_{\nddp_\cp}(u)-\Psi_{\nddp_{\cp^*}}(u)| & \le \xi (1-u)^2\, .
\end{align*}
Together with $\Psi''_{\nddp_\cp}(1)<0$, this implies that, if 
$\delta$ is small enough, $u=1$ is a local maximum of $ \Psi_{\nddp_\cp}(u)$.
It follows from the hypotheses on $\partial_{\cp}\Psi_{\nddp_{\cp^*}}(u)$,
$u\in\cW$, that it is also a global maximum.
\eproof

\begin{figure}
\centering
\setlength{\unitlength}{0.5bp}
\begin{picture}(210,230) 
\put(19,10)
{
\put(0,0){\includegraphics[scale=0.62]{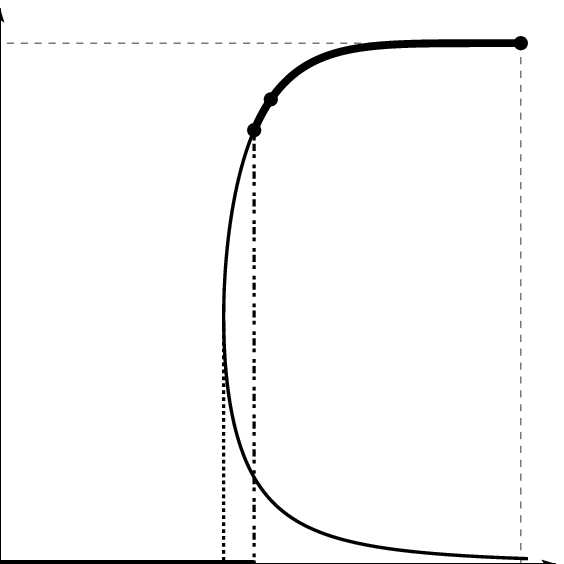}}
\setlength{\unitlength}{0.62bp}
\put(7,177){\makebox(0,0){\small{$\xh^\EBP$}}}
\put(170,0){\makebox(0,0){\small{$\ih$}}}
\put(-8,-8){\makebox(0,0){\small{$0$}}}
\put(83,-8){\makebox(0,0){\small{$\ih^\MAP$}}}
\put(153,-8){\makebox(0,0){\small{$1$}}}
\put(-8,154){\makebox(0,0){\small{$1$}}}
\put(150,160){\makebox(0,0){\small{A}}}
\put(90,132){\makebox(0,0){\small{B}}}
\put(69,130){\makebox(0,0){\small{C}}}
}
\end{picture}
\caption{\label{fig:36ebpexit} 
(E)BP \exitentropy function $\xh^\EBP(\ih)$.}
\vspace{2bp}
\end{figure}

The conditions in the above corollary are relatively easy to verify.
Let us demonstrate this by means of two examples.
\bex[Ensemble LDPC($x^2,x^5$)] Consider the  
$(3,6)$-regular LDPC ensemble. For convenience of
the reader its EBP EXIT curve is
repeated in Fig.~\ref{fig:36ebpexit}. 

\begin{figure}[htp]
\centering
\setlength{\unitlength}{1bp}
\begin{picture}(240,70)
\put(10,0){\includegraphics[width=60bp]{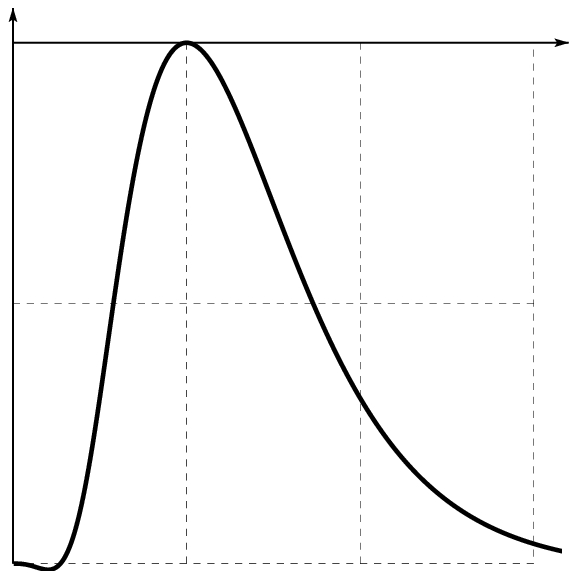}}
\put(90,0){\includegraphics[width=60bp]{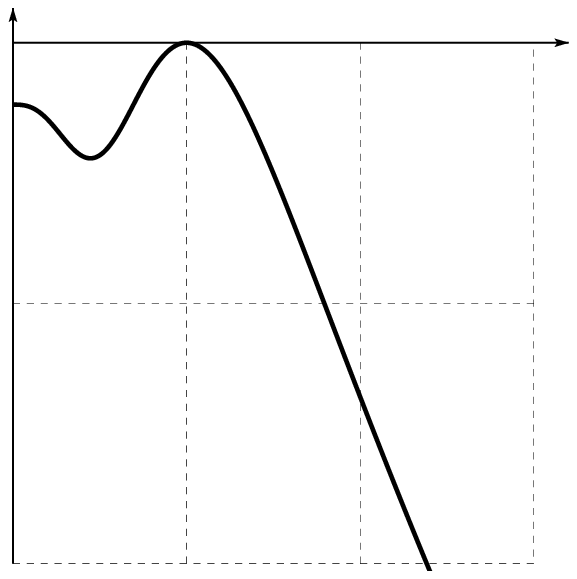}}
\put(170,0){\includegraphics[width=60bp]{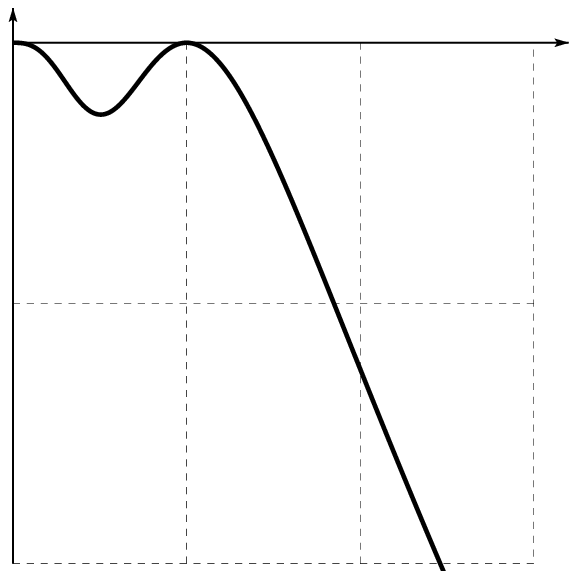}}
\put(10,0)
 {
 \put(30,20){\makebox(0,0){A}}
 \put(58,-3){\makebox(0,0){$\scriptstyle{3}$}}
 \put(-3,58){\makebox(0,0){$\scriptstyle{0}$}}
 \put(19,-3){\makebox(0,0){$\scriptstyle{1}$}}
 \put(-6,29){\makebox(0,0){$\scriptstyle{-\frac14}$}}
 \put(-6,0){\makebox(0,0){$\scriptstyle{-\frac12}$}}
 }
\put(90,0)
 {
 \put(30,20){\makebox(0,0){B}}
 \put(58,-3){\makebox(0,0){$\scriptstyle{3}$}}
 \put(-3,58){\makebox(0,0){$\scriptstyle{0}$}}
 \put(19,-3){\makebox(0,0){$\scriptstyle{1}$}}
 \put(-6,29){\makebox(0,0){$\scriptstyle{-\frac14}$}}
 \put(-6,0){\makebox(0,0){$\scriptstyle{-\frac12}$}}
 }
\put(170,0)
 {
 \put(30,20){\makebox(0,0){C}}
 \put(58,-3){\makebox(0,0){$\scriptstyle{3}$}}
 \put(-3,58){\makebox(0,0){$\scriptstyle{0}$}}
 \put(19,-3){\makebox(0,0){$\scriptstyle{1}$}}
 \put(-6,29){\makebox(0,0){$\scriptstyle{-\frac14}$}}
 \put(-6,0){\makebox(0,0){$\scriptstyle{-\frac12}$}}
 }
\end{picture}
\caption{
Function $\Psi_{\nddp}(u)$ for the \ddp formed by the residual 
ensemble in A, B and C.
}
\label{fig:wd36}
\end{figure}

Let us apply Theorem \ref{CountingTheorem}. 
We start with $\ih_A=1$ (point A).
The residual degree distribution at this point 
corresponds of course to the $(3, 6)$-ensemble itself.
As shown in the left-most picture in Fig.~\ref{fig:wd36}, the
corresponding function $\Psi_{\nddp}(u)$ has only a single maximum
at $u=1$ and one can verify that 
$\Psi_{\nddp}''(1)<0$. Therefore, by Lemma \ref{WELemma}
we know that with high probability the rate of a randomly chosen element from this
ensemble is close to the design rate. 
Next, consider the point $\ih_B=0.52$ (point B). Again, the
conditions are verified, and therefore the conditional entropy
at this point is given by equation (\ref{equ:conditonalentropy}).
We get $H(X \mid Y(\ih_B)) \approx 0.02755$.
Finally, consider the ``critical'' point $\ih_C \approx 0.48815$.
As one can see from the right-most picture in Fig.~\ref{fig:wd36},
this is the point at which a second global maximum appears. Just to the right
of the point the conditions of Theorem \ref{CountingTheorem}
are still fulfilled, whereas to the left of it they are violated.
Further, at this point Eq. (\ref{equ:conditonalentropy})
states that $H(X \mid Y(\ih_C)) = 0$. We conclude that
$\ih^\MAP = \ih_C \approx 0.48815$, confirming our result from 
Example \ref{exa:36upperbound}. Since the bound is tight at the MAP
threshold it follows that $\xh^\MAP = \xh^\BP$ for all points ``to the right''
of the MAP threshold (this is true since $\xh^\MAP \leq \xh^\BP$ always,
and the tightness of the bound at the MAP threshold
shows that the area under $\xh^\BP$ is exactly equal to the rate).
We see that in this simple case Theorem \ref{CountingTheorem}
allows us to construct the complete MAP \exitentropy curve.
\eex

\begin{figure}
\centering
\setlength{\unitlength}{0.5bp}
\begin{picture}(210,230) 
\put(19,10)
{
\put(0,0){\includegraphics[scale=0.62]{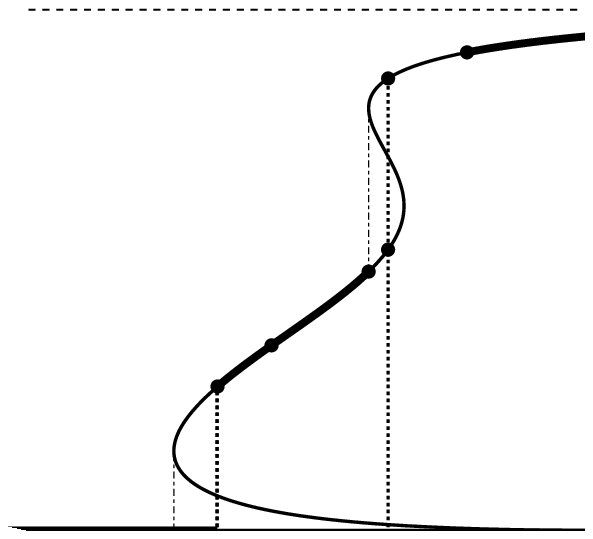}}
\setlength{\unitlength}{0.62bp}
\put(127,105){\makebox(0,0){\small{$\xh^\EBP$}}}
\put(170,0){\makebox(0,0){\small{$\ih$}}}
\put(-8,-8){\makebox(0,0){\small{$0$}}}
\put(60,-8){\makebox(0,0){\small{$\ih^\MAP$}}}
\put(106,-8){\makebox(0,0){\small{$\ih^{\MAP,2}$}}}
\put(153,-8){\makebox(0,0){\small{$1$}}}
\put(-8,154){\makebox(0,0){\small{$1$}}}
\put(154,160){\makebox(0,0){\small{A$\rightarrow1$}}}
\put(128,130){\makebox(0,0){\small{B}}}
\put(105,142){\makebox(0,0){\small{C}}}
\put(111,77){\makebox(0,0){\small{D}}}
\put(90,83){\makebox(0,0){\small{E}}}
\put(69,70){\makebox(0,0){\small{F}}}
 \put(53,54){\makebox(0,0){\small{G}}}
}
\end{picture}
\caption{\label{fig:2Jebpexit} 
(E)BP \exitentropy function $\xh^\EBP(\ih)$.
}
\vspace{2bp}
\end{figure}

\bex[Ensemble LDPC($\frac{3x+3x^2+4x^{13}}{10},x^6$)] 
\label{equ:doublejumpexample}
Consider the ensemble described in Fig.~\ref{fig:multijump}. Its 
EPB EXIT curve is repeated for the convenience of the reader in 
Fig.~\ref{fig:2Jebpexit}. 
The corresponding BP EXIT curve is shown in detail in Fig.~\ref{fig:epsilon}.
A further discussion of this ensemble can be found in Example \ref{ex:running2jumps}. 
\begin{figure}[htp]
\centering
\setlength{\unitlength}{1bp}
\begin{picture}(240,150)
\put(0,80)
{
\put(10,0){\includegraphics[width=60bp]{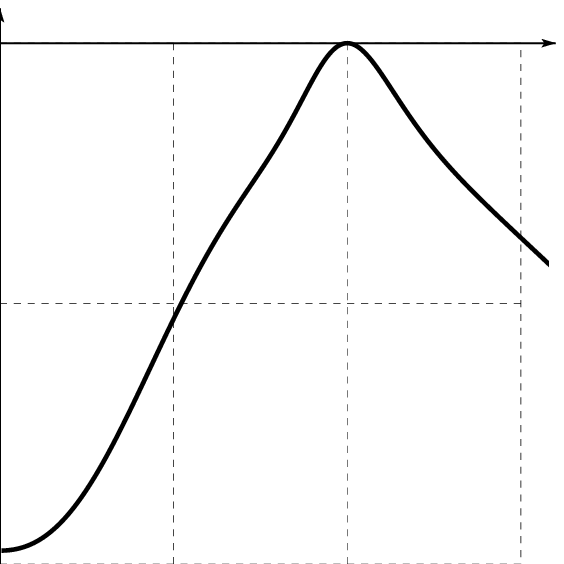}}
\put(90,0){\includegraphics[width=60bp]{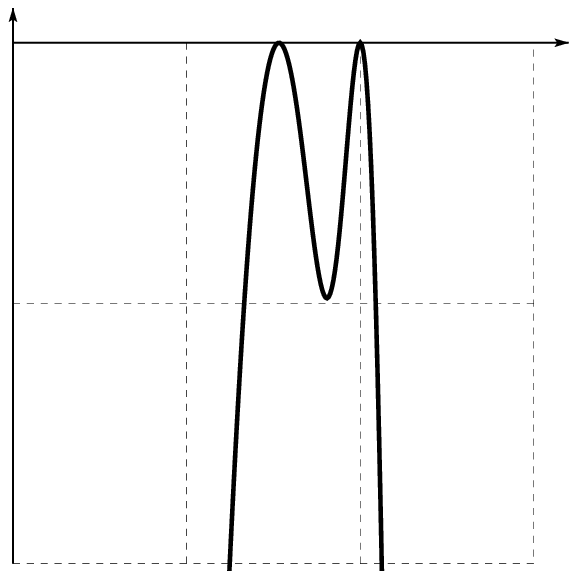}}
\put(170,0){\includegraphics[width=60bp]{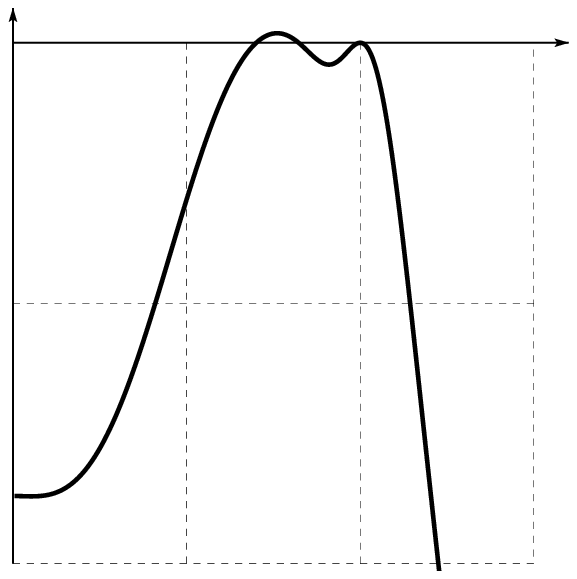}}

\put(10,0)
 {
 \put(30,20){\makebox(0,0){A}}
 \put(58,-3){\makebox(0,0){$\scriptstyle{\frac32}$}}
 \put(-3,58){\makebox(0,0){$\scriptstyle{0}$}}
 \put(38,-3){\makebox(0,0){$\scriptstyle{1}$}}
 \put(-6,29){\makebox(0,0){$\scriptstyle{-\frac14}$}}
 \put(-6,0){\makebox(0,0){$\scriptstyle{-\frac12}$}}
 }
\put(90,0)
 {
 \put(30,20){\makebox(0,0){B}}
 \put(58,-3){\makebox(0,0){$\scriptstyle{\frac32}$}}
 \put(-3,58){\makebox(0,0){$\scriptstyle{0}$}}
 \put(38,-3){\makebox(0,0){$\scriptstyle{1}$}}
 \put(-6,29){\makebox(0,0){$\scriptstyle{-\frac{1}{400}}$}}
 \put(-6,0){\makebox(0,0){$\scriptstyle{-\frac{1}{200}}$}}
 }
\put(170,0)
 {
 \put(30,20){\makebox(0,0){C}}
 \put(58,-3){\makebox(0,0){$\scriptstyle{\frac32}$}}
 \put(-3,58){\makebox(0,0){$\scriptstyle{0}$}}
 \put(38,-3){\makebox(0,0){$\scriptstyle{1}$}}
 \put(-6,29){\makebox(0,0){$\scriptstyle{-\frac{1}{40}}$}}
 \put(-6,0){\makebox(0,0){$\scriptstyle{-\frac{1}{20}}$}}
 }
}
\put(0,0)
{

\put(10,0)
 {
 \put(0,0){\includegraphics[width=60bp]{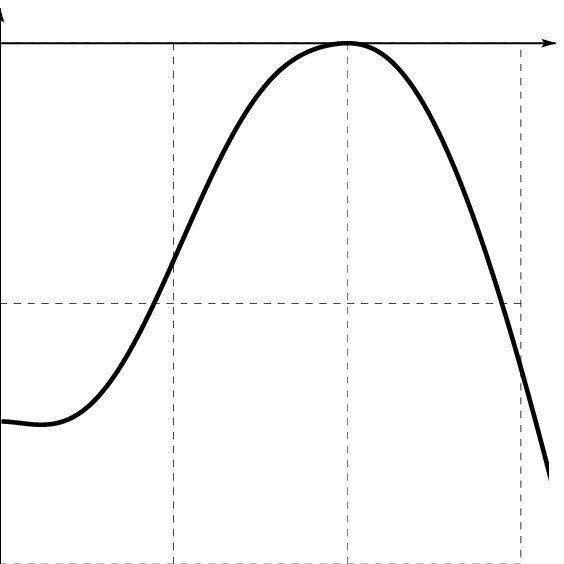}}
 \put(30,20){\makebox(0,0){E}}
 \put(58,-3){\makebox(0,0){$\scriptstyle{\frac32}$}}
 \put(-3,58){\makebox(0,0){$\scriptstyle{0}$}}
 \put(38,-3){\makebox(0,0){$\scriptstyle{1}$}}
 \put(-6,29){\makebox(0,0){$\scriptstyle{-\frac{1}{40}}$}}
 \put(-6,0){\makebox(0,0){$\scriptstyle{-\frac{1}{20}}$}}
 }
\put(90,0)
 {
 \put(0,0){\includegraphics[width=60bp]{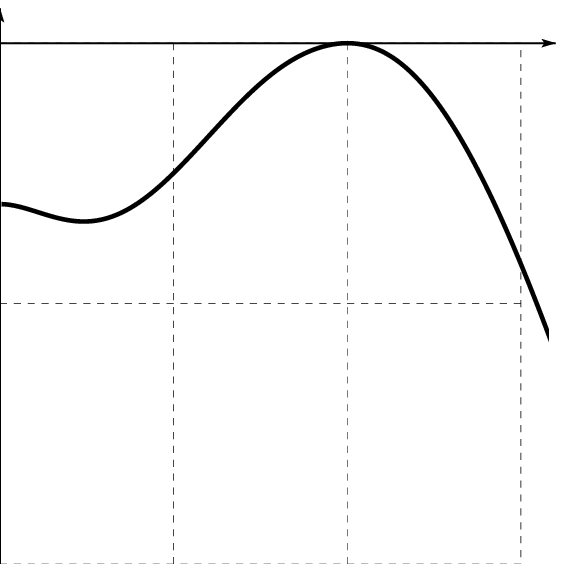}}
 \put(30,20){\makebox(0,0){F}}
 \put(58,-3){\makebox(0,0){$\scriptstyle{\frac32}$}}
 \put(-3,58){\makebox(0,0){$\scriptstyle{0}$}}
 \put(38,-3){\makebox(0,0){$\scriptstyle{1}$}}
 \put(-6,29){\makebox(0,0){$\scriptstyle{-\frac{1}{40}}$}}
 \put(-6,0){\makebox(0,0){$\scriptstyle{-\frac{1}{20}}$}}
 }
\put(170,0)
 {
\put(0,0){\includegraphics[width=60bp]{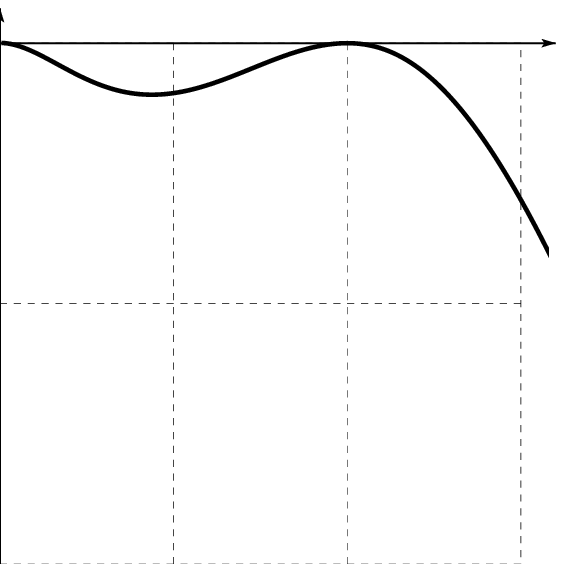}}
\put(30,20){\makebox(0,0){G}}
\put(58,-3){\makebox(0,0){$\scriptstyle{\frac32}$}}
\put(-3,58){\makebox(0,0){$\scriptstyle{0}$}}
\put(38,-3){\makebox(0,0){$\scriptstyle{1}$}}
\put(-6,29){\makebox(0,0){$\scriptstyle{-\frac{1}{40}}$}}
\put(-6,0){\makebox(0,0){$\scriptstyle{-\frac{1}{20}}$}}
 }
}
\end{picture}
\caption{
Function $\Psi_{\nddp}(u)$ for the \ddp formed by the residual 
ensemble in A, B, C, E, F and G.
}
\label{fig:2JWD}
\end{figure}
Let us again apply Theorem \ref{CountingTheorem}.
We start with $\ih_A=1$ (point A). The residual
degree distribution corresponds of course to the ensemble itself.
As the top left-most picture in Fig.~\ref{fig:2Jebpexit} shows,
the hypotheses are fulfilled and we conclude again that
with high probability the rate of a randomly chosen element from this
ensemble is close to the design rate which is equal to $\drate\approx0.4872$.
Now decrease $\ih$ smoothly. The conditions of Theorem \ref{CountingTheorem} stay
fulfilled until we get to $\epsilon_B \approx 0.5313$ (point B). At this point
a second global maximum of the function $\Psi_{\nddp}(u)$ occurs. 
As the pictures in the bottom row of
Fig.~\ref{fig:2Jebpexit} show, the hypotheses of Theorem \ref{CountingTheorem}
are again fulfilled over the whole segment from E (the
first threshold of the BP decoder corresponding to
$\ih_E \approx0.5156$) till G. In particular,
at the point G, which corresponds to $\ih_G = \ih^\MAP \approx 0.4913$,
the trial entropy reaches zero, which shows that this
is the MAP threshold.

We see that for this example Theorem \ref{CountingTheorem} allows
us to construct the MAP EXIT curve for the segment from A to B and the
segment from E to G. Over both these segments we have $\xh^\MAP=\xh^\BP$.
In summary, we can determine the MAP threshold and we see that the
balance condition applies ``at the jump G'' (the MAP threshold).
But the straightforward application of Theorem \ref{CountingTheorem}
does not provide us with a means of determining $\xh^\MAP$
between the points B and D. Intuitively, $\xh^\MAP$ should go from
B to C (which corresponds to $\ih^C \approx 0.5156$).
At this point one would hope that a local balance condition
again applies and that the MAP EXIT curve jumps
to the ``lower branch'' to point D. It should then continue smoothly
until the point G (the MAP threshold) at which it finally jumps to zero.
As we will discuss in more detail in Example \ref{ex:running2jumps}, after our analysis
of the \maxwell decoder, this is indeed true, and $\xh^\MAP$ is
as shown in Fig.~\ref{fig:multijump}.

Assuming Theorem \ref{CountingTheorem} applies,
we know that at the MAP threshold the matrix corresponding to the residual
graph becomes a full rank square matrix. What happens at the jump
at point C? At this point the matrix corresponding to the residual
graph takes, after some suitable swapping of
columns and rows, the form
\begin{align*}
\left(
\begin{array}{cc}
U & V \\
0 & W
\end{array}
\right),
\end{align*}
where $W$ is a full rank square matrix of dimension 
$\epsilon_C (\lnode(\xr_C)-\lnode(\xr_D))$.
The MAP decoder can therefore solve the part of the equation corresponding to the
submatrix $W$.
\eex


%
%
\section{Maxwell Construction}
\label{MaxwellSection}
\begin{figure}[htp]
\centering
\setlength{\unitlength}{1bp}
\begin{picture}(240,140)
\put(5,28)
{
\put(5,5){\includegraphics[width=100bp]{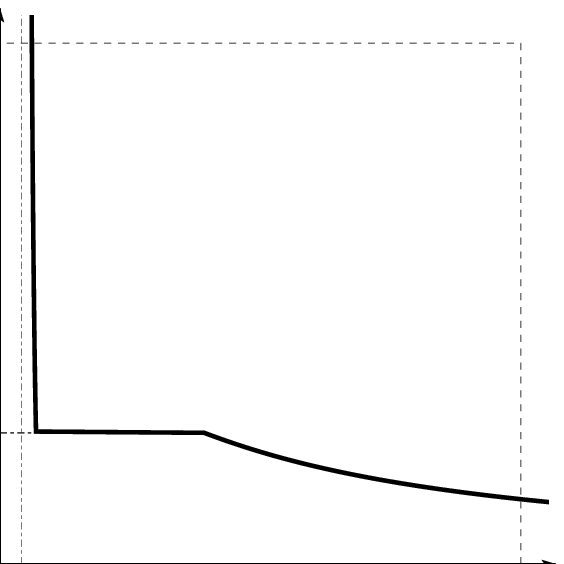}}
\put(135,5){\includegraphics[width=100bp]{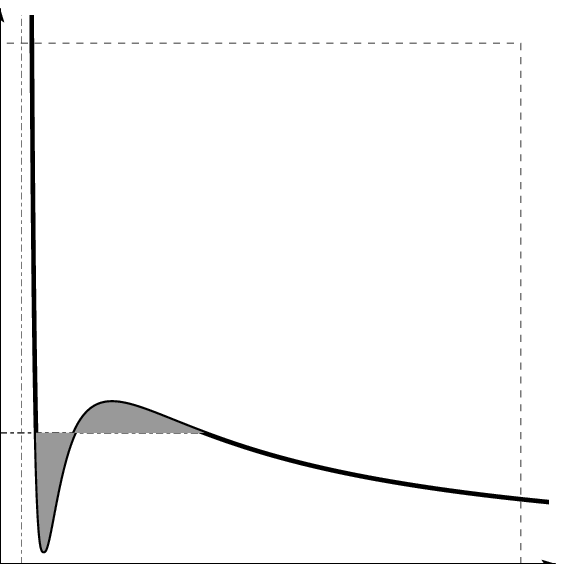}}
\put(0,0){\makebox(0,0){\small{$0$}}}
\put(100,0){\makebox(0,0){\small{${\scriptstyle{8}}$}}}
\put(9,0){\makebox(0,0){\small{${\scriptscriptstyle{\frac13}}$}}}
\put(0,100){\makebox(0,0){\small{$\scriptstyle2$}}}
\put(130,0){\makebox(0,0){\small{$0$}}}
\put(230,0){\makebox(0,0){\small{$\scriptstyle8$}}}
\put(139,0){\makebox(0,0){\small{${\scriptscriptstyle{\frac13}}$}}}
\put(130,100){\makebox(0,0){\small{$\scriptstyle2$}}}
\put(89,47){\makebox(0,0){\includegraphics[scale=0.16]{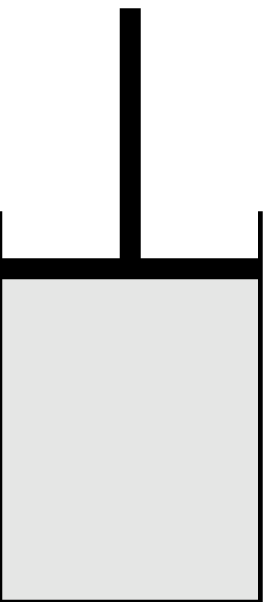}}}
\put(72,25){$\swarrow$}
\put(39,55){\makebox(0,0){\includegraphics[scale=0.16]{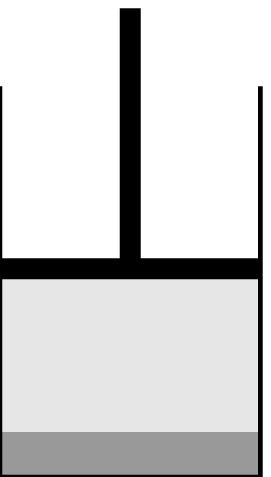}}}
\put(38,33){$\downarrow$}
\put(22,55){\makebox(0,0){\includegraphics[scale=0.16]{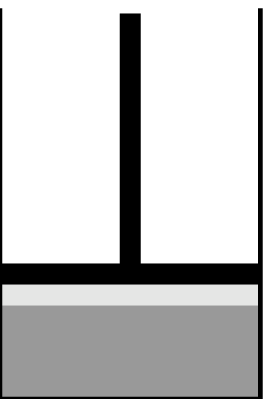}}}
\put(18,33){$\downarrow$}
\put(20,92){\makebox(0,0){\includegraphics[scale=0.16]{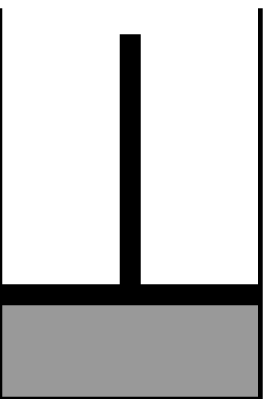}}}
\put(11,70){$\swarrow$}
\put(4,110){\makebox(0,0){$P$}}
\put(-4,30){\makebox(0,0){$P_c$}}
\put(110,4){\makebox(0,0){$V$}}
\put(179,-5){\makebox(0,0){\small{$~$}}}
\put(200,80){\makebox(0,0){\small{$~$}}}
\put(130,0)
  {
  \put(4,110){\makebox(0,0){$P$}}
  \put(-4,30){\makebox(0,0){$P_c$}}
  \put(110,4){\makebox(0,0){$V$}}
  }
}
\put(60,7){\makebox(0,0){{\small{(a)}}}}
\put(190,7){\makebox(0,0){{\small{(b)}}}}
\end{picture}
\caption{
Maxwell construction in thermodynamics.  
(a) Pressure-volume diagram for
the liquid-vapor phase transition (b) Van der Waals 
curve (using reduced variables, given by 
$(p+\frac{2}{V^2})(3V-1)=8T$ at the reduced temperature $T=0.85$ ) and
the Maxwell construction. 
Consider the case of a
liquid-gas phase transition of water. 
If a small amount of liquid is placed in a completely empty 
(and hermetically closed)
large container at room temperature, the water evaporates.
The vapor exerts pressure on the walls of the container.
By gradually reducing the volume of the container, 
 we increase the vapor pressure $P$ until it reaches a {\em critical} value $P_c$
(which depends on the temperature). At this point the vapor 
condensates into liquid water. The pressure stays constant throughout this
transformation.
When there is no space left for the vapor,
the pressure starts to rise again, and as shown in (a) it does so 
very quickly (since it is difficult to compress water).
In many theoretical descriptions of this phenomenon, a
non-monotonic  pressure-volume curve is obtained like in (b) with   
the Van Der Waals model.  The Maxwell construction
allows to modify the ``unphysical'' part of this curve and to obtain a
consistent result. We want to join the two decreasing branches of
the theoretical curve with a constant-pressure line, as observed in experiments.
 At which height should we placed the horizontal
line? The basic idea of the Maxwell construction is that, at the critical
pressure $P_c$, the vapor and the liquid are in {``{equilibrium}''}.
This means that we can transform an infinitesimal quantity of vapor
into liquid (or vice versa) without doing any {``{work}''} on the system.
Because of this reason, the vapor begins its transformation into
liquid at $P_c$. The work done on the system in an infinitesimal
transformation is $P\, \text{d}V$, where $\text{d}V$ represents the
variation of the volume. Using this fact, it can be shown that the
above equilibrium condition implies the equality of the  areas
of the two regions between the horizontal line and the original
non-monotonous pressure-volume curve. See, e.g.,~\cite{KiK02}. 
}
\label{fig:VanDerWaals}
\end{figure}

The balance condition described in Section \ref{OverviewSection}
and Section \ref{UpperSection} is strongly reminiscent of the 
well-known ``Maxwell construction'' in 
the theory of phase transitions. This is described briefly
in Fig.~\ref{fig:VanDerWaals}.

%
%
\subsection{Maxwell Decoder}
\label{MaxwellIntro}

Inspired by the statistical mechanics analogy, we will explain the
balance condition (shown on the right in Fig.~\ref{fig:exitcurve}) which determines
the MAP threshold by analyzing a ``BP decoder with guessing''. 
The state of the algorithm can be associated to a point moving
along the EBP \exitentropy curve. The evolution starts at the
point of full entropy and ends at zero entropy.
The analysis of this algorithm is also most conveniently 
phrased in terms of the EBP \exitentropy curve and implies a proof of
Theorem~\ref{ImprovedUB}.
Because of this balance condition we term this decoding algorithm
the Maxwell (M) decoder.  
 Note that a similar algorithm is discussed in \cite{PiF04} 
although it is motivated by some more practical concerns.

 Analogously to the usual BP decoder for the erasure channel,
the \maxwell 
decoder admits two equivalent descriptions: either as a {\em sequential} 
(i.e., bit-by-bit in the spirit of \cite{LMSS01b}) or as a {\em message-passing}
algorithm. While the former approach is more intuitive,
the latter allows for a simpler analysis.
We shall first describe the \maxwell decoder as a sequential 
procedure and sketch the main features of its behavior. In the next section
we will turn to a message-passing setting and complete its analysis. 

\begin{figure*}[hbt]
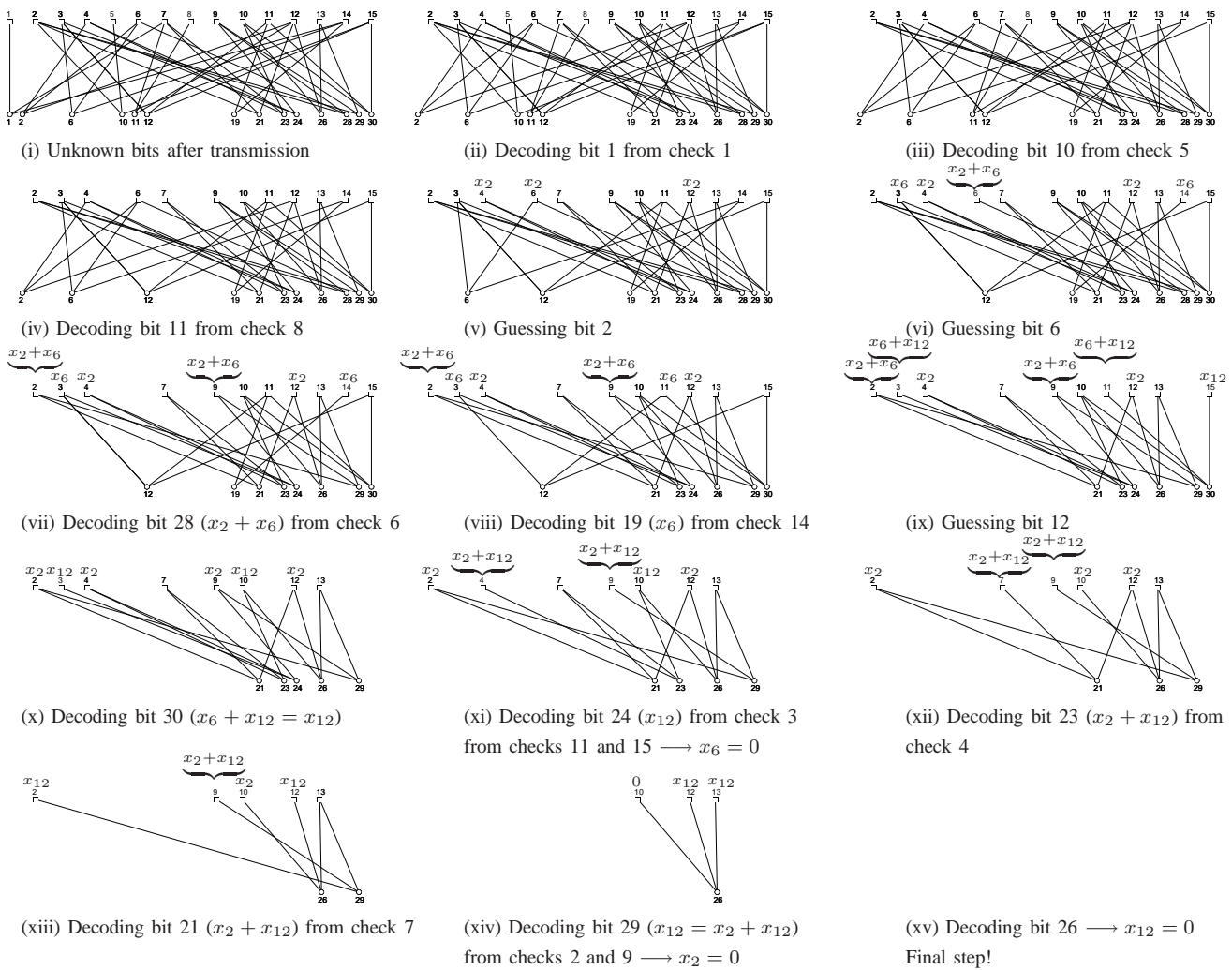

\begin{center}
\hfil
\newlength{\FigLength}
\setlength{\FigLength}{150bp}
\setlength{\unitlength}{0.625bp}
\begin{tabular}{lll}
\input{./maxwelldecodertutorial/residu1} &
\input{./maxwelldecodertutorial/residu2} &
\input{./maxwelldecodertutorial/residu3} \\
{\footnotesize (i) Unknown bits after transmission} 
& {\footnotesize (ii) Decoding  bit 1 from check 1}  
& {\footnotesize(iii)  Decoding  bit 10 from check 5}  \\
\input{./maxwelldecodertutorial/residu4} &
\input{./maxwelldecodertutorial/residu5} &
\input{./maxwelldecodertutorial/residu6}  \\
 {\footnotesize(iv)  Decoding  bit 11 from check 8} 
& {\footnotesize(v) Guessing bit 2} 
& {\footnotesize(vi) Guessing bit 6} \\
\input{./maxwelldecodertutorial/residu7} &
\input{./maxwelldecodertutorial/residu8} &
\input{./maxwelldecodertutorial/residu9} \\
{\footnotesize(vii) Decoding  bit 28 ({\footnotesize{$x_2+x_6$}}) from check 6} 
& {\footnotesize(viii) Decoding  bit 19 ({\footnotesize{$x_6$}}) from check 14} 
& {\footnotesize(ix) Guessing bit 12} \\
\input{./maxwelldecodertutorial/residu10} &
\input{./maxwelldecodertutorial/residu11} &
\input{./maxwelldecodertutorial/residu12}  \\
 {\footnotesize(x)  Decoding  bit 30 ({\footnotesize{$x_6+x_{12}=x_{12}$}})}
& {\footnotesize(xi)  Decoding  bit 24 ({\footnotesize{$x_{12}$}}) from check 3} 
& {\footnotesize(xii)   Decoding  bit 23 ({\footnotesize{$x_2+x_{12}$}}) from } \\
  {\footnotesize ~}               
&   {\footnotesize from checks 11 and 15 $\longrightarrow x_6=0$}  
&  {\footnotesize check 4}  \\
\input{./maxwelldecodertutorial/residu13} &
\input{./maxwelldecodertutorial/residu14} & \\
{\footnotesize(xiii)  Decoding  bit 21 ({\footnotesize{$x_2+x_{12}$}})  from check 7} 
& {\footnotesize(xiv)  Decoding  bit 29  ({\footnotesize{$x_{12}=x_2+x_{12}$}}) } 
& {\footnotesize(xv) Decoding  bit 26 $\longrightarrow  x_{12}=0$ }  \\
 {\footnotesize ~}                
&  {\footnotesize from checks 2 and 9 $\longrightarrow x_2=0$}  
& {\footnotesize Final step! } \\
\end{tabular}
\caption{\label{fig:maxwelldecoder} \maxwell decoder applied to a simple 
example: a $(3,6)$ LDPC code of length $\n=30$. Assume that the all-zero 
codeword has been transmitted. At the decoder, the 
received (i.e., known and equal to $0$) bits are removed from the 
bipartite graph. The remaining graph is shown in (i). The first 
phase is the standard BP algorithm: in the first three steps, the decoder 
proceeds as the standard BP decoder and determines the bits $1$, $10$ and 
$11$, until it gets stuck in a stopping set shown in (iv). The second 
phase is distinct to the \maxwell decoder: it is the guessing/contradiction 
phase. The decoder guesses the (randomly chosen) bit $2$: this means that
it creates two simultaneously running copies, one which proceeds under
the assumption that bit $2$ takes the value $0$, the other which assumes
that this bit takes the value $1$. The decoder
then proceeds as the standard BP algorithm. Any time it gets stuck, it 
guesses a new bit and duplicates the number of simultaneously running copies. 
This process continues until a contradiction occurs, e.g., at the 
9$^{\text{th}}$ step $(ix)$: the variable node $x_{30}$ (either 
$x_{30}=0$ or $x_{30}=1$ depending of which copy we are considering) 
is connected to two check nodes
of degree one. The incoming messages from those nodes are $x_6+x_{12}$ 
and $x_{12}$, respectively. 
Consistency now requires that $x_6+x_{12}=x_{12}$, i.e., that $x_6=0$, 
such that only the decoding copies corresponding to $x_6=0$ survive. 
Phases of guessing and phases of standard BP decoding might alternate. 
Decoding is successful (in the sense that a MAP decoder would
have succeeded) if only a single copy survives at the very end
of the decoding process. ``Contradictions'' can be seen as ``confirmations'' 
or ``conditions'' in this message-passing setting. 
}
\end{center}
\hrulefill
\vspace*{4pt}
\end{figure*}
Given the received word which was
transmitted over  BEC$(\ih)$, the decoder proceeds iteratively as does
the standard BP decoder. At each time step a parity-check equation
involving a single undetermined variable is chosen and used to 
determine the value of the variable. This value is substituted in any 
parity-check equation involving the same variable.
If at any time the iterative
decoding process gets stuck in a non-empty stopping set,
a position $i\in [\n]$ is chosen uniformly at random. The decoder is 
said to {\em guess} a bit. 
If the bit associated to this position is not known yet,
the decoder replicates\footnote{Here we describe the decoder 
as a `breadth-first' search procedure: at each bifurcation we explore 
in parallel all the available options. 
One can easily construct an equivalent `depth-first' search:
first take a complete sequence of choices and, if no codeword is 
found, backtrack.} any running copy of the decoding process, 
and it proceeds by running one copy of each process
under the assumption
that $x_i=\0$ and the other one under the assumption that  $x_i=\1$. 

\begin{figure}[htp]
\centering
\setlength{\unitlength}{0.8bp}
\newlength{\longueur}
\setlength{\longueur}{240bp}
\newcounter{Subdi}
\setcounter{Subdi}{20}
\newcounter{Step}
\setcounter{Step}{19} 
\newcounter{Shift}
\setcounter{Shift}{0}
\begin{picture}(300,120)
\put(4,0)
{
 \put(0,0){\includegraphics[width=\longueur]{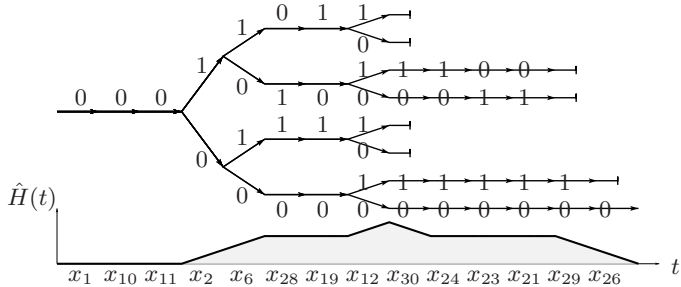}}
\put(-3,38){\begin{turn}{0}\makebox(0,0){{\small $\entropy(\iter)$} }\end{turn}}
\put(300,5){\begin{turn}{0}\makebox(0,0){{\small $\iter$} }\end{turn}}
\put(\value{Subdi},0){\begin{turn}{0}\makebox(0,0){{\small $x_1$} }\end{turn}}
\addtocounter{Subdi}{\value{Step}}
\put(\value{Subdi},0){\begin{turn}{0}\makebox(0,0){{\small $x_{10}$} }\end{turn}}
\addtocounter{Subdi}{\value{Step}}
\put(\value{Subdi},0){\begin{turn}{0}\makebox(0,0){{\small $x_{11}$} }\end{turn}}
\addtocounter{Subdi}{\value{Step}}
\put(\value{Subdi},0){\begin{turn}{0}\makebox(0,0){{\small $x_{2}$} }\end{turn}}
\addtocounter{Subdi}{\value{Step}}
\put(\value{Subdi},0){\begin{turn}{0}\makebox(0,0){{\small $x_{6}$} }\end{turn}}
\addtocounter{Subdi}{\value{Step}}
\put(\value{Subdi},0){\begin{turn}{0}\makebox(0,0){{\small $x_{28}$} }\end{turn}}
\addtocounter{Subdi}{\value{Step}}
\put(\value{Subdi},0){\begin{turn}{0}\makebox(0,0){{\small $x_{19}$} }\end{turn}}
\addtocounter{Subdi}{\value{Step}}
\put(\value{Subdi},0){\begin{turn}{0}\makebox(0,0){{\small $x_{12}$} }\end{turn}}
\addtocounter{Subdi}{\value{Step}}
\put(\value{Subdi},0){\begin{turn}{0}\makebox(0,0){{\small $x_{30}$} }\end{turn}}
\addtocounter{Subdi}{\value{Step}}
\put(\value{Subdi},0){\begin{turn}{0}\makebox(0,0){{\small $x_{24}$} }\end{turn}}
\addtocounter{Subdi}{\value{Step}}
\put(\value{Subdi},0){\begin{turn}{0}\makebox(0,0){{\small $x_{23}$} }\end{turn}}
\addtocounter{Subdi}{\value{Step}}
\put(\value{Subdi},0){\begin{turn}{0}\makebox(0,0){{\small $x_{21}$} }\end{turn}}
\addtocounter{Subdi}{\value{Step}}
\put(\value{Subdi},0){\begin{turn}{0}\makebox(0,0){{\small $x_{29}$} }\end{turn}}
\addtocounter{Subdi}{\value{Step}}
\put(\value{Subdi},0){\begin{turn}{0}\makebox(0,0){{\small $x_{26}$} }\end{turn}}
\addtocounter{Subdi}{\value{Step}}
\setcounter{Subdi}{20}
\addtocounter{Shift}{85}
\put(\value{Subdi},\value{Shift}){\begin{turn}{0}\makebox(0,0){{\small $0$} }\end{turn}}
\addtocounter{Subdi}{\value{Step}}
\put(\value{Subdi},\value{Shift}){\begin{turn}{0}\makebox(0,0){{\small $0$} }\end{turn}}
\addtocounter{Subdi}{\value{Step}}
\put(\value{Subdi},\value{Shift}){\begin{turn}{0}\makebox(0,0){{\small $0$} }\end{turn}}
\addtocounter{Subdi}{\value{Step}}
\addtocounter{Shift}{15}
\put(\value{Subdi},\value{Shift}){\begin{turn}{0}\makebox(0,0){{\small $1$} }\end{turn}}
\addtocounter{Shift}{-44}
\put(\value{Subdi},\value{Shift}){\begin{turn}{0}\makebox(0,0){{\small $0$} }\end{turn}}
\addtocounter{Subdi}{\value{Step}}
\addtocounter{Shift}{62}
\put(\value{Subdi},\value{Shift}){\begin{turn}{0}\makebox(0,0){{\small $1$} }\end{turn}}
\addtocounter{Shift}{-28}
\put(\value{Subdi},\value{Shift}){\begin{turn}{0}\makebox(0,0){{\small $0$} }\end{turn}}
\addtocounter{Shift}{-24}
\put(\value{Subdi},\value{Shift}){\begin{turn}{0}\makebox(0,0){{\small $1$} }\end{turn}}
\addtocounter{Shift}{-27}
\put(\value{Subdi},\value{Shift}){\begin{turn}{0}\makebox(0,0){{\small $0$} }\end{turn}}
\addtocounter{Subdi}{\value{Step}}
\addtocounter{Shift}{86}
\put(\value{Subdi},\value{Shift}){\begin{turn}{0}\makebox(0,0){{\small $0$} }\end{turn}}
\addtocounter{Shift}{-40}
\put(\value{Subdi},\value{Shift}){\begin{turn}{0}\makebox(0,0){{\small $1$} }\end{turn}}
\addtocounter{Shift}{-13}
\put(\value{Subdi},\value{Shift}){\begin{turn}{0}\makebox(0,0){{\small $1$} }\end{turn}}
\addtocounter{Shift}{-40}
\put(\value{Subdi},\value{Shift}){\begin{turn}{0}\makebox(0,0){{\small $0$} }\end{turn}}
\addtocounter{Subdi}{\value{Step}}
\addtocounter{Shift}{93}
\put(\value{Subdi},\value{Shift}){\begin{turn}{0}\makebox(0,0){{\small $1$} }\end{turn}}
\addtocounter{Shift}{-40}
\put(\value{Subdi},\value{Shift}){\begin{turn}{0}\makebox(0,0){{\small $0$} }\end{turn}}
\addtocounter{Shift}{-13}
\put(\value{Subdi},\value{Shift}){\begin{turn}{0}\makebox(0,0){{\small $1$} }\end{turn}}
\addtocounter{Shift}{-40}
\put(\value{Subdi},\value{Shift}){\begin{turn}{0}\makebox(0,0){{\small $0$} }\end{turn}}
\addtocounter{Subdi}{\value{Step}}
\addtocounter{Shift}{93}
\put(\value{Subdi},\value{Shift}){\begin{turn}{0}\makebox(0,0){{\small $1$} }\end{turn}}
\addtocounter{Shift}{-15}
\put(\value{Subdi},\value{Shift}){\begin{turn}{0}\makebox(0,0){{\small $0$} }\end{turn}}
\addtocounter{Shift}{-12}
\put(\value{Subdi},\value{Shift}){\begin{turn}{0}\makebox(0,0){{\small $1$} }\end{turn}}
\addtocounter{Shift}{-13}
\put(\value{Subdi},\value{Shift}){\begin{turn}{0}\makebox(0,0){{\small $0$} }\end{turn}}
\addtocounter{Shift}{-13}
\put(\value{Subdi},\value{Shift}){\begin{turn}{0}\makebox(0,0){{\small $1$} }\end{turn}}
\addtocounter{Shift}{-12}
\put(\value{Subdi},\value{Shift}){\begin{turn}{0}\makebox(0,0){{\small $0$} }\end{turn}}
\addtocounter{Shift}{-15}
\put(\value{Subdi},\value{Shift}){\begin{turn}{0}\makebox(0,0){{\small $1$} }\end{turn}}
\addtocounter{Shift}{-13}
\put(\value{Subdi},\value{Shift}){\begin{turn}{0}\makebox(0,0){{\small $0$} }\end{turn}}
\addtocounter{Subdi}{\value{Step}}
\addtocounter{Shift}{93}
\put(\value{Subdi},\value{Shift}){\begin{turn}{0}\makebox(0,0){{\small $~$} }\end{turn}}
\addtocounter{Shift}{-15}
\put(\value{Subdi},\value{Shift}){\begin{turn}{0}\makebox(0,0){{\small $~$} }\end{turn}}
\addtocounter{Shift}{-12}
\put(\value{Subdi},\value{Shift}){\begin{turn}{0}\makebox(0,0){{\small $1$} }\end{turn}}
\addtocounter{Shift}{-13}
\put(\value{Subdi},\value{Shift}){\begin{turn}{0}\makebox(0,0){{\small $0$} }\end{turn}}
\addtocounter{Shift}{-13}
\put(\value{Subdi},\value{Shift}){\begin{turn}{0}\makebox(0,0){{\small $~$} }\end{turn}}
\addtocounter{Shift}{-12}
\put(\value{Subdi},\value{Shift}){\begin{turn}{0}\makebox(0,0){{\small $~$} }\end{turn}}
\addtocounter{Shift}{-15}
\put(\value{Subdi},\value{Shift}){\begin{turn}{0}\makebox(0,0){{\small $1$} }\end{turn}}
\addtocounter{Shift}{-13}
\put(\value{Subdi},\value{Shift}){\begin{turn}{0}\makebox(0,0){{\small $0$} }\end{turn}}
\addtocounter{Shift}{93}
\addtocounter{Subdi}{\value{Step}}
\put(\value{Subdi},\value{Shift}){\begin{turn}{0}\makebox(0,0){{\small $~$} }\end{turn}}
\addtocounter{Shift}{-15}
\put(\value{Subdi},\value{Shift}){\begin{turn}{0}\makebox(0,0){{\small $~$} }\end{turn}}
\addtocounter{Shift}{-12}
\put(\value{Subdi},\value{Shift}){\begin{turn}{0}\makebox(0,0){{\small $1$} }\end{turn}}
\addtocounter{Shift}{-13}
\put(\value{Subdi},\value{Shift}){\begin{turn}{0}\makebox(0,0){{\small $0$} }\end{turn}}
\addtocounter{Shift}{-13}
\put(\value{Subdi},\value{Shift}){\begin{turn}{0}\makebox(0,0){{\small $~$} }\end{turn}}
\addtocounter{Shift}{-12}
\put(\value{Subdi},\value{Shift}){\begin{turn}{0}\makebox(0,0){{\small $~$} }\end{turn}}
\addtocounter{Shift}{-15}
\put(\value{Subdi},\value{Shift}){\begin{turn}{0}\makebox(0,0){{\small $1$} }\end{turn}}
\addtocounter{Shift}{-13}
\put(\value{Subdi},\value{Shift}){\begin{turn}{0}\makebox(0,0){{\small $0$} }\end{turn}}
\addtocounter{Subdi}{\value{Step}}
\addtocounter{Shift}{93}
\put(\value{Subdi},\value{Shift}){\begin{turn}{0}\makebox(0,0){{\small $~$} }\end{turn}}
\addtocounter{Shift}{-15}
\put(\value{Subdi},\value{Shift}){\begin{turn}{0}\makebox(0,0){{\small $~$} }\end{turn}}
\addtocounter{Shift}{-12}
\put(\value{Subdi},\value{Shift}){\begin{turn}{0}\makebox(0,0){{\small $0$} }\end{turn}}
\addtocounter{Shift}{-13}
\put(\value{Subdi},\value{Shift}){\begin{turn}{0}\makebox(0,0){{\small $1$} }\end{turn}}
\addtocounter{Shift}{-13}
\put(\value{Subdi},\value{Shift}){\begin{turn}{0}\makebox(0,0){{\small $~$} }\end{turn}}
\addtocounter{Shift}{-12}
\put(\value{Subdi},\value{Shift}){\begin{turn}{0}\makebox(0,0){{\small $~$} }\end{turn}}
\addtocounter{Shift}{-15}
\put(\value{Subdi},\value{Shift}){\begin{turn}{0}\makebox(0,0){{\small $1$} }\end{turn}}
\addtocounter{Shift}{-13}
\put(\value{Subdi},\value{Shift}){\begin{turn}{0}\makebox(0,0){{\small $0$} }\end{turn}}
\addtocounter{Subdi}{\value{Step}}
\addtocounter{Shift}{93}
\put(\value{Subdi},\value{Shift}){\begin{turn}{0}\makebox(0,0){{\small $~$} }\end{turn}}
\addtocounter{Shift}{-15}
\put(\value{Subdi},\value{Shift}){\begin{turn}{0}\makebox(0,0){{\small $~$} }\end{turn}}
\addtocounter{Shift}{-12}
\put(\value{Subdi},\value{Shift}){\begin{turn}{0}\makebox(0,0){{\small $0$} }\end{turn}}
\addtocounter{Shift}{-13}
\put(\value{Subdi},\value{Shift}){\begin{turn}{0}\makebox(0,0){{\small $1$} }\end{turn}}
\addtocounter{Shift}{-13}
\put(\value{Subdi},\value{Shift}){\begin{turn}{0}\makebox(0,0){{\small $~$} }\end{turn}}
\addtocounter{Shift}{-12}
\put(\value{Subdi},\value{Shift}){\begin{turn}{0}\makebox(0,0){{\small $~$} }\end{turn}}
\addtocounter{Shift}{-15}
\put(\value{Subdi},\value{Shift}){\begin{turn}{0}\makebox(0,0){{\small $1$} }\end{turn}}
\addtocounter{Shift}{-13}
\put(\value{Subdi},\value{Shift}){\begin{turn}{0}\makebox(0,0){{\small $0$} }\end{turn}}
\addtocounter{Shift}{93}
\addtocounter{Subdi}{\value{Step}}
\put(\value{Subdi},\value{Shift}){\begin{turn}{0}\makebox(0,0){{\small $~$} }\end{turn}}
\addtocounter{Shift}{-15}
\put(\value{Subdi},\value{Shift}){\begin{turn}{0}\makebox(0,0){{\small $~$} }\end{turn}}
\addtocounter{Shift}{-12}
\put(\value{Subdi},\value{Shift}){\begin{turn}{0}\makebox(0,0){{\small $~$} }\end{turn}}
\addtocounter{Shift}{-13}
\put(\value{Subdi},\value{Shift}){\begin{turn}{0}\makebox(0,0){{\small $~$} }\end{turn}}
\addtocounter{Shift}{-13}
\put(\value{Subdi},\value{Shift}){\begin{turn}{0}\makebox(0,0){{\small $~$} }\end{turn}}
\addtocounter{Shift}{-12}
\put(\value{Subdi},\value{Shift}){\begin{turn}{0}\makebox(0,0){{\small $~$} }\end{turn}}
\addtocounter{Shift}{-15}
\put(\value{Subdi},\value{Shift}){\begin{turn}{0}\makebox(0,0){{\small $1$} }\end{turn}}
\addtocounter{Shift}{-13}
\put(\value{Subdi},\value{Shift}){\begin{turn}{0}\makebox(0,0){{\small $0$} }\end{turn}}
\addtocounter{Shift}{93}
\addtocounter{Subdi}{\value{Step}}
\put(\value{Subdi},\value{Shift}){\begin{turn}{0}\makebox(0,0){{\small $~$} }\end{turn}}
\addtocounter{Shift}{-15}
\put(\value{Subdi},\value{Shift}){\begin{turn}{0}\makebox(0,0){{\small $~$} }\end{turn}}
\addtocounter{Shift}{-12}
\put(\value{Subdi},\value{Shift}){\begin{turn}{0}\makebox(0,0){{\small $~$} }\end{turn}}
\addtocounter{Shift}{-13}
\put(\value{Subdi},\value{Shift}){\begin{turn}{0}\makebox(0,0){{\small $~$} }\end{turn}}
\addtocounter{Shift}{-13}
\put(\value{Subdi},\value{Shift}){\begin{turn}{0}\makebox(0,0){{\small $~$} }\end{turn}}
\addtocounter{Shift}{-12}
\put(\value{Subdi},\value{Shift}){\begin{turn}{0}\makebox(0,0){{\small $~$} }\end{turn}}
\addtocounter{Shift}{-15}
\put(\value{Subdi},\value{Shift}){\begin{turn}{0}\makebox(0,0){{\small $~$} }\end{turn}}
\addtocounter{Shift}{-13}
\put(\value{Subdi},\value{Shift}){\begin{turn}{0}\makebox(0,0){{\small $0$} }\end{turn}}
\addtocounter{Subdi}{\value{Step}}
}
\end{picture}
\caption{ \maxwell decoder applied to the simple
example shown in Fig. \ref{fig:maxwelldecoder}. The all-zero codeword is decoded. 
The initial phase 
coincides with standard message-passing BP algorithm:
a single copy of the process decodes a bit at a time. 
After three steps, the BP decoder gets stuck in a stopping 
set and  several steps of guessing follow. During this phase 
the associated entropy $\entropy(\iter)$  increases. 
After this guessing phase, the standard message
passing phase resumes. More and more copies terminate due to 
inconsistent messages arriving at variable nodes.
At the end only one copy survives. This shows that this example 
has a unique MAP solution.
}
\label{fig:maxwelltwodecoder} 
\end{figure}

It can happen that during the decoding process a variable
receives non-erased messages from several check nodes.
In such a case, these messages can be distinct and, therefore,
inconsistent.
Such an event is termed a {\em contradiction}.
Any running copy of the decoding process which encounters a contradiction 
terminates. 
The decoding process finishes once all bits have been determined.
At this point, each surviving copy outputs the determined word.
Each such word is by construction a codeword which is
compatible with the received information.
Vice versa, for each codeword which is compatible with
the received information, there will be a surviving copy. 
In other words, the \maxwell decoder performs a complete {\em list decoding} of
the received message. Fig.~\ref{fig:maxwelldecoder} shows 
the workings of the \maxwell decoder by means of a specific example.
The corresponding instance of the decoding process is depicted in
Fig.~\ref{fig:maxwelltwodecoder} from the
perspective of the various simultaneous copies. 

Let us briefly describe how the analysis of the above algorithm 
is related to the balance condition and the proof of
Theorem~\ref{ImprovedUB}.
Instead of explaining the balance between the areas as shown in 
Fig.~\ref{fig:exitcurve}, we  consider the balance of the two areas shown 
in Fig.~\ref{fig:exitbalance}. Note that these two areas differ from the
previous ones only by a common term, so that the condition for balance stays 
unchanged.
From the above description it follows that at any given time $\iter$ 
there are $2^{\entropy(\iter)}$ copies running, 
where $\entropy(\iter)$ is a natural number which evolves with time.
In fact, each time a bit is guessed, the number of copies is doubled,
while it is halved each time a contradiction occurs.
Call $\iter_{\text{out}}$ the time at which all transmitted bits have 
been determined and the list of decoded words is output 
($\iter_{\text{out}}$ does not depend upon the particular copy of the process in
consideration). Since the \maxwell decoder is a complete list decoder
and since all output codewords have equal posterior probability,
$H(X|Y) = \entropy(\iter_{\text{out}})$. On the other hand, $\entropy(\iter_{\text{out}})$
is equal to the total number of guesses minus the total number of 
contradictions which occurred during the evolution of the algorithm. 
As we will see in greater detail in the next section,
the total number of guesses divided by $n$ converges to the area of the 
dark gray region
in Fig.~\ref{fig:exitbalance} (a), while the total number of contradictions
divided by $n$ is asymptotically not larger than the dark gray area
in Fig.~\ref{fig:exitbalance} (b). Therefore, as long as 
$\cp$ is strictly larger than the value at which we have balance, 
call this value $\overline{\cp}^\MAP$,
$\lim_{\n\to\infty}\frac{\expectation[H(X|Y(\ih))]}{\n}>0$. This implies that $\overline{\cp}^\MAP\ge\cp^\MAP$.

We expect that the number of 
contradictions divided by $\n$ is indeed asymptotically equal to 
the dark gray area in Fig.~\ref{fig:exitbalance} (b).
Although we are not able to prove this statement in full generality,
it follows from Theorem~\ref{CountingTheorem}, whenever the hypotheses hold.

\subsection{Message-Passing Setting}
\label{MessagePassingSection}
We describe now a message-passing algorithm that is equivalent 
to the above sequential formulation. First note that because
of the code linearity, the symmetries of the channel and the decoding algorithm,
we can simplify our analysis by making the all-zero codeword assumption,
see \cite{RiU01}.

We assign a label $\chm{i}$ to the variable node of 
index $i$. The label can take three possible values 
$\chm{i}\in\{\0, \?, \g\}$. 
It can be viewed as the output of some fictitious channel, 
and indicates how the algorithm is going to treat 
that variable node. The fictitious channel is memoryless: each 
variable node is assigned a 
$\0$ with probability $1-\ih$, a $\?$ with probability $\ih(1-\gp)$
and a $\g$ with probability $\ih \gp$. 
The parameter
$\gp$ represents the fraction of 
{\em guesses} ventured so far.

The new message-passing algorithm employs left-to-right messages $\lrm{}$ 
and right-to-left messages $\rlm{}$, all of which take values in $\{\0,\?,\g\}$.
The meaning of the $0$ message and the $\?$ message is the same as for the
BP algorithm. A $\g$ message indicates that either the bit from which this
message emanates has been guessed or that the value of this bit 
can be expressed as a linear combination of other bit values which have
been guessed.
Operationally, we can think of the message $\mu_i=\g$ 
as being a shorthand
for a non-empty list of indices $\List_i = \{j_1,\dots, j_k\}$.
This list indicates that $x_i$ is expressible as $x_i= x_{j_1}+\cdots +x_{j_k}$,
where $\{x_{j_1}, \cdots, x_{j_k}\}$ is a set of guessed bits.

\begin{figure}[hbt]
\vspace{5bp}
\centering
\setlength{\unitlength}{0.5bp}
\begin{picture}(360,180)
\put(0,20){\includegraphics[scale=0.5]{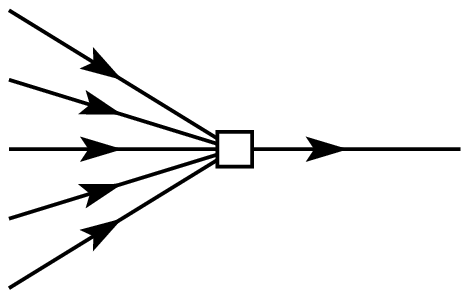}}
\put(200,20){\includegraphics[scale=0.5]{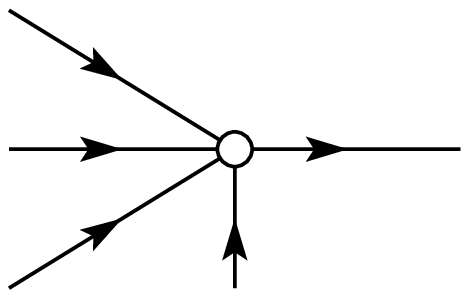}}
\put(155,100){\makebox(0,0){$\rlm{}$}}
\put(355,100){\makebox(0,0){$\lrm{}$}}
\put(5,140){\makebox(0,0){$\mu_{1}$}}
\put(5,120){\makebox(0,0){$\mu_{2}$}}
\put(5,60){\makebox(0,0){$\mu_{\rdegree-1}$}}
\put(205,140){\makebox(0,0){$\mu_{1}$}}
\put(205,60){\makebox(0,0){$\mu_{\ldegree-1}$}}
\put(280,50){\makebox(0,0){$\chm{}$}}
\put(50,0){\makebox(0,0){\footnotesize{(i)} }}
\put(250,0){\makebox(0,0){\footnotesize{(ii)} }}
\end{picture}
\caption{Update rule for parity-check nodes (i) and variable nodes (ii).}
\label{fig:updatecnodemaxwell}
\end{figure}

This motivates the following update 
rules for the parity-check and variable nodes shown in Fig.~\ref{fig:updatecnodemaxwell}. 

(i) Update rule for a parity-check node of degree $\rdegree$:
Assume that the index set for the $(\rdegree-1)$ messages which enter
the check node is ${\cal{R}}= [\rdegree-1]$.  
Then
\begin{align*}
\rlm{} =
\begin{cases}
\0, & \text{if~$\forall i\in {\cal{R}}, ~\m_{i}=\0$}, \\
\?, & \text{if~$\exists i\in {\cal{R}}, ~\m_{i}=\?$}, \\
\g, & \text{if~$\forall j\in {\cal{R}}, ~\m_{j} \neq \?$, and
$\exists i\in {\cal{R}}, ~\m_{i}=\g$}.\\
\end{cases}
\end{align*}
With respect to the BP decoder, the only new rule is the one which
leads to  $\rlm{} = \g$. It is motivated as follows. 
Assume that for all $i \in {\cal{R}}$ we have
$\lrm{i}=0\backslash \g$ and that at least one such message is $\g$.
This means that the connected variables $x_i$, $i \in {\cal{R}}$, are either
known, have been guessed themselves, or can be expressed as 
a linear combination of guessed bits (and at least one such value
is indeed either a guess itself or expressible as a linear combination
of guesses).
Since the variable connected to the outgoing edge is the sum of the
variables connected to the incoming edges, it follows that this variable
is also expressible as a linear combination of guesses. Therefore,
$\rlm{}=\g$ in this case.
Operationally, we have $\rdegree-1$
lists $\List_1,\dots,\List_{\rdegree-1}$ (at least one of
which is non-empty) entering the  check node.  
The outgoing list $\List^{\tt y}$ is obtained as the {\em union} of the incoming lists, 
where indices which occur an even number of times in the incoming lists are eliminated.
The list $\List^{\tt y}$  provides a resolution rule for
$x_{1}+\cdots +x_{\rdegree-1}$, and therefore for the variable connected to the
outgoing edge.

In the above description and the definition of the message-passing rules
we have ignored the possibility that the union
of the incoming lists (at least one of which is non-empty) is empty.
This can happen if a complete cancellation occurs (every index appears
an even number of times in the incoming lists). 
Fortunately, as we shall see, this assumption
has no influence on the proof of Theorem~\ref{ImprovedUB}.

(ii) Update rule for a variable node of degree $\ldegree$: 
Assume that the index set for the $\ldegree-1$ messages
which enter the variable node is ${\cal{L}}= [\ldegree-1]\cup \{ {\tt \ih}\}$. 
 Then
\begin{align*}
\lrm{} =
\begin{cases}
\0, & \text{if~$\exists i\in {\cal{L}}, ~\m_{i}=\0$}, \\
\?, & \text{if~$\forall i\in {\cal{L}}, ~\m_{i}=\?$}, \\
\g, & \text{if~$\forall i\in {\cal{L}}, ~\m_{i} \neq \0$ and $\exists j\in {\cal{L}}, ~\m_{j}=\g$}.
\end{cases}
\end{align*}
Once again, it should be enough to motivate the rule which leads to 
$\lrm{} = \g$. Recall that $\g$ indicates that the bit is not known but
that it has either been guessed or that the bit is expressible as a linear combination
of guessed bits. Therefore, if none of the incoming messages is a $0$,
and at least one is a $\g$, then the outgoing message is a $\g$.
Operationally, this means that the outgoing list is equal to {\em one}
of the incoming non-empty lists. E.g., if the bit itself has been guessed
(i.e., $\chm{i}=\g$) and all other incoming 
messages are $\?$ then the outgoing message is $\{i\}$.


From the messages we can obtain estimates $\nu_i$, $\in [n]$,
of the transmitted bits (the $\nu_i$'s are node- rather than 
edge-quantities).
In order to obtain these estimates we apply
the same rule as for the variable node update, see (ii)
above, with incoming messages corresponding to {\em all} of the 
neighboring check nodes. In other words, for a degree $\ldegree$ variable
node, we have ${\cal L} = [\ldegree]\cup\{\epsilon\}$
instead of ${\cal L} = [\ldegree-1]\cup\{\epsilon\}$.

The consistency of the estimates
implies a set of linear {\em conditions} on the guessed variables.
Consider all the messages $\mu_i$ entering a fixed variable node
and the associated (possibly empty) lists $\List_{i}=\{j^i_1, \cdots, j^i_k\}$.
Let ${\cal L}_{\m}$, $\m\in\{\0,\g,\?\}$ denote the subsets of indices
$i$ with $\m_i=\m$. 
\begin{enumerate}
\item If ${\cal L}_{\0}\neq\emptyset$ and ${\cal L}_{\g}\neq\emptyset$,
then, for any $i\in {\cal L}_{\g}$, we have the condition
\begin{eqnarray}
x_{j^i_1}+\cdots +x_{j^i_k} = 0\, ,\;\;\;\;\;\;\;\;{\rm mod}\; 2\, .
\label{ConditionA}
\end{eqnarray}
%
The total number
of resulting conditions is $|{\cal L}_{\g}|$.
\item If ${\cal L}_{\0} = \emptyset$ and $|{\cal L}_{\g}|\ge 2$,
then fix $i\in {\cal L}_{\g}$. For any $l\in {\cal L}_{\g}\backslash \{i\}$,
we have the condition 
\begin{eqnarray}
x_{j^i_1}+\cdots +x_{j^i_k} = x_{j^l_1}+\cdots +x_{j^l_k}\, ,\;\;\;\;\;\;{\rm mod}\; 2\, .
\label{ConditionB}
\end{eqnarray}
%
The total number of resulting conditions is $|{\cal L}_{\g}|-1$.
\end{enumerate}
The algorithm stores in memory each new condition produced during its
execution. Notice that each conditions involves uniquely bits $x_i$
for which $\mu_i^{\epsilon}=\g$. It can happen that 
a particular condition is either linearly dependent upon previous 
ones or empty. The last case occurs if the corresponding lists are empty,
which in turn may be the consequence of a previous parity-check node update
(see the description of the check-node update rule above).
Given a set of guesses, any subset of them whose values can be chosen 
freely without violating any of the conditions produced by 
the \maxwell decoder, is said to be {\em independent}.
Of course, the maximal number of independent guesses is equal to the number of 
guesses minus the number of linearly independent conditions.

{\em Conditions} are equivalent, in the present setting 
to what have been called contradictions in the description of 
of Sec.~\ref{MaxwellIntro}. In fact, if one thinks of guessed bits
as i.i.d. uniformly random in $\{\0,\1\}$ then each new, 
independent condition, cf. 
Eqs.~(\ref{ConditionA}), (\ref{ConditionB}) is satisfied with 
probability $1/2$.

It is useful to estabilish the following convention for denoting the 
successive message passing iterations. At  the $t^{\text{th}}$ iteration 
(with $t=0,1,\dots$) we first update all the left-to-right messages and 
then all the right-to-left messages. We have therefore
$\dots\to\rlm{}(t-1)\to\lrm{}(t)\to\rlm{}(t)\to\lrm{}(t+1)\to\dots $.
Notice that, as the number of iterations increases, a given 
message can change its status according to one of the transitions
$\?\to\g$, $\g\to\0$ or $\?\to\0$. Therefore the algorithms surely stops 
after a finite number of iterations (at most twice the number of edges 
in the graph). We shall denote the fixed point as $\lrm{}(\infty)$,
$\rlm{}(\infty)$.
At the $t^{\text{th}}$ iteration the algorithm deliver an estimate
$\nu_i(t)$, $i\in[n]$ of the $i^{\text{th}}$ transmitted bit. 
%
%
\subsection{The Case of Tree Graphs and Some Simple Consequences}
\label{sec:treecase}
As for other message-passing algorithms, it is instructive
to study the behavior of the \maxwell decoder on trees. In particular,
we will show that:
(a) On a tree the {\em sequential} \maxwell decoder guesses exactly as many 
variables as there are degrees of freedom in the system (implying that
all these guesses are independent);
(b) on a tree the number of {\em independent} guesses ventured
by  the (not necessarily sequential) \maxwell decoder
by end of the decoding process is equal to the number of degrees
of freedom of the system and it can be computed in a {\em local} way;
(c) the same local counting formula gives in general (for Tanner graphs that
are not necessarily trees) an {\em upper bound} on the number of independent guesses
which remain at the end of the decoding process.


We have already 
explained that, for the purpose of analysis, we can
make the all-zero codeword assumption. Therefore, in the sequel
we only have to consider linear systems of equations with a zero right side.
We say that the \maxwell decoder is 
{\em bit-by-bit} (or {\em sequential}) if any time the BP phase comes to
a halt, the decoder guesses a {\em single} unknown bit and then proceeds
by processing all consequences until no further progress is achieved.

\blemma[Number Of Guesses of Sequential \maxwell Decoder]
\label{lem:numberofsequentialguesses}
Consider a binary linear system of equations with right side equal to zero
and $k$ degrees of freedom (i.e., $k$ is equal to the number of variables minus
the rank of the system).
Assume that the Tanner graph associated to this system is a tree.
Then the sequential \maxwell decoder ventures exactly $k$ guesses during the
decoding process and all these guesses are independent.
\elemma
\bproof
Without loss of generality
we can assume that there are no check leaf nodes. 
In fact, whenever degree-one check nodes are present, 
the standard BP decoder can be run until
all such nodes have been removed.
For each variable node which is removed in this fashion, the rank of the
system is decreased by exactly one as well.

We claim that the resulting system of equations has full rank.
To see this, assume to the contrary
that there is a non-zero linear combinations of  
equations that yields
zero. Look at the Tanner graph corresponding to this subset of equations:
 all variable nodes have (even) degree at least two
and all check nodes have degree at least two (as argued above).
It is well known that
a graph with minimum degree at least two contains at least one cycle,
contradicting the hypothesis that the initial graph was a tree. 

Consider therefore a Tanner graph which is a tree and all 
of its leaf nodes are variables.
Let $\ldegree_i$, $i \in [n]$, ($\rdegree_i$, $i \in [m]$)
denote the degree of variable (check) node $i$.
By our remarks above,
the corresponding system of equations has $n-m$ degrees of freedom.  
Therefore, it is clear that the \maxwell decoder has to guess {\em at least}
$n-m$ bits before it stops.
We claim that it ventures exactly $n-m$ guesses, i.e.,
that on a tree the sequential guesses are {\em independent}.

At the start of the decoding process all messages are
erasures. We will show that at the end of the decoding process
each edge carries exactly one $\g$ message in one direction 
and a $\?$ message
in the other direction. This proves our claim:  it implies that
a variable node which has been guessed, and hence all of its outgoing messages
carry a $\g$ message, has no incoming $\g$ message.  
It is therefore not constrained 
by any of the other guesses, i.e., it is independent.
Clearly, at the end of the decoding process
each edge has to carry a $\g$ message in {\em at least one} direction;
otherwise the connected bit has not been determined yet,
contradiction the assumption that the \maxwell decoder has halted.

\begin{figure}[htp]
\centering
\setlength{\unitlength}{1bp}
\begin{picture}(240,140)
\put(5,5){\includegraphics[width=100bp]{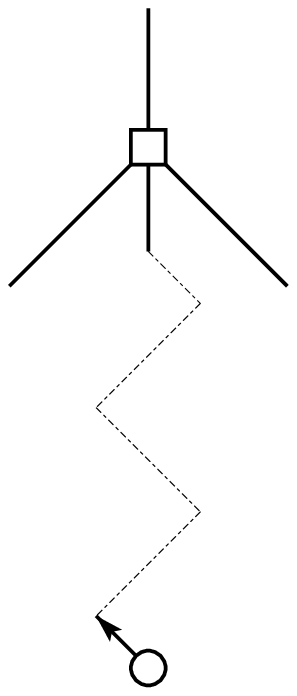}}
\put(30,100){\begin{turn}{45}\makebox(0,0){{$\underset{\longrightarrow}{\g}$} }\end{turn}}
\put(59,100){\begin{turn}{90}\makebox(0,0){{$\rightarrow$} }\end{turn}}
\put(59,130){\begin{turn}{90}\makebox(0,0){{$\rightarrow$} }\end{turn}}
\put(52,100){\begin{turn}{90}\makebox(0,0){{$\leftarrow$} }\end{turn}}
\put(52,130){\begin{turn}{90}\makebox(0,0){{$\leftarrow$} }\end{turn}}
\put(56,143){\makebox(0,0){$\edge$}}
\put(56,85){\makebox(0,0){$\tilde{\edge}$}}
\put(62,93){\makebox(0,0){$\g$}} 
\put(62,123){\framebox(10,10){${\tt \mathbf{g}}$}}
\put(48,90){\makebox(0,0){$\g$}} 
\put(48,120){\makebox(0,0){$\g$}}
\put(82,98){\begin{turn}{-45}\makebox(0,0){{$\underset{\longleftarrow}{\g}$} }\end{turn}}
\put(130,0)
   {
   \put(56,143){\makebox(0,0){$\edge$}}
   \put(56,85){\makebox(0,0){$\tilde{\edge}$}}
   \put(60,100){\begin{turn}{90}\makebox(0,0){{$\rightarrow$} }\end{turn}}
   \put(60,130){\begin{turn}{90}\makebox(0,0){{$\rightarrow$} }\end{turn}}
   \put(51,100){\begin{turn}{90}\makebox(0,0){{$\leftarrow$} }\end{turn}}
   \put(51,130){\begin{turn}{90}\makebox(0,0){{$\leftarrow$} }\end{turn}}
   \put(62,93){\makebox(0,0){$\g$}} 
   \put(62,123){\framebox(10,10){${\tt \mathbf{g}}$}}
   \put(48,90){\makebox(0,0){$\g$}} 
   \put(48,120){\makebox(0,0){$\g$}}
   }
\put(135,5){\includegraphics[width=100bp]{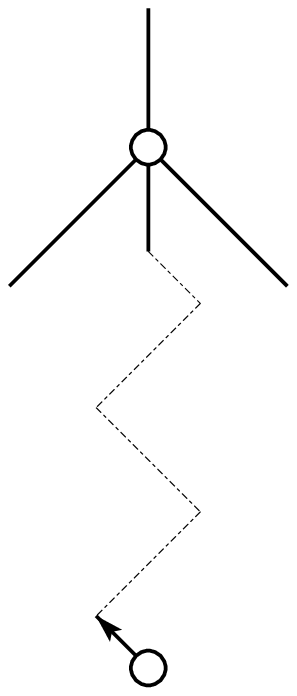}}
\put(55,-4){\makebox(0,0){\small{(a)}}}
\put(190,-4){\makebox(0,0){\small{(b)}}}
\put(63,12){\makebox(0,0){$\g$}}
\put(193,12){\makebox(0,0){$\g$}}
\end{picture}
\caption{
In (a) consider the messages flowing along edge $\edge$. Assume that
the {\em outgoing} message (shown in a frame) switches as a consequence of a newly
guessed bit from $\?$ to $\g$. Assume further that the {\em incoming} 
message flowing in the opposite direction
is $\g$ as well. This provides the induction step from odd levels to even levels.
As indicated in the figure, it then follows that both messages
along edge $\tilde{\edge}$ are $\g$ as well. The case of an edge exiting a variable
node is shown in (b) and follows by essentially the same argument. 
}
\label{fig:independentproof}
\end{figure}

Let us show that it can not carry a $\g$ message in both directions.
Initially all messages are $\?$. The sequential \maxwell decoder
proceeds in phases, guessing a bit and then determining all
consequences of this guess during the BP phase until it gets stuck again.
Let us call one such guess followed by the BP phase one {\em iteration}.
Let us agree that during the BP phase the consequence of a newly guessed bit
are computed {\em in order of increasing distance} from the guessed bit. This means,
that we first process all edges directly connected to this bit
(call this {\em level zero}), then all edges at distance one (call this {\em level one})
and so on. Assume that when we process level $\iter$, $\iter \geq 1$,
we encounter an edge whose
outgoing (away from the newly guessed bit) message switches from $\?$ to $\g$
and whose incoming message already is $\g$. We claim that then the same
must have occurred at level $\iter-1$. This is quickly verified by checking
explicitly both cases: 
an edge which goes from a check node to a variable node
(odd levels $\iter$; left picture in Fig.~\ref{fig:independentproof}) 
and the case of an edge which 
goes from a variable node to a check node
(even levels $\iter$; left picture in Fig.~\ref{fig:independentproof}).
If we apply this argument inductively, we see that the guessed variable node
must have had an incoming message which was $\g$, contradicting the fact
that the \maxwell decoder decided to guess this bit.
\eproof

What happens if we run the \maxwell decoder in a non-sequential way, i.e.,
if we guess many/several bits each time we get stuck? In this
case it can happen that some of the guesses are dependent.
Nevertheless, the number of independent guesses remaining at the end
of the process is
still equal to the degrees of freedom of the system of equations.
More importantly, on a tree this number of independent guesses
can be computed in a {\em local} way.
\blemma[Number of Independent Guesses]
\label{lem:independentguesses}
Consider a binary linear system of equations with right side equal to zero
and $k$ degrees of freedom (i.e., $k$ is equal to the number of variables minus
the rank of the system).
Assume that the Tanner graph associated to this system is a tree
and that it contains no check nodes of degree one.
Then the number of {\em independent} guesses ventured by the \maxwell decoder
at the end of the decoding process is equal to $k$.
Further, let ${\mathbb{G}}$ denote the total number of  guesses of the \maxwell decoder, 
denote by $\ldegree^{\g}_{i}$ the number of
incoming $\g$ messages at variable node $i$ (including, if applicable, the guess 
of the bit itself), and by ${\cal C}_{\g}$ the subset of 
all check nodes all of its incoming messages are $\g$. Then
\begin{align}
\label{equ:countingontreeone}
k & = {\mathbb{G}}-
\sum_{i \in {\cal V}} (\ldegree^{\g}_i-1)+
\sum_{i \in {\cal C}_{\g}} (\rdegree_i-1) .
\end{align}
\elemma
\bproof
By definition of the algorithm, at the end of the decoding
process all bits have been determined (i.e., guessed or expressed
in terms of guessed bits). This means that among the guesses ventured
by the \maxwell decoder there must be $k$ independent such guesses.
Now note that the final state of the messages is independent of the
order in which the guesses are taken. It is convenient to imagine
that we first venture the $k$ independent guesses and then apply the BP decoder.
At the end of this phase all bits are known. Further, from
Lemma \ref{lem:numberofsequentialguesses}
we know that $\ldegree^{\g}_i=1$ for all $i \in [n]$
and ${\cal C}_{g}$ is the empty set.
Therefore, the stated counting formula is correct at this stage.
Assume now we proceed in iterations, adding one guess at a time
and propagating all its consequences. We will verify that
the counting formula stays valid.
Assume therefore that the counting formula is
correct at the start of an iteration and add a further guess, lets say of variable $i$.
This extra guess increases $\ldegree^{\g}_i$ by one and increases the
number of guesses by one, keeping the counting formula intact.
Consider now the ensuing BP phase. Consider an edge $\edge$ emanating from
a variable node $i$, the check node connected to it, call it $j$
and all the edges and variable nodes connected to this check node. 
Assume that the message from $i$ to $j$ is $\?$ (in the case
that this message is already $\g$, the message does not change
and there is nothing to prove).
As a consequence the
message from $j$ to $i$  must be a $\g$ because of the argument above.
Also, all the incoming messages into $j$ but the one form $i$
must be $\g$ as well (otherwise the update rule would have been violated
at node $j$). 
Update all the corresponding
edge messages. If the  message from $i$ to $j$ does not change, 
then neither does any of the messages outgoing at the check node
and the counting formula stays valid. If, on the other hand, the outgoing
message along edge $\edge$ flips to $\g$ then 
so do all the messages outgoing from the check node $j$.
  Assume that the check node has degree $\rdegree_j$.
Then, ${\cal C}_{g}$ now contains $j$. This increases the right hand side
of the counting formula by $\rdegree_j-1$. On the other hand it also increases
$\ldegree^{\g}_l$ by one for all $l \in {\cal V}$ which are connected to check node $j$, but for node $i$ (the corresponding message was already a $\g$).
In total this decreases the right hand side of the counting formula by $\rdegree_j-1$.
\eproof

\begin{figure}[hbt]
\vspace{5bp}
\centering
\setlength{\unitlength}{0.5bp}
\begin{picture}(360,180)
\put(0,20){\includegraphics[scale=0.5]{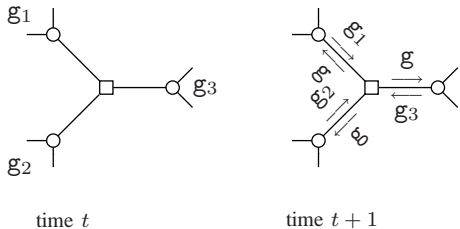}}
\put(200,20){\includegraphics[scale=0.5]{correctionterm}}
\put(155,100){\makebox(0,0){$\g_3$}}
\put(15,155){\makebox(0,0){$\g_1$}}
\put(15,40){\makebox(0,0){$\g_2$}}
\put(251,91){\begin{turn}{45}\makebox(0,0){{$\underset{\longrightarrow}{\g_2}$} }\end{turn}}
\put(269,69){\begin{turn}{45}\makebox(0,0){{$\overset{\longleftarrow}{\g}$} }\end{turn}}
\put(266,133){\begin{turn}{-45}\makebox(0,0){{$\underset{\longrightarrow}{\g_1}$} }\end{turn}}
\put(248,112){\begin{turn}{-45}\makebox(0,0){{$\overset{\longleftarrow}{{\g}}$} }\end{turn}}
\put(310,114){\begin{turn}{0}\makebox(0,0){{$\underset{\longrightarrow}{\g}$} }\end{turn}}
\put(310,86){\begin{turn}{0}\makebox(0,0){{$\overset{\longleftarrow}{\g_3}$} }\end{turn}}
\put(50,0){\makebox(0,0){\footnotesize{time $t$} }}
\put(250,0){\makebox(0,0){\footnotesize{time $t+1$}}}
\end{picture}
\caption{Computation of the number of linearly 
independent conditions. To each of the incoming edges corresponds a list.
To keep things simple and without essential loss of generality, assume that
$\List_i=\{i\}$. The three outgoing lists are then $\List_1=\{2, 3\}$,
$\List_2=\{1, 3\}$, and $\List_3=\{1, 2\}$. Compare the incoming and
outgoing list at node 1: we get the condition $x_1=x_2+x_3$. But
exactly the same condition appears at node $2$ and node $3$. In general, a
check node of degree $\rdegree$, all of its incoming messages are $g$, generates 
$\rdegree-1$ linearly dependent conditions.}
\label{fig:correctionterm}
\end{figure}

Each part of the counting equation (\ref{equ:countingontreeone}) has a pleasing
interpretation. As stated, ${\mathbb{G}}$ is the total number of ventured guesses.
If a variable node has $\ldegree^{\g}$ incoming $\g$ messages then
these correspond to $\ldegree^{\g}$ linear equations, each of which
determines the same bit. This gives rise to $(\ldegree^{\g}-1)$ linear
conditions which the ${\mathbb{G}}$ guesses have to fulfill. But not all these conditions
are linearly independent.
Consider Fig.~\ref{fig:correctionterm}. If a check node of degree $\rdegree$
has all of its incoming messages equal to $\g$ then the $\rdegree$ equations
which correspond to the $\rdegree$ outgoing messages are identical, i.e.,
$\rdegree-1$ of them are linearly dependent. The last term in the counting
formula (\ref{equ:countingontreeone}) therefore corrects the over-counting
of dependent conditions.

\bex
Consider a code whose Tanner graph is a 
tree and all leaves are variable nodes. Let the set
of variables (checks) be indexed by $[n]$ ($[m]$, and let
$\ldegree_i$, $i \in [n]$, ($\rdegree_i$, $i \in [m]$) 
be the degree of variable (check) node $i$. Assume that the \maxwell
decoder guesses all leaf (variable) nodes and then proceeds by
message passing. It is not very hard to see that in this setting the decoder
proceeds with the message-passing phase (starting from the leaf nodes) until all variables have
been determined and that no further guesses have to be made. Further, at the end of the decoding
process {\em all} messages are $\g$.

Let us determine the number of independent guesses at
the end of the decoding process using the counting formula
(\ref{equ:countingontree}).
Note that for each leaf node we have $\ldegree^{\g}=2$ (one guess and
one additional incoming $\g$ message. For all internal variable nodes
we have $\ldegree^{\g}=\ldegree$. Finally, ${\cal C}_{\g}={\cal C}$.
If we let $n_l$ denote the number leaf nodes, so that ${\mathbb{G}}=n_l$, we get 
that the number of independent guesses is equal to
\begin{align*}
& n_l - \sum_{i \in \text{leaves} } (2-1) - 
\sum_{i \in [n] \setminus \text{leaves} } (\ldegree_i-1) + 
\sum_{i \in [m]} (\rdegree_i-1) \\
= & 
-\sum_{i \in [n] } (\ldegree_i-1) +
\sum_{i \in [m]} (\rdegree_i-1) = n-m.
\end{align*}
This is of course the expected result since the
system has exactly $n-m$ degrees of freedom.
\eex

So far we have only considered sets of equations whose Tanner graph is a
tree. What happens if we run the \maxwell decoder on a general system of
equations.  For a general Tanner graph, the above counting of the total
number of independent guesses is not necessarily tight.  The counting
of the total number of conditions generated by the \maxwell decoder is
always correct. But it can happen that besides the obvious over-counting
at check nodes, there are other dependencies generated by loops in the
graph which are not considered in the counting formula.  Therefore,
in general we only get a lower bound. Let us state this explicitly.
\begin{lemma}[Lower Bound on Independent Guesses]
\label{LemmaConditions}
Consider a binary linear system of equations with right side equal to zero
and $k$ degrees of freedom (i.e., $k$ is equal to the number of variables minus
the rank of the system).
Assume that the Tanner graph associated to this system 
contains no check nodes of degree one.
Let ${\mathbb{G}}$ denote the number of all guesses of the \maxwell decoder, denote
by $\ldegree^{\g}_{i}$ the number of
incoming $\g$ messages at variable node $i$ (including the guess
if this node has been guessed), and by ${\cal C}_{\g}$ the subset 
of all check nodes all of whose incoming messages are  $\g$. Then
\begin{align}
\label{equ:countingontree}
k & \geq {\mathbb{G}}-
\sum_{i \in {\cal V}} (\ldegree^{\g}_i-1)+
\sum_{i \in {\cal C}_{\g}} (\rdegree_i-1) .
\end{align}
%
%
%
\end{lemma}

\subsection{Density Evolution Analysis}
\label{sec:densityevolution}

Let us now perform the usual DE analysis.
Let $\lra^t_{\lrm{}}$ denote the probability that a left-to-right message at time $t$
is equal to  $\lrm{}\in\{\0,\?,\g\}$, and let 
$\rla^t_{\rlm{}}$ denote the corresponding probability for a
right-to-left message.

(i) At the check node side the DE relations read
\begin{align*}
\rla_{\0}^{t} & =\redge(\lra_{\0}^{t}),\\
\rla_{\?}^{t} & = 1-\redge(    \lra_{\0}^{t}+ \lra_{\g}^{t}     )=1-\redge(1-\lra^t_{\?}),\\
\rla_{\g}^{t} & =1-\rla_{\0}^{t}-\rla_{\?}^{t}=\redge(\lra_{\0}^{t}+ \lra_{\g}^{t})-   \redge(\lra_{\0}^{t}).
\end{align*}

(ii) At the variable node side the DE relations are
\begin{align*}
\lra_{\0}^{t+1} & = 1-\ih \ledge( \rla_{\g}^t  + \rla_{\?}^t  ),\\
\lra_{\?}^{t+1} & = (1-\gamma) \ih \ledge(\rla_{\?}^t),\\
\lra_{\g}^{t+1} & = \ih \ledge( \rla_{\g}^t  + \rla_{\?}^t  ) - (1-\gamma)\ih\ledge(\rla_{\?}^t).
\end{align*}
According to our convention, the iteration counter is 
increased only in the variable node operation. Moreover,
the variables $\lra_{\?}^{t}$ ($\rla_{\?}^{t}$) and 
$\lra_{\?}^{t}+\lra_{\g}^{t}$ ($\rla_{\?}^{t}+\rla_{\g}^{t}$) 
satisfy the same equations as the fractions of erased messages
in the standard BP decoder with erasure probabilities
$\cp(1-\gamma)$ and $\gamma$, respectively. This is an immediate consequence of the 
update rules defined in section \ref{MessagePassingSection}. 

When the time $t$ tends to $\infty$, DE converges to the fixed-point 
probability distribution. 
To settle our notation, we write 
$(\lra_{\0}^{t},\lra_{\?}^{t},\lra_{\g}^{t})\underset{t\to\infty}{\longrightarrow}\left(\lra_{\0}^{\infty}(\ih,\gp),\lra_{\?}^{\infty}(\ih,\gp)  ,\lra_{\g}^{\infty}(\ih,\gp)\right)$ 
and equivalently 
$(\rla_{\0}^{t},\rla_{\?}^{t},\rla_{\g}^{t})\underset{t\to\infty}{\longrightarrow}\left(\rla_{\0}^{\infty}(\ih,\gp),\rla_{\?}^{\infty}(\ih,\gp)  ,\rla_{\g}^{\infty}(\ih,\gp)\right)$ . 
Observe that $\lra_{\?}^{\infty}(\ih,\gp)$ satisfies the equation $\lra = \ih(1-\gp)\ledge(1-\redge(1-\lra))$,
while  $\lra_{\0}^{\infty}(\ih,\gp)=\lra_{\0}^{\infty}(\ih)$ 
satisfies the equation $(1-\lra) = \ih \ledge(1-\redge(1-(1-\lra)))$.  

Notice that the asymptotic state of the algorithm has the following structure.
The variable nodes such that $\nu_i(\infty)=\?$ {\em or} $\nu_i(\infty) = \g$,
form a stopping set: in fact this is the largest stopping set contained in the
set of variable nodes for which $\mu_i^{\cp} = \?$ {\em or} 
$\mu_i^{\cp} = \g$. Further, the set of variable nodes such that 
$\nu_i(\infty)=\?$ form a stopping set contained in the previous one:
this is the largest stopping set  contained in the set $\mu_i^\cp = \?$.

In the analysis below we shall repeatedly use the following trick.
We shall compute expectations with respect to asymptotic ($t=\infty$) 
incoming messages  in a given node. In such computations,
we shall treat such messages as i.i.d. with distribution
$\left(\lra_{\0}^{\infty},\lra_{\?}^{\infty},\lra_{\g}^{\infty} \right)$,
(for left-to-right messages) or 
$\left(\rla_{\0}^{\infty},\rla_{\?}^{\infty},\rla_{\g}^{\infty} \right)$,
(for right-to-left messages). 
As long as  $(\cp,\gp)$
take non-exceptional values, i.e., at continuity points of 
$\left(\lra_{\0}^{\infty}(\ih,\gp),\lra_{\?}^{\infty}(\ih,\gp),
\lra_{\g}^{\infty}(\ih,\gp) \right)$, cf. Section \ref{sec:Counting},
this is justified as follows. First consider messages after a finite 
number of iterations $t$. For $n$ large enough these are independent because
the Tanner graph is locally a tree. But, if $(\cp,\gamma)$
is non-exceptional the number of message which change between the
$t^{\text{th}}$ iteration and the asymptotic state is bounded by $n\delta(t)$
with $\delta(t)\to 0$ as $t\to\infty$. This argument is essentially the same 
as the one of App.~\ref{ProofCountingDegree}.
%
%
\subsection{Guessing Strategy}
\label{sec:guessstrat}

In the analysis of the \maxwell decoder, we can chose the order 
of guesses at our convenience. As long as the message is completely 
decoded and the final estimates are 
$\nu_i(\infty)\in\{\0,\g\}$ for any bit $i$,
the algorithm realizes a complete list decoding.

We shall adopt the following strategy: we perform $n_{\text{rounds}}$
``decoding rounds''. Our progress will be measured by
the parameter $\gamma$, which is initially set to zero
and which advances by $\Delta \gamma = 1/n_{\text{rounds}}$ 
in each round.

Set $\gamma=0$. Start with the messages received via 
BEC$(\ih)$ and apply BP decoding until the algorithm gets stuck. 
Then consider each of the bits not yet determined  
and set $\mu_i^{\ih} = \g$ independently for each of them with probability
$\Delta\gamma /(1-\gamma)$. (In the first round this probability
is equal to $\Delta \gamma$.)
Set $\gamma\defas\gamma+\Delta\gamma$.\footnote{Note
that if a bit is first selected with probability $\gamma$
and then independently selected with probability $\Delta \gamma/(1-\gamma)$,
then the probability that it was selected at least once is equal to $\gamma+\Delta \gamma$.
This is the rational for our choice of parameters.}
Apply the \maxwell decoder until it
gets stuck.
This is repeated $n_{\text{rounds}}$ times until $\gamma=1$.
If at any earlier phase complete decoding is achieved, the algorithm 
is halted and the current set of decoded codewords output.

The analysis becomes simpler (and the algorithm more efficient)
if we take $\Delta\gamma\to 0$. We shall always think of this limit being taken
after $n\to\infty$. We will see that in this limit the appearance of contradictions
is sharply concentrated to those rounds which include  a discontinuity of
the EXIT curve. In other words, we will see that the algorithm
alternates between the following two phases which are
well separated: in the ``guessing phase'' the algorithm guesses  
a small fraction of bits and the processes the
consequences but theses consequences do not propagate too far
and essentially stay local; in the ``contradiction phase'' on
the other hand the algorithm suddenly 
discovers many relationships (finds many contradictions) and
the size of the residual graph changes by a constant fraction
which is independent of the step size $\Delta \gamma$.
%
%
\subsection{Analysis: Guess Work}
\label{sec:guesswork}

Consider a non-exceptional point $(\cp,\gp)$ and let
$\n\Delta {\mathbb{G}}$ be the number of newly guessed variables when $\gamma$
is changed by an amount $\Delta\gamma>0$.

The process can be described as follows. For each $i\in[n]$, $i$
is selected independently with probability $\Delta\gamma/(1-\gamma)$.
For each selected bit, we consider the present estimate provided by
the \maxwell decoder: $\nu_i(\infty)\in\{\0,\g,\?\}$.
If $\nu_i(\infty) = \?$, the observation on $i$ is changed
from $\m^\cp_i=\?$ to $\m^\cp_i = \g$: the counter of newly guessed variables is
increased by one.
By linearity of expectation, we get
\begin{align*}
\expectation [
\Delta{\mathbb{G}}] & = \frac{1}{\n}\sum_{i\in [n]}\pr(\mbox{$i$ is selected})\, 
\pr(\nu_i(\infty) = \?)  \\
& = \frac{\Delta\gamma}{1-\gp}\, 
\cp (1-\gp) \lnode(\rla_{\?}^{\infty}) 
 =  \cp \lnode(\rla_{\?}^{\infty})\Delta\gamma\, .
\end{align*}
Notice that, in this computation we assumed $n\to\infty$ and $t\to\infty$
afterwards.

Recall that, after $\gamma$ is changed to $\gamma+\Delta\gamma$
and the $n \Delta{\mathbb{G}}$ new guesses are introduced, the message
passing \maxwell decoder is started again until a new fixed point is reached.
%
%
\subsection{Analysis: Confirmation Work}

At each step of the above algorithm, it may happen that several 
$\g$ messages are transmitted to the same variable node $x_i$. 
Each of these lists corresponds to
a distinct resolution rule for $x_i$.
Their convergence on the same node imposes some non-trivial 
condition on the variables which appear in the resolution rules.
Here we estimate the number of independent such conditions by 
exploiting Lemma~\ref{LemmaConditions} above. 
Notice that in Lemma~\ref{LemmaConditions} we assume $\m^{\cp}_i\in\{\g,\?\}$.
In order to make contact with this assumption we could
first run the classical BP decoder until
no further progress can be made.
We could now directly apply Lemma~\ref{LemmaConditions}
to the {\em residual} graph. The disadvantage of this strategy
is that in this scheme it is not so straightforward to 
relate the progress of the \maxwell decoder on the residual
graph to the original DE equations.

Alternatively we can apply Lemma~\ref{LemmaConditions} directly
to the original graph if (i) we do not count contradictions generated
at variable nodes which receive at least one $0$ message (either from
the channel or from the graph) and (ii) we count towards the degree of a check
node only those edges whose incoming messages are not $0$.
With these two conventions one can check that Lemma~\ref{LemmaConditions}
holds for a general graph including degree-one check nodes as well
as variable nodes which are known.

Let $(\cp,\gp)$ be a non-exceptional point and denote by $n{\mathbb C}$
the number of contradictions as estimated by the right-hand side of 
(\ref{equ:countingontree}). The first term counts
the number of conditions arising at that node.
We get
\begin{multline*}
\expectation\left\{\frac{1}{n}\sum_{i\in\cal V}\max(|{\cal L}_{i,\g}|-1,0)
\right\} =\\
 \cp(1-\gp)\sum_{\ldegree}\lnode_{\ldegree}
\expectation_{\ldegree}\left\{\max(n_\g-1,0)\,\ind_{n_\0 = 0}\right\}  \\
+\cp\gp \sum_{\ldegree}\lnode_{\ldegree}
\expectation_{\ldegree}\left\{\max(n_\g,0)\,\ind_{n_\0 = 0}\right\}  \, ,
\end{multline*}
where $\ind_A$ is the indicator function for the event $A$ and
where $n_\g$, $n_\0$, and $n_\?$ count the number of incoming $\g$, $\0$,
and $?$ messages.
Here the limits $n\to\infty$ and $t\to\infty$ are understood and
$\expectation_{\ldegree}$ denotes expectation with respect the multinomial 
variables $n_\0,n_\g,n_\?$ with sum $\ldegree$ and parameters
$\rla_{\0}^{\infty},\rla_{\*}^{\infty},\rla_{\?}^{\infty}$. 
Note that we have the indicator function $\ind_{n_\0 = 0}$
since by our remarks above we should only consider nodes
``in the residual graph'', i.e., nodes which were not
already determined in the BP phase as a consequence of the received
bits.
Throughout this section we shall adopt 
the shorthands  $\rla_{\0},\rla_{\*},\rla_{\?}$ for
$\rla_{\0}^{\infty},\rla_{\g}^{\infty},\rla_{\?}^{\infty}$
(and analogous ones for left-to-right messages).
By computing these expectations we get
\begin{multline}
\expectation\left\{\frac{1}{n}\sum_{i\in\cal V}\max(|{\cal L}_{i,\g}|-1,0)
\right\} =\\
 \cp(1-\gp)\left\{\lnode'(\rla_\?+\rla_\g)\rla_\g -\lnode(\rla_\?+\rla_\g)
+\lnode(\rla_\?)\right\}  \\
+\cp\gp  \lnode'(\rla_\?+\rla_\g)\rla_\g\, .\label{Cond1}
\end{multline}
We must now evaluate the correction term in (\ref{equ:countingontree}).
Consider a check node $a$. Assume that its ``residual'' degree is $\rdegree'_a$. I.e.,
$\rdegree'_a$ counts the number of edges whose incoming messages are not zero. 
If the corresponding $\rdegree{}'_a$ outgoing messages 
are all $\g$ (equivalently, the $\rdegree'_a$ ingoing messages 
are all $\g$), then the same condition has been overcounted $\rdegree{}'_a-1$
times. We denote the set of such check nodes as ${\cal C}$ and obtain
\begin{multline*}
\expectation\left\{\frac{1}{n}\sum_{a\in{\cal C}}(\rdegree{}'_{a}-1)
\right\} = \\
\frac{\lnode'(1)}{\Gamma'(1)}\sum_{\rdegree}\Gamma_{\rdegree}\,
\expectation_{\rdegree}\left\{\max(n_{\g}-1, 0)\,\ind_{n_\?=0}\right\}\, ,
\end{multline*}
where $\expectation_{\rdegree}$ denotes expectation with respect the 
multinomial variables $n_\0,n_\g,n_\?$ with sum $\rdegree$ and parameters
$\lra_{\0}^{\infty},\lra_{\*}^{\infty},\lra_{\?}^{\infty}$. 
Once again, it is quite easy to compute the above expectations. One obtains
\begin{multline}
\expectation\left\{\frac{1}{n}\sum_{a\in{\cal C}}(\rdegree{}'_{a}-1)
\right\} =
\frac{\lnode'(1)}{\Gamma'(1)}\{
\Gamma'(1-\lra_{\?})\lra_{\g}-\\
-\Gamma(1-\lra_{\?})+\Gamma(1-\lra_{\?}-\lra_{\g})\}\, . \label{Cond2}
\end{multline}

By taking the difference of Eqs.~(\ref{Cond1}) and (\ref{Cond2}),
and after a few algebraic manipulations, we finally get
the desired result
\begin{eqnarray*}
\expectation [{\mathbb C}] = F(\lra,\cp,\gp)\, ,
\end{eqnarray*}
where
\begin{multline*}
F(\lra,\cp,\gp) \defas \lnode'(1)[\lra_\?(1-\rla_\?)-(\lra_\?+\lra_\g)
(1-\rla_\?-\rla_\g)]-\\
-\cp(1-\gp)[\lnode(\rla_\?+\rla_\g)-\lnode(\rla_\?)] + \\
+\frac{\lnode'(1)}{\Gamma'(1)}[\Gamma(1-\lra_\?)-\Gamma(1-\lra_\?-\lra_\g)]\, .
\end{multline*}
Here we used the shorthand $\lra$ for the vector
$(\lra_\?,\lra_\g,\lra_\0,\rla_\?,\rla_\g,\rla_\0)$.

Imagine now changing $\gamma\to\gamma+\Delta\gamma$ and computing 
the number of new conditions on the newly guessed variables
(whose expected number was computed in the previous section).
Call $\Delta{\mathbb C}$
the upper bound on their number provided by  Lemma \ref{LemmaConditions}.
It is clear that, repeating the above derivation, we get
\begin{multline*}
\expectation [\Delta {\mathbb C}] = 
F(\lra^{\infty}(\cp,\gp+\Delta\gp),\cp,\gp+\Delta\gp)-\\
-F(\lra^{\infty}(\cp,\gp),\cp,\gp+\Delta\gp)\, ,
\end{multline*}
Consider now two separate possibilities. In the first case
$\lra^{\infty}(\cp,\gp')$ is continuous (and therefore analytic) in the
interval $\gp'\in [\gp,\gp+\Delta\gp]$. By Taylor expansion we get
\begin{eqnarray*}
\expectation [\Delta {\mathbb C}] = -
\frac{\partial F}{\partial\lra}(\lra',\cp,\gp+\Delta\gp)
\cdot 
\frac{\partial \lra^{\infty}(\cp,\gp)}{\partial\gp} \, \Delta\gamma
+ O((\Delta\gp)^2)\, .
\end{eqnarray*}
with the gradient of $F$ being evaluated at
$\lra' =  \lra^{\infty}(\cp,\gp+\Delta\gp)$. A direct calculation
shows that the gradient vanishes at this point leading to 
$\expectation [\Delta {\mathbb C}] = O((\Delta\gp)^2)$.

In the second case, the interval $[\gp,\gp+\Delta\gp]$ includes a 
discontinuity point (a jump) $\gp_{\rm j}$. 
Let $\lra_{{\rm j}+}\defas \lxl^{j+1} = \lim_{\gp\downarrow\gp_{\rm j}}
\lra^{\infty}(\cp,\gp)$ and
$\lra_{{\rm j}-} \defas \uxl^{j}  = \lim_{\gp\uparrow\gp_{\rm j}}\lra^{\infty}(\cp,\gp)$. We have
\begin{eqnarray*}
\expectation [\Delta {\mathbb C}] = 
F(\lra_{{\rm j}+},\cp,\gp_{\rm j})- F(\lra_{{\rm j}-},\cp,\gp_{\rm j})+ O(\Delta\gp)\, .
\end{eqnarray*}
\subsection{Finishing the proof}

Consider now the guessing strategy explained in Section \ref{sec:guessstrat}. First 
the received message is decoded with the usual iterative decoder.
At this point $\gamma = 0$.
Then each bit is selected independently
with $\Delta\gamma/(1-\gamma)$ and guessed if its valued was not determined
(eventually in terms of former guesses) at previous stages.
The \maxwell decoder is then run until a fixed point is reached.
The number of new guesses at this stage is $\Delta{\mathbb G}_{\gamma}$ and 
the number of new conditions is upper bounded by $\Delta{\mathbb C}_{\gamma}$.
This operation is repeated until $\nu_i(\infty) \in\{\0, \g\}$ for each $i$.
Without loss of generality, we may imagine this to happen at
$\gp = 1$.

At this point each realization of the guesses compatible with the 
conditions yields a codeword compatible with the received message.
We have 
\begin{multline*}
\lim_{n\to\infty}\frac{1}{n}\expectation_{\graph}[ H_{\graph}(X|Y)] \ge
\sum_{\gamma}\expectation[\Delta{\mathbb G}_{\gamma}] -
\sum_{\gamma}\expectation[\Delta{\mathbb C}_{\gamma}] \\
 =\int_{0}^{1} \cp\lnode(\rla_\?(\gp,\cp))\, \de\gp 
-\sum_{\gamma_{\rm j}}\Delta F_{\rm j}
+ O(\Delta\gp)\, ,
\end{multline*}
where the last sums runs over the jump positions $\gamma_{{\rm j}}$ and
$\Delta F_{\rm j}\le
F(\lra_{{\rm j}+},\cp,\gp_{\rm j})- F(\lra_{{\rm j}-},\cp,\gp_{\rm j})$
is the discontinuity of $F$ at those positions.
In order to finish the proof of Lemma \ref{ImprovedUB}, notice that
$H(X|Y)$ does not depend upon $\Delta\gp$ and we can therefore take
the limit $\Delta\gp\to 0$ discarding $O(\Delta\gp)$ terms.
Moreover $\rla_{\?}(\gp,\cp) = \rla(\cp(1-\gp))$ (the last
quantity being the fixed point of DE for the usual BP decoder at 
erasure probability $\cp$), and therefore
\begin{align*}
\int_{0}^{1} \cp\lnode(\rla_\?(\gp,\cp)) \, \de\gp = 
\int_{0}^{\cp} \lnode(\rla(\cp')) \, \de\cp'\, 
\end{align*}
is just the area under the BP \exitentropy curve 
(dark gray in Fig.~\ref{fig:exitcurve}, (a)). Finally, 
let $\cp_{\rm j} = (1-\gp_{\rm j})\cp$ and 
$(\lra(\cp_{\rm j}+),\rla(\cp_{\rm j}+))$ 
and $(\lra(\cp_{\rm j}-),\rla(\cp_{\rm j}-))$ be the fixed point of 
DE for the usual iterative decoder just above and below the jump.
Then
\begin{eqnarray*}
\Delta F_{\rm j} = P_{\cp_{\rm j}}(\lra(\cp_{\rm j}-),\rla(\cp_{\rm j}-)
-P_{\cp_{\rm j}}(\lra(\cp_{\rm j}+),\rla(\cp_{\rm j}+))\, ,
\end{eqnarray*}
where $P_\cp(\lra,\rla)$ is the trial entropy,  cf. Def.~\ref{TrialDef}.
Because of Lemma~\ref{TrialLemma}, $\Delta F_{\rm j}$ is just the area
delimited by the EBP \exitentropy curve and a vertical line through the jump,
(dark gray in Fig.~\ref{fig:exitcurve}, (b)).
%
%
\subsection{Maxwell Decoder: Illustration and Implementation}
The Maxwell decoder provides an {\em interpretation} for the balance of areas 
which we described in Sections \ref{UpperSection} and \ref{sec:Counting}. 
For many ensembles, e.g., the $(3,6)$-regular ensemble, 
Theorem \ref{CountingTheorem} gives a complete characterization
of the MAP EXIT function and therefore a complete justification 
of the Maxwell construction. In some other cases we are not quite
as lucky, see e.g. the ensemble discussed in Example \ref{equ:doublejumpexample}, 
and we can only conjecture that the parts
of the MAP EXIT function which are not covered by Theorem \ref{CountingTheorem}
also follow the Maxwell construction. Let us now review some
typical case. 

\begin{figure}[htp]
\begin{center}
\begin{picture}(240,130)
\put(50,5)
{
\put(-67,-10){\includegraphics[scale=0.62]{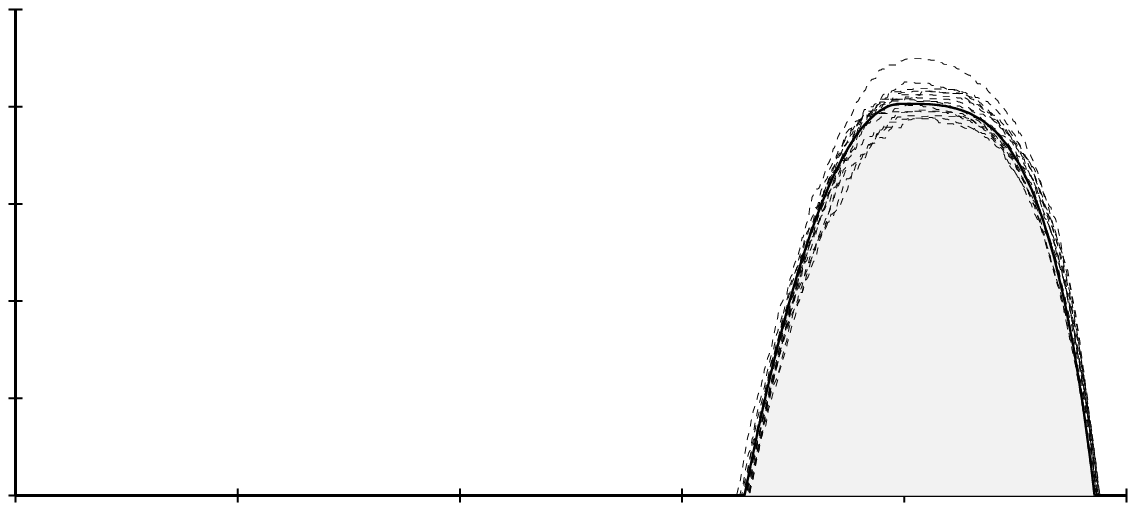}}
\put(10,15){\includegraphics[scale=0.3]{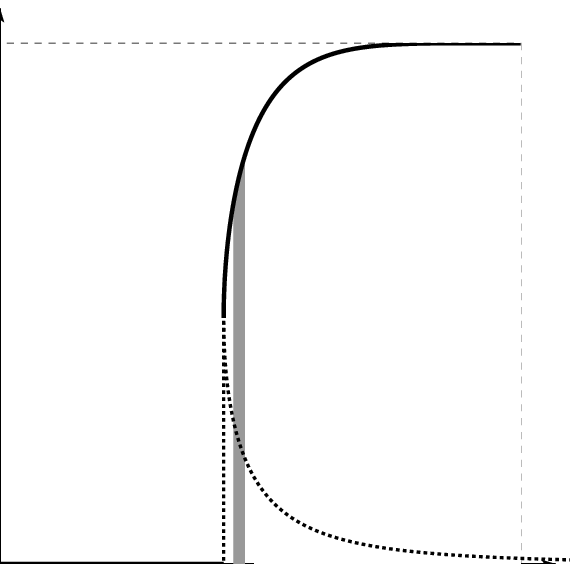}}
\put(30,80){\includegraphics[scale=0.3]{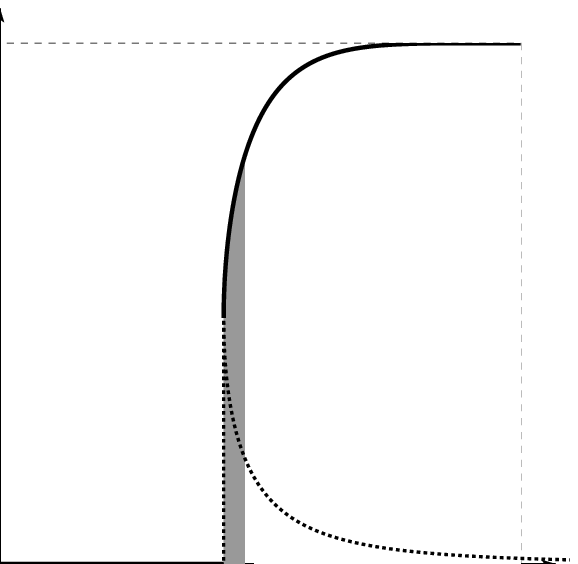}}
\put(135,70){\includegraphics[scale=0.3]{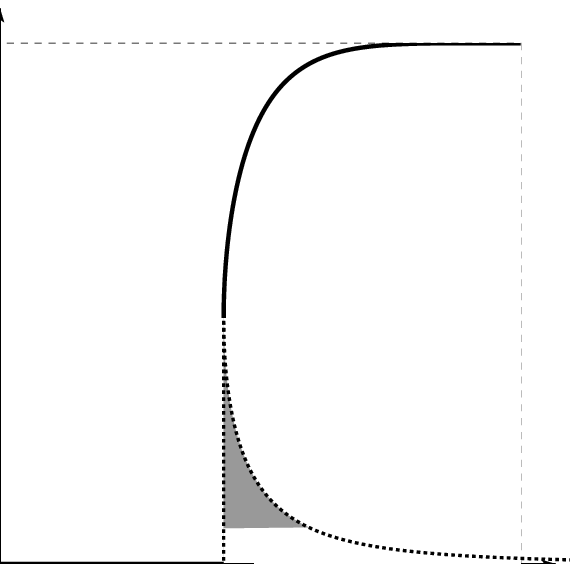}}
\put(149,11){\includegraphics[scale=0.3]{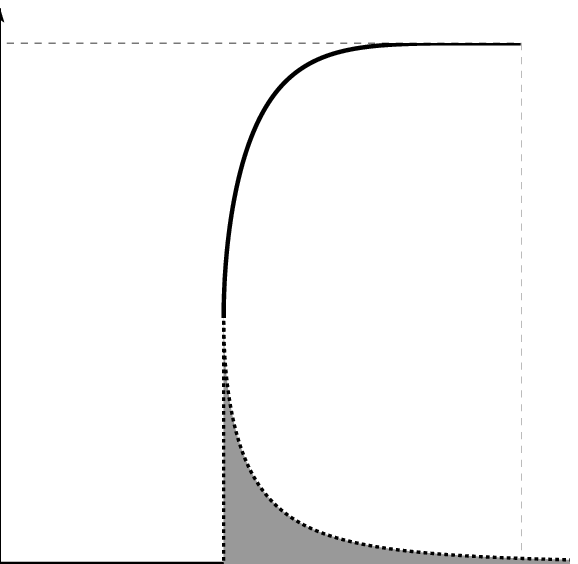}}
\put(68,40){\makebox(0,0){$\rightarrow$}}
\put(91,94){\makebox(0,0){$\searrow$}}
\put(138,64){\makebox(0,0){$\swarrow$}}
\put(144,8){\makebox(0,0){$\swarrow$}}
\put(-55,101){\makebox(0,0){$\entropy$}}
\put(70,8){\makebox(0,0){A}}
\put(103,82){\makebox(0,0){B}}
}
\put(0,0)
{
\put(2,79){\makebox(0,0)[l]{{$680$}}}
\put(2,61){\makebox(0,0)[l]{{$510$}}}
\put(2,43){\makebox(0,0)[l]{{$340$}}}
\put(2,25){\makebox(0,0)[l]{{$170$}}}
\put(-5,1){\makebox(0,0)[l]{{$0.0$}}}
\put(30,1){\makebox(0,0)[l]{{$0.2$}}}
\put(68,1){\makebox(0,0)[l]{{$0.4$}}}
\put(110,1){\makebox(0,0)[l]{{$0.6$}}}
\put(148,1){\makebox(0,0)[l]{{$0.8$}}}
\put(190,1){\makebox(0,0)[l]{{$1.0$}}}
}
\end{picture}
\caption{\maxwell decoder applied to the $(3,6)$-regular LDPC ensemble. 
Asymptotic entropy of the \maxwell decoder $\entropy$ 
(logarithm of the number of running copies) as a function of the
fraction of determined bits.  15
channel and code realizations with  $\ih=0.46$ and blocklength 
$\n=34\cdot 10^3$ are shown (dashed curves) together with the analytic 
asymptotic curve (solid curve). The inserts show how the entropy 
curve can be constructed from the \exitentropy curve. The fraction of guesses 
is shown in the 2 left-most inserts while the fraction of 
contradictions is shown in the 2 right inserts. }
\label{fig:guesscont} 
\end{center}
\end{figure}

\bex[$(3,6)$ LDPC ensemble] \label{ex:prof36} Consider the 
\ddp $(\ledge,\redge)=(x^2,x^5)$ and the  
corresponding $\ldpc$ ensemble with design rate one-half. 
Its BP and MAP \exitentropy functions are depicted in 
Fig.~\ref{fig:exitcurve} together with
the balance conditions. 
Fig.~\ref{fig:guesscont} shows the evolution of the
entropy $\entropy(\iter)$, i.e., the logarithm of the number of running copies 
as discussed in Fig. \ref{fig:maxwelltwodecoder}, 
as a function of the fraction of bits determined by the decoding 
process for the $(3, 6)$-regular LDPC ensemble. 
Transmission takes 
place over BEC($\ih = 0.46$), i.e., we fix the channel parameter $\ih$ so that 
$\ih^\BP\approx0.4294<\ih<\ih^\MAP\approx0.4882$. After transmission,
a fraction $1-\ih=0.54$ of bits is known. The classical BP algorithm  
proceeds until it gets stuck at the fixed point 
$(\xl^\ih \approx 0.3789, \xr^\ih \approx 0.9076)$
of DE. At this point (point A in the figure), 
a fraction $1-\ih\lnode(\xr^\ih)\approx0.6561$ of bits has been determined. 
Now the guessing phase of the \maxwell decoder starts.
It ends at point B, which corresponds to the BP threshold 
$(\xl^\BP \approx 0.2606, \xr^\BP \approx 0.7790)$.
The total fraction of guesses that the \maxwell decoder has to venture is 
$\int_{\xl^\BP}^{\xl^\ih}\xh(\ih(\xl))\text{d}\ih(\xl)=P(\xl^\ih,\xr^\ih)-
P(\xl^\BP,\xr^\BP)$.  
For our specific example we have
$P(\xl,\xr(\xl))=-\frac{5\xl^2}{2}+10\xl^3-\frac{25\xl^4}{2}+7\xl^5-\frac{3\xl^6}{2}$,
so that the total fraction of guesses is equal to $0.0201509$.
For a blocklength of $\n=34000$ this corresponds to roughly $685$ guesses.
At this point the BP decoding phase resumes. More and more 
guesses are confirmed. Since we are operating below the MAP threshold,
(essentially) all guesses are eventually confirmed and the 
\maxwell decoder comes to a halt.

\begin{figure}[htp]
\begin{center}
\begin{picture}(240,100)
\put(50,0) 
{
\put(10,0){\includegraphics[scale=0.5]{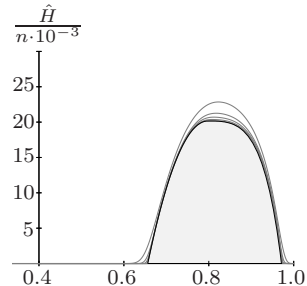}}
\put(10,101){\makebox(0,0)[l]{$\frac{\entropy}{\n\cdot10^{-3}}$ }}
\put(10,78){\makebox(0,0)[l]{\footnotesize{$25$}}}
\put(10,64){\makebox(0,0)[l]{\footnotesize{$20$}}}
\put(10,51){\makebox(0,0)[l]{\footnotesize{$15$}}}
\put(10,37){\makebox(0,0)[l]{\footnotesize{$10$}}}
\put(14,23){\makebox(0,0)[l]{\footnotesize{$5$}}}
\put(14,5){\makebox(0,0)[l]{\footnotesize{$0.4$}}}
\put(46,5){\makebox(0,0)[l]{\footnotesize{$0.6$}}}
\put(78,5){\makebox(0,0)[l]{\footnotesize{$0.8$}}}
\put(110,5){\makebox(0,0)[l]{\footnotesize{$1.0$}}}
%
}
\end{picture}
\caption{\maxwell decoder applied to the $(3,6)$ LDPC ensemble:  
Expected symptotic entropy  as a function of the fraction of determined bits at $\ih=0.46$ (solid curve) 
and  empirical average entropy curves (gray curves). Simulations are shown for $\n=780$ (average over $6\cdot10^4$ realizations), $\n=3125$ (average over $16\cdot10^3$ realizations), $\n=12500$ (average over $4\cdot10^3$ realizations), 
$\n=50000$ (average over $10^3$ realizations), $\n=200000$ (average over $150$ realizations).
}
\label{fig:guesscontbis} 
\end{center}
\end{figure}
\eex

\bex[Typical Double ``Jump''] \label{ex:prof2j}
\label{ex:running2jumps} 
Consider the 
\ddp $(\ledge,\redge)=(\frac{3x + 3 x^2+4x^{13}}{10},x^6)$ and the  
corresponding $\ldpc$ ensemble with design rate $\drate=\frac{19}{39}\approx0.4872$. 
Its BP \exitentropy function is depicted in 
Fig.~\ref{fig:epsilon}, its EBP \exitentropy curve together with
the balance conditions is shown in 
Fig.~\ref{fig:multijump}. Finally, in Example \ref{equ:doublejumpexample} we have
discussed how large parts of the MAP \exitentropy curve can be constructed
based on Theorem \ref{CountingTheorem}. 
%
The MAP threshold is $\ih^\MAP\approx0.4913$ 
(at $\xl^\MAP\approx0.1434$). According to the Maxwell costruction,
the second MAP discontinuities occurs at
$\ih^{\MAP,2}\approx0.5186$ (at $\lxl^{\MAP,2}\approx0.2378$, $\uxl^{\MAP,2}\approx0.4121$) .

\begin{figure}[htp]
\begin{center}
\begin{picture}(240,100)
\put(-5,0){
\put(10,0){\includegraphics[scale=0.5]{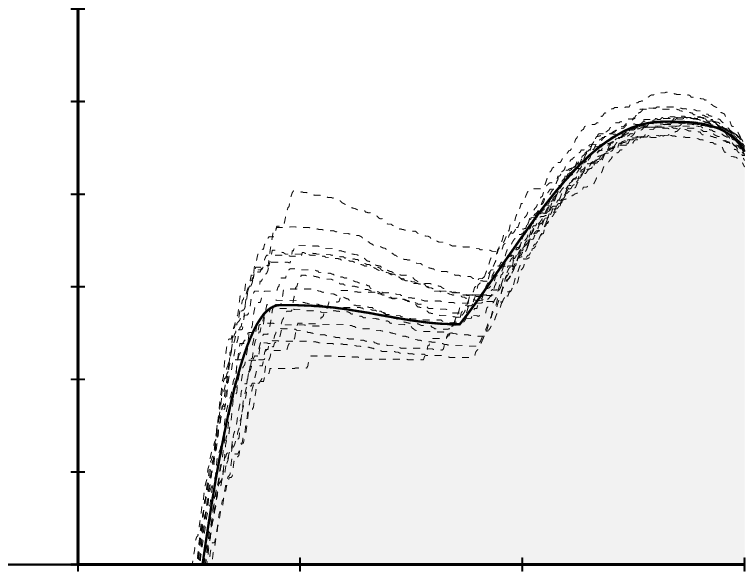}}
\put(14,101){\makebox(0,0)[l]{\footnotesize$\entropy$ }}
\put(6,78){\makebox(0,0)[l]{\footnotesize{$850$}}}
\put(6,64){\makebox(0,0)[l]{\footnotesize{$680$}}}
\put(6,51){\makebox(0,0)[l]{\footnotesize{$510$}}}
\put(6,37){\makebox(0,0)[l]{\footnotesize{$340$}}}
\put(6,23){\makebox(0,0)[l]{\footnotesize{$170$}}}
\put(14,5){\makebox(0,0)[l]{\footnotesize{$0.4$}}}
\put(46,5){\makebox(0,0)[l]{\footnotesize{$0.6$}}}
\put(78,5){\makebox(0,0)[l]{\footnotesize{$0.8$}}}
\put(110,5){\makebox(0,0)[l]{\footnotesize{$1.0$}}}
\put(68,-3){\makebox(0,0){{\small{(a)}}}}

}

\put(120,0) 
{
\put(10,0){\includegraphics[scale=0.5]{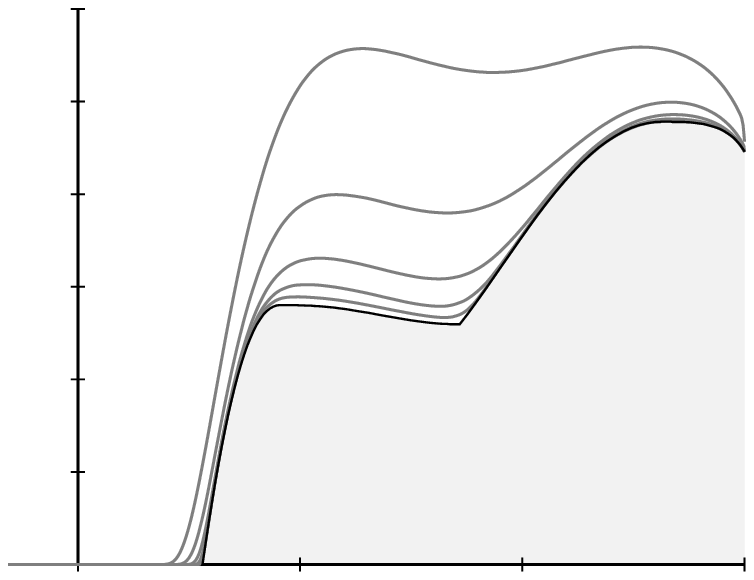}}
\put(10,101){\makebox(0,0)[l]{$\frac{\entropy}{\n\cdot10^{-3}}$ }}
\put(10,78){\makebox(0,0)[l]{\footnotesize{$25$}}}
\put(10,64){\makebox(0,0)[l]{\footnotesize{$20$}}}
\put(10,51){\makebox(0,0)[l]{\footnotesize{$15$}}}
\put(10,37){\makebox(0,0)[l]{\footnotesize{$10$}}}
\put(14,23){\makebox(0,0)[l]{\footnotesize{$5$}}}
\put(14,5){\makebox(0,0)[l]{\footnotesize{$0.4$}}}
\put(46,5){\makebox(0,0)[l]{\footnotesize{$0.6$}}}
\put(78,5){\makebox(0,0)[l]{\footnotesize{$0.8$}}}
\put(110,5){\makebox(0,0)[l]{\footnotesize{$1.0$}}}
\put(68,-3){\makebox(0,0){{\small{(b)}}}}

}
\end{picture}
\caption{\maxwell decoder applied to the irregular ``double-jump'' LDPC ensemble shown in Fig.~\ref{fig:epsilon}:  
Asymptotic entropy  as a function of the fraction of determined bits at $\ih=0.5313$ (point $B$). (a) 15 channel 
and code realizations of blocklength $\n=34000$ are shown (dashed curves) together with the analytic 
asymptotic curve (solid curve). (b) Convergence of the average entropy curves (gray curves) to the analytic 
expected curve (solid curve). Simulations are shown for $\n=780$ (average over $6\cdot10^4$ realizations), $\n=3120$ (average over $16\cdot10^3$ realizations), $\n=12480$ (average over $4\cdot10^3$ realizations), 
$\n=50017$ (average over $10^3$ realizations), $\n=200500$ (average over $250$ realizations).
}
\label{fig:guesscont2J} 
\end{center}
\end{figure}

Fig.~\ref{fig:guesscont2J} shows the evolution of the
entropy $\entropy(\iter)$ for $\ih=0.5313$. This corresponds
to the point $C$ in Fig.~\ref{equ:doublejumpexample}, the
first point at which the counting argument no longer applies. 
By comparing the result of the simulations to the analytic curve,
corresponding to the Maxwell construction we can see that
at least emperically the Maxwell construction seems to be 
valid over the whole range.

\eex

\section{Some Further Examples} 
\subsection{Special Cases}
Although (for sake of simplicity) we did not 
discuss this case in the previous sections, 
other curious (but frequent) examples are those 
when the number of discontinuities $J^\BP$ 
of the BP \exitentropy curves is not equal to the number of 
discontinuities $J^\MAP$ of the MAP \exitentropy curve. 
Examples \ref{ex:mapleqbp} and \ref{ex:bpleqmap} show two such cases. 

\bex[$J^\MAP<J^\BP$] \label{ex:mapleqbp} Consider the 
\ddp $(\ledge,\redge)=(\frac{x^{10}+x^{60}}{2},\frac{3 x^{10}+17 x^{80}}{20})$ 
 and the  corresponding $\ldpc$ ensemble with rate $\drate=\frac{3209}{5832}\approx0.5502$.

\begin{figure}[htp]
\centering
\setlength{\unitlength}{1bp}
\begin{picture}(240,140)
\put(5,28)
{
\put(5,5){\includegraphics[width=100bp]{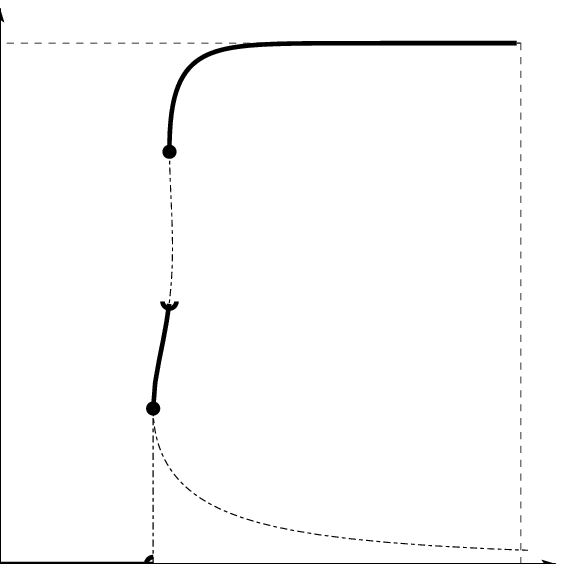}}
\put(135,5){\includegraphics[width=100bp]{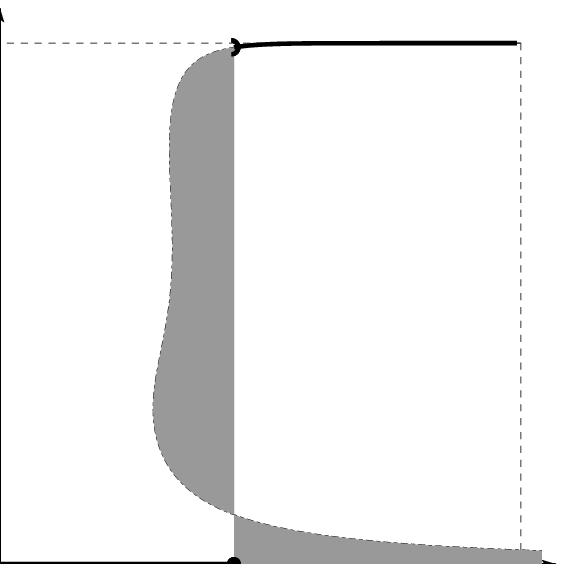}}
\put(0,0){\makebox(0,0){\small{$0$}}}
\put(100,0){\makebox(0,0){\small{$1$}}}
\put(0,100){\makebox(0,0){\small{$1$}}}
\put(130,0){\makebox(0,0){\small{$0$}}}
\put(230,0){\makebox(0,0){\small{$1$}}}
\put(130,100){\makebox(0,0){\small{$1$}}}
\put(40,-5){\makebox(0,0){\small{$\ih^\BP$}}}
\put(179,-5){\makebox(0,0){\small{$\ih^\MAP$}}}
\put(200,80){\makebox(0,0){\small{$~$}}}
}
\put(60,7){\makebox(0,0){{\small{(a)}}}}
\put(190,7){\makebox(0,0){{\small{(b)}}}}
\end{picture}
\caption{
When the numbers of BP and MAP ``jumps'' (respectively, 
$J^\BP$ and $J^\MAP$) are different:  
(a) BP \exitentropy function with $J^\BP=2$ (b) MAP \exitentropy function 
with $J^\MAP=1$ and Maxwell constriction. 
}
\label{fig:malicious2}
\end{figure}
The MAP \exitentropy curve has a single ``jump'' at 
$\ih^\MAP\approx0.4493$ ($\xl^\MAP\approx0.4425$) 
whereas the BP \exitentropy curve has two such 
singularities at $\ih^\BP\approx0.2941$ ($\xl^\BP\approx0.05738$) 
and $\ih^{(\BP,2)}\approx0.3254$ ($\uxl^{(\BP,2)}\approx0.2117$) as 
shown in Fig.~\ref{fig:malicious2}.
%
%
\begin{figure}[htp]
\centering
\setlength{\unitlength}{1bp}
\begin{picture}(80,70)
\put(10,0){\includegraphics[width=60bp]{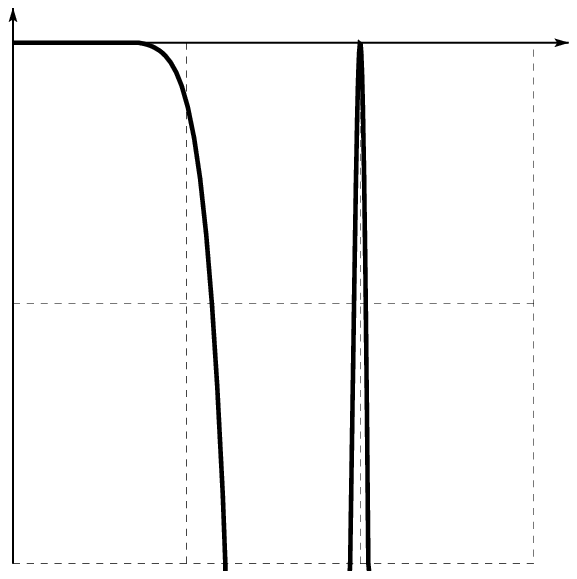}}
\put(10,0)
 {
 \put(58,-3){\makebox(0,0){$\scriptstyle{\frac32}$}}
 \put(-3,58){\makebox(0,0){$\scriptstyle{0}$}}
 \put(19,-3){\makebox(0,0){$\scriptstyle{\frac12}$}}
 \put(-6,29){\makebox(0,0){$\scriptstyle{-\frac{1}{40}}$}}
 \put(-6,0){\makebox(0,0){$\scriptstyle{-\frac{1}{20}}$}}
 }
\end{picture}
\caption{
Function $\Psi_{\nddp}(u)$ for the \ddp formed by the residual 
ensemble at $\ih^\MAP=0.4493$.
}
\label{fig:wd1bp2map}
\end{figure}
As shown in Fig.~\ref{fig:wd1bp2map},
Theorem \ref{CountingTheorem} applies at the MAP
threshold and so the whole MAP EXIT curve is determined by the counting
argument in this case. The Maxwell construction is therefore confirmed
in this case. 
\eex

\bex[$J^\BP<J^\MAP$] \label{ex:bpleqmap} 
Consider the \ddp $(\ledge,\redge)=(\frac{3x+3x^2+14x^{50}}{20},x^{15})$ 
and the  corresponding $\ldpc$ ensemble with design rate $\drate=\frac{311}{566}\approx0.5495$. 
\begin{figure}[htp]
\centering
\setlength{\unitlength}{1bp}
\begin{picture}(240,140)
\put(5,28)
{
\put(5,5){\includegraphics[width=100bp]{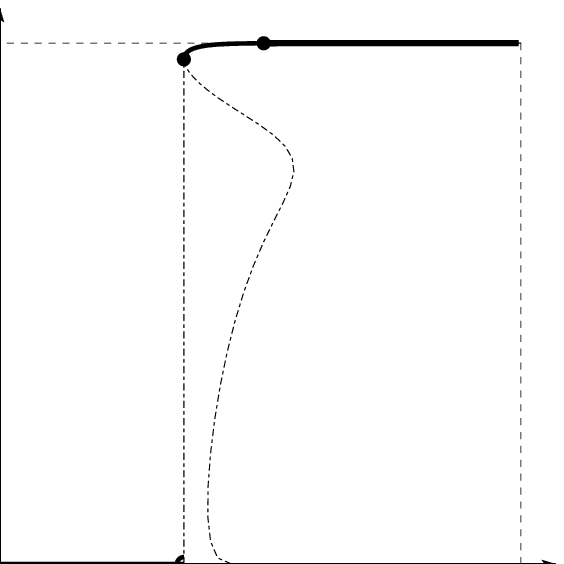}}
\put(135,5){\includegraphics[width=100bp]{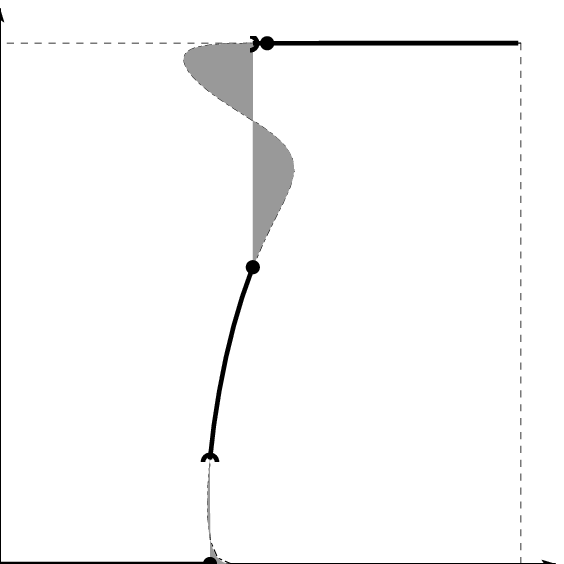}}
\put(0,0){\makebox(0,0){\small{$0$}}}
\put(100,0){\makebox(0,0){\small{$1$}}}
\put(0,100){\makebox(0,0){\small{$1$}}}
\put(130,0){\makebox(0,0){\small{$0$}}}
\put(230,0){\makebox(0,0){\small{$1$}}}
\put(130,100){\makebox(0,0){\small{$1$}}}
\put(40,-5){\makebox(0,0){\small{$\ih^\BP$}}}
\put(54,105){\makebox(0,0){\small{A}}}
\put(179,-5){\makebox(0,0){\small{$\ih^\MAP$}}}
\put(200,80){\makebox(0,0){\small{$~$}}}
\put(185,105){\makebox(0,0){\small{A}}}
}
\put(60,7){\makebox(0,0){{\small{(a)}}}}
\put(190,7){\makebox(0,0){{\small{(b)}}}}
\end{picture}
\caption{
When the numbers of BP and MAP ``jumps'' (respectively, $J^\BP$ and $J^\MAP$) are different:  
(a) BP \exitentropy function with $J^\BP=1$ (b) MAP 
\exitentropy function with $J^\MAP=2$ and Maxwell construction. 
}
\label{fig:malicious1}
\end{figure}
The BP \exitentropy curve has a single ``jump'' at 
$\ih^\BP\approx0.3531$ ($\xl^\BP\approx0.3008$).
 
Unfortunately, Theorem \ref{CountingTheorem} shows
the tightness of the \maxwell construction
only up to point A (at $\ih\approx0.5063$, see Fig.~\ref{fig:malicious1}) .
But it is
quite natural to conjecture that the MAP \exitentropy curve 
has two singularities, namely 
at $\ih^\MAP\approx0.3986$ ($\xl^\MAP\approx0.0340$) 
and at $\ih^{(\MAP,2)}\approx0.4855$ ($\uxl^{(\MAP,2)}\approx0.1096$) 
as shown in Fig.~\ref{fig:malicious1}. This is validated by the \maxwell 
 decoder. Namely the \maxwell decoder 
gives a residual entropy (as a fraction of the blocklength) of 
 $\frac{\entropy}{\n}\approx0.0121$ at $\ih=0.44$. This value is 
exactly the value of the area (between $\ih=0$ and $\ih=0.44$) under 
the conjectured MAP EXIT curve. This shows that, between the two 
conjectured MAP phase transitions, the \maxwell decoder follows 
the part of the EBP EXIT function which is ``hidden'' from the 
BP decoder.
%
%
%
The Maxwell construction is  conjectured to hold in this case.
\eex

\subsection{Difference Between MAP and BP Threshold}
Let $\drate<1$ be the design rate. Consider a sequence of degree distribution pairs 
$\{(\ledge(x),\redge(x))=(x^{\ldegree-1},x^{\frac{\ldegree}{1-\drate}-1})\}_{\ldegree\geq2}$ 
with fixed design rate $\drate$. Ensembles associated to this sequence are regular LDPC 
code ensembles. We have seen in
Fact~\ref{fact:regular} that such ensembles have at most one jump and
therefore we expect our bound on the MAP threshold to be tight.
It was shown already in \cite{Mac97}, that if $\ldegree$ is increased then
the weight distribution of such ensembles converges to the one
of Shannon's random ensemble and, hence, the MAP threshold of
such ensembles converges to the Shannon limit.
Using the replica method,
an explicit asymptotic expansion of the MAP threshold was given
in \cite{Andrea}.

Let us give here an alternative proof of this fact using our
machinery.
That the MAP threshold $\ih^\MAP(\ldegree)$ converges to the Shannon threshold is shown
in Fact~\ref{theo:randomregularml}.
On the other hand, as stated 
in Fact~ \ref{theo:randomregular}, the 
BP threshold  $\ih^\MAP(\ldegree)$ goes to 0 when $\ldegree\to\infty$.
This shows that the two thresholds can be arbitrarily far apart, and nevertheless
the MAP \exitentropy curve can be constructed from the corresponding (E)BP \exitentropy curve!

\begin{figure}[htp]
\centering
\setlength{\unitlength}{1bp}%
\begin{picture}(240,140)
\put(5,20)
{ 
\put(5,5){\includegraphics[width=100bp]{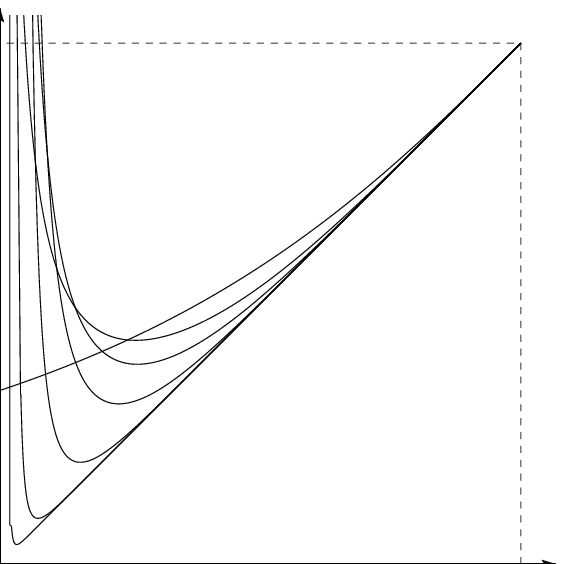}} 
\put(135,5){\includegraphics[width=100bp]{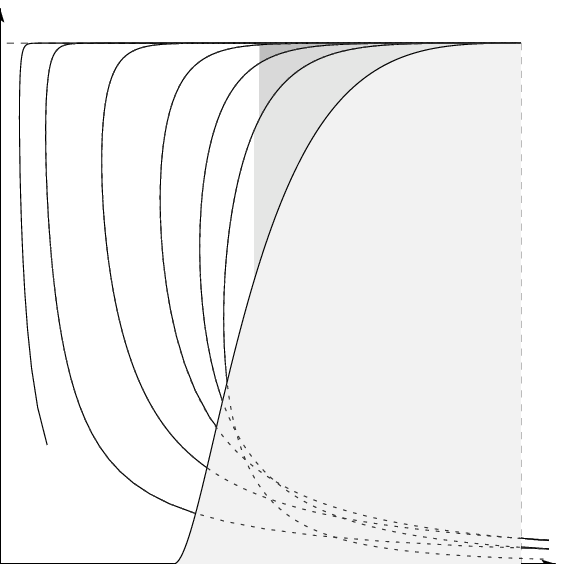}}
\put(0,0){\makebox(0,0){\small{$0$}}}
\put(100,0){\makebox(0,0){\small{$1$}}}
\put(0,100){\makebox(0,0){\small{$1$}}}
\put(130,0){\makebox(0,0){\small{$0$}}}
\put(230,0){\makebox(0,0){\small{$1$}}}
\put(130,100){\makebox(0,0){\small{$1$}}}
\put(-7,42){\makebox(0,0){\scriptsize{$\ih^{\BP}(2)$}}}
\put(-9,33){\makebox(0,0){\scriptsize{$=\ih^{\text{SC}}$}}}
\put(33,48){\makebox(0,0){\scriptsize{$\ih^\BP(3)$}}}
\put(31,28){\makebox(0,0){\scriptsize{$\ih^\BP(12)$}}}
\put(31,17){\makebox(0,0){\scriptsize{$\ih^\BP(35)$}}}
\put(24,9){\makebox(0,0){\scriptsize{$\ih^\BP(70)$}}}
\put(60,80){\makebox(0,0){\small{$\ih(\xl)$}}}
\put(170,-2){\makebox(0,0){\scriptsize{$\ih^\BP(2)$}}}
\put(185,45){\makebox(0,0){\scriptsize{$\ih^\BP(3)$}}}
\put(178,54){\makebox(0,0){\scriptsize{$\ih^\BP(4)$}}}
\put(170,65){\makebox(0,0){\scriptsize{$\ih^\BP(6)$}}}
\put(162,75){\makebox(0,0){\scriptsize{$\ih^\BP(12)$}}}
\put(153,85){\makebox(0,0){\scriptsize{$\ih^\BP(35)$}}}
\put(148,94){\makebox(0,0){\scriptsize{$\ih^\BP(70)$}}}
\put(185,105){\makebox(0,0){\small{$\ih^\MAP$}}}
\put(210,80){\makebox(0,0){\small{$\xh(\ih)$}}}
}
\put(60,7){\makebox(0,0){{\small{(a)}}}}
\put(190,7){\makebox(0,0){{\small{(b)}}}}
\end{picture}
\caption{Regular BP \exitentropy entropy curves with design rate $\drate=\frac{1}{2}$. (a) Channel entropy function $\xl\mapsto\ih^{(\ldegree)}(\xl)$ (b) \exitentropy curve $\xh^{(\ldegree)}(\ih)\longleftrightarrow\ih^{(\ldegree)}(\xh)$. The depicted ensembles are, in decreasing order, the $(100,200)$, the $(35,70)$, the $(12,24)$, the $(6,12)$, the $(4,8)$, the $(3,6)$ and the $(2,4)$ regular ensemble. While the BP threshold goes to 0, the bit MAP threshold goes to the Shannon limit $0.5$.
  }
 \label{fig:regcodes}
 \end{figure}

This is illustrated in Fig.~\ref{fig:regcodes} and the proofs are given in the sequel.  

\blemma \label{lemma:randomregularepsilon}
For a fixed non-negative $\xl\in(0,1]$, denoting $\ih^{(\ldegree)}(\xl)\defas\frac{x}{(1-(1-\xl)^{\frac{\ldegree}{1-\drate}-1})^{\ldegree-1}}$, we get $\ih^{(\ldegree)}(\xl)\underset{\ldegree\to\infty}{\longrightarrow}\xl$.
\elemma
\bproof
This limit is  classically obtained with $(\ldegree-1)\log[1-(1-\xl)^{\frac{\ldegree}{1-\drate}-1}]\sim-(\ldegree-1)(1-\xl)^{\frac{\ldegree}{1-\drate}-1}$ which gives $(1-(1-\xl)^{\frac{\ldegree}{1-\drate}-1} )^{\ldegree-1}  \underset{\ldegree\to\infty}{\longrightarrow}1^{-}$.
\eproof
\bprop\label{theo:randomregular}
Consider the sequence $(x^{\ldegree-1},x^{\frac{\ldegree}{1-\drate}-1})\}_{\ldegree\geq2}$ with fixed
rate $\drate<1$, then the BP threshold $\ih^\BP(\ldegree)\underset{\ldegree\to\infty}{\longrightarrow}0$.
\eprop
\bproof
Consider first the BP threshold  $\ih^\BP(\ldegree)\defas\min_\xl\{\ih^{(\ldegree)}(\xl)\}$. 
Fix $\xi>0$ (very small). Clearly $0\leq\ih^\BP(\ldegree)\leq\ih^{(\ldegree)}(\frac{\xi}{2})$, and, since
$\ih^{(\ldegree)}(\frac{\xi}{2})\underset{\ldegree\to\infty}{\longrightarrow}\frac{\xi}{2}$ with 
Lemma~\ref{lemma:randomregularepsilon}, we 
can state
\[
\exists \ldegree_0 \in \naturals, ~~ \forall \ldegree\geq\ldegree_0 ~~ \ih^{(\ldegree)}(\frac{\xi}{2})\leq \frac{\xi}{2}+\frac{\xi}{2}.
\]
This gives that, for all $\ldegree\geq\ldegree_0$, the statement $0\leq\ih^\BP(\ldegree)\leq \xi$ holds. This is true for any fixed $\xi$ meaning $\ih^\BP(\ldegree) \underset{\ldegree\to\infty}{\longrightarrow} 0$.
\eproof
Instead of studying the parameterized \exitentropy quantity  $\xh(\xl)\defas(1-(1-\xl)^{\rdegree-1})^{\ldegree}$, 
it is often more convenient to work directly with the inverse mapping $\xh \mapsto \xl(\xh)\defas1-[1-\xh^{\frac{1}{\ldegree}}]^{\frac{1}{\rdegree-1}}$ such that we can eventually use $\ih(\xh)=\frac{1-[1-\xh^{\frac{1}{\ldegree}}]^{\frac{1}{\rdegree-1}}}{\xh^{\frac{\ldegree-1}{\ldegree}}}$ for $\xh\in(0,1]$.
\blemma
For a fixed $\xh\in(0,1)$, we have $\ih(\xh)=\frac{1-(1-\xh^{\frac{1}{\ldegree}})^{\frac{\drate-1}{\ldegree-\drate+1}}}{\xh^{\frac{\ldegree-1}{\ldegree}}} \underset{\ldegree\to\infty}{\longrightarrow}0.$
\elemma
\bproof
The second term of the numerator goes  to 1 since, $\log(1-\xh^{\frac{1}{\ldegree}})=\frac{\log\xh}{\ldegree}+\log(\frac{1}{\xh^{\frac{1}{\ldegree}}}-1)=\frac{\log\xh}{\ldegree} + \log(\frac{-\log\xh}{\ldegree}+o(\frac{1}{\ldegree}))$ such that  ${     { \frac{\drate-1}{\ldegree-1+\drate}}} [ \frac{\log\xh}{\ldegree} + \log(\frac{-\log\xh}{\ldegree}+o(\frac{1}{\ldegree})) ] ~\underset{\ldegree\to\infty}{\longrightarrow}0$.
 The lemma follows from $\xh^{\frac{\ldegree-1}{\ldegree}}\sim\xh>0$.
\eproof
\bprop\label{theo:randomregularml}
Consider again the sequence $(x^{\ldegree-1},x^{\frac{\ldegree}{1-\drate}-1})\}_{\ldegree\geq2}$ with fixed
rate $\drate<1$, then $\ih^\MAP(\ldegree)\underset{\ldegree\to\infty}{\longrightarrow}\ih^{\Sh}=1-\drate>0$. 
\eprop
\bproof
First, the inequality  $0\leq\ih^\Sh-\ih^\MAP(\ldegree)$ holds from  the Area Theorem.\footnote{An alternative way is to show it via the Shannon Coding Theorem!}
Second, \[\ih^\Sh-\ih^\MAP(\ldegree) = (1-\drate)-\ih^\MAP(\ldegree) = {\cal{A}}^{(\ldegree)} \leq {\tilde{\cal{A}}}^{(\ldegree)} \] where, in short, ${\cal{A}}^{(\ldegree)}$ represents the closed area between $\{\ih(\xh)\}_{\ih^\BP\leq\ih\leq1}$, the horizontal axis $\{\ih=\ih^\MAP\}$ and the vertical axis $\{\xh=1\}$. The area ${\tilde{\cal{A}}}^{(\ldegree)}$ is the surface of the unit square which lies under $\{\ih(\xh)\}_{0\leq\ih\leq1}$. Now, consider the function $\tilde{\ih}^{(\ldegree)}(\xh)=\min\{\ih(\xh)^{(\ldegree)},1\}\leq1$. The Dominated Convergence Theorem\footnote{Observe that $\ih(\xh)$ does not uniformly converge to $0$ on $(0,1)$ since $\int_0^1\ih(\xh)\text{d}\xh=1-\drate\neq0$.} 
applied to the sequence $\tilde{\ih}^{(\ldegree)}$ gives that 
$\lim_{\ldegree\to\infty}{\cal{\tilde{A}}}^{(\ldegree)} =0$, which concludes the proof. 
\eproof
\subsection{Application to other Iterative Coding Schemes}
Although LDPC ensembles have been used to present the discussed concepts, 
 the picture is not limited to such ensembles.
Equivalent statements are expected to hold in large generality.

To give just one example, consider generalized LDPC (GLDPC) ensembles:
Part of our results can be directly applied like, e.g., 
Lemma~\ref{theo:exactmlthreshold}.
Consider a GLDPC ensemble: Equivalently to the \ddp $(\ledge,\redge)$, 
the pair 
$(\ledge(\xl),\xr(\xl))\defas(\ledge(\xl),1-\redge(1-\xl))$ 
suffices to describe the BP decoding of the ensemble in the asymptotic limit. 
The left (right) component of the pair $(\ledge(\xl),\xr(\xl))$ gives 
the \exitentropy entropy outgoing from the left (right) nodes during the BP decoding. 
To be more precise, at a fixed channel parameter $\ih$,  
 the function  $\xl(\xr)\defas\ih\ledge(\xl)$ is the \exitentropy entropy outgoing from the 
left
and  $\xr(\xl)\defas\xr(\xl,\ih)$ is the \exitentropy 
entropy outgoing from the right.\footnote{Contrary to the left nodes which stay  
 simple repetition codes, the right nodes can be more complex linear codes. 
Therefore, $\xr(\xl)$ often depends on the edge type. 
For GLDPC ensembles, we consider 
the average over all types of node. For Turbo codes, one usually distinguish between systematic 
versus parity bits.}
 A few calculus or computations lead, in general, to an  
expression for the right component \exitentropy entropy (see, e.g, \cite{AKtB04,MeU03a}). 
\bex[GLDPC Codes]
Generalized LDPC codes (see, e.g., \cite{Tan81,BPZ99,LZ98}) are LDPC codes whose check nodes are replaced by some 
more complex linear constraints. Such constraints are viewed as
 component codes which typically have minimum distance $d_{\text{min}}\geq 3$: they are bit MAP 
decoded and the component \exitentropy entropy $\xr(\xl)$ has  smallest  degree $d_{\text{min}}-1$ (see, e.g., \cite{MeU03a}). 
The \exitentropy entropy  $\xr(\xl)$ is the function $ \xr(\xl)\defas\expectation\frac{1}{\rdegree}\sum_{i=1}^{\rdegree} \xr_i(\xl)$, 
where $\rdegree$ is the length of a particular component code and where the expectation is taken with respect to 
all such component codes. 
The distribution $\ledge$ can be freely chosen but must satisfy the design rate constraint $\drate=1-\frac{1-\int \xr}{\int\ledge}$ 
where $\int \xr$ is the rate of the average component code (Area Theorem). For example, consider GLDPC ensembles using 
$[2^p-1,2^p-p-1,3]$ binary Hamming codes as component codes. Then, when $\expectation d_{\text{min}}\geq3$, the BP \exitentropy entropy has 
at least one discontinuity at the BP threshold. It is given as,
\[
(\ih,\xh) =\Big(\frac{\xl}{\ledge(\xr(\xl))},\lnode(\xr(\xl))\Big). 
\]
Theorem \ref{theo:exactmlthreshold} shows that, in general, $\ih^\BP\neq\ih^\MAP$ (The BP threshold being not given by the stability condition whenever the right component code has  $d_{\text{min}}\geq 3$). In the next table, the first example  uses $[7,4,3]$ Hamming codes such that its design rate is $\drate=\frac{1}{7}$ with the pair $(\ledge,\xr)=(\xl,3\xl^2 + 4\xl^3 - 15\xl^4 + 12\xl^5 - 3\xl^6)$ whereas the second example uses the $[15,11,3]$ Hamming code. It can be observed that this classical GLDPC have relatively bad BP threshold compared
to its  MAP upper-bound. In the third example, 
$d_{\text{min}}$ is no longer $>2$ since we choose, in the node perspective, 
a mixture of 40 percent of $[7,6,2]$ Single Parity-Check codes, 
40 percent of $[7,4,3]$ Hamming codes and 20 percent of $[15,11,3]$ Hamming codes. 
The BP \exitentropy function has however still a discontinuity at the BP threshold. 
\begin{center}
{
\tablespace
\small{
\begin{tabular}{ccccc}
$\ledge(\xl)$  & $\xr(\xl)$ & $\ih^\BP$ & $\overline{\ih}^\MAP$  & $\ih^\Sh$  \\  \hline
$\xl$     & $[7,4,3]$   &  0.75645    &   0.85616   &  0.85714  \\ 
$\xl$     & $[15,11,3]$  &  0.46785    &   0.52780    &  0.53333   \\ 
$\frac{3\xl+7\xl^8}{10}$     & mixture  &   0.70483    &   0.71301   & 0.72801   \\
\end{tabular}
}
\tablespace
}
\end{center}
\eex

\section{Conclusion}

We have shown that there is
a close connection between the BP and
the MAP decoder. While this connection is quite general, we focused
in this paper on communication over the binary erasure channel.
In this case, the relation is furnished by the so-called Maxwell
decoder which gives an operational meaning to the various areas
under the EBP \exitentropy curve as number of guesses and  number
of confirmations.
Unfortunately, this paper falls slightly short on several accounts
of proving this relationship in the most general case. Let us
summarize what seem to be the most important issues that still need
to be addressed.


First, there is currently no direct proof which establishes the
existence of the asymptotic MAP \exitentropy curve. Rather, the existence 
follows from the explicit characterization of this limit. This occurs 
 via Theorem \ref{CountingTheorem} in all those cases where
the conditions of the theorem are fulfilled. Although theses conditions
apply to a large class of ensembles, it would be pleasing to show
the existence of the limit in the general case.

A further point that needs some clarification is the restriction we
had to impose in the second proof of Theorem \ref{theo:ebpareatheorem}.
Recall that the argument on the computation tree via the Area Theorem
required that the underlying ensemble has a non-trivial stability condition,
since otherwise part of the EBP \exitentropy curve lies ``outside the unit box,''
i.e., part of the curve corresponds to ``erasure probabilities above one.''
While an analytical prove of Theorem \ref{theo:ebpareatheorem}
is possible, it would be interesting (especially in view of generalizations)
to have a conceptual proof valid for  unconditionally stable ensembles. 

Without doubt the most important challenge is to assert the correctness of
Conjecture \ref{conj:mainconjecture}. This would yield an easy and 
geometrically
pleasing way of constructing the MAP \exitentropy curve from the EBP \exitentropy curve in
the general case.

Finally, an interesting research direction consists in the analysis
of more general combinatorial search problems through a suitable
`Maxwell construction'. An example (extremely close to the topic of this paper)
consists in the problem of satisfiability of random sparse linear systems
(`XORSAT')
considered in ~\cite{MRTZ02,CDMM02}. The counting argument presented
in Section \ref{sec:Counting} is indeed closely related to the approach
of these papers. The ideas presented here can probably be used to analyze the 
behavior of simple resolution algorithms for this problem (see
\cite{BLRZ02} for a numerical exploration).
%
%
\section*{Acknowledgment}
The authors would like to thank Nicolas Macris, Changyan Di, Gerhard Kramer, 
and Tom Richardson for useful discussions.

A.M. has been partially supported by EVERGROW
(i.p. 1935 in the
complex systems initiative of the Future and Emerging Technologies
directorate of the IST Priority, EU Sixth Framework).

\appendices
\section{Proofs for Concentration Theorems} \label{app:concentrationproofs}
Throughout this section, we use the shorthand $H_n =H_{\graph}(X|Y)$ 
to denote the conditional entropy under transmission over the BMS channel 
$p_{Y_{[\n]} \mid X_{[\n]}}(\cdot \mid \cdot)$ using a code  $\graph$ 
 chosen uniformly at random from $\ldpc(\n,\ledge,\redge)$.
\label{app:ConcentrationProofs}
\subsection{Concentration of the Conditional Entropy}
Fix an arbitrary order for the $m= (1-\rate)\n$ parity-check nodes, and let 
$\graph_t$, $t\in [m]$, be a random variable describing the first $t$ parity-check 
equations. 
Furthermore, let $\graph_0$ be a trivial (empty)
random variable. Define the Doob martingale $Z_t \equiv \expectation[H_\n \mid \graph_t]$.
The martingale property $\expectation[Z_{t+1} \mid Z_0, \dots, Z_t] = Z_t$ follows
by construction. 
In order to stress that $Z_t$ is a (deterministic) function of the random variable
$\graph_t$, we will write $Z_t = Z(\graph_t)$.
Obviously, $Z_0 = \expectation[H_\n]$ is the expected conditional entropy over the
code ensemble, and $Z_m = H_n \equiv H_{\graph}(X \mid Y)$ is the conditional entropy for a 
random code $\graph$. Theorem \ref{theo:concentrationml} follows therefore 
from the Hoeffding-Azuma inequality, 
once we bound the differences $|Z_{t+1}-Z_t|$. This is
our aim in the remaining of this subsection.

Assume, for the sake of definiteness, that parity-checks 
have been ordered by increasing
degree. The first $m_1$ of them have degree $\rdegree_1$, the successive
$m_2$ have degree $\rdegree_2$, and so on, with $\rdegree_1<\rdegree_2<\dots$. 
The $(t+1)^{\text{th}}$ parity-check will therefore have a well defined degree, to be denoted
by $\rdegree$.
Consider two realizations $\graph_{t+1}$ and $\graph'_{t+1}$ 
of the first $(t+1)$ parity-checks which differ uniquely in
the $(t+1)^{\text{th}}$ check. 
Let $\graph$ be a code uniformly distributed over $\ldpc(\ledge,\redge,\n)$ 
whose restriction to the first $(t+1)$ parity-checks coincides with $\graph_{t+1}$.
Construct a new code $\graph'$ whose restriction to the first $(t+1)$ parity-checks 
is $\graph'_{t+1}$, and which differs from $\graph$ in at most 
$(\rdegree+1)$ parity-checks. This can be done by the `switching' procedure of 
\cite{RiU01}. This switching procedure results in a ``pairing up'' of graphs.
In order to obtain the desired result, it is now enough to show
that $|H_{\graph}(X \mid Y)-H_{\graph'}(X \mid Y)|\le \alpha$, 
for some $n$-independent constant $\alpha$. 

Let us focus on the variation in conditional entropy under the addition
of a single parity-check. Let $\graph$ be a generic linear code and let $\graph+1$,
be the same code with the added constraint that
$x_{i_1}\oplus \cdots\oplus x_{i_{\rdegree}}= \0$. 
Define the corresponding parity bit
$\tilde{x}=x_{i_1}\oplus \cdots\oplus x_{i_{\rdegree}}$,
Then
\begin{align*}
H_{\graph}(X \mid Y) & =  H_{\graph}(X \mid \tilde{X},Y)+ 
H_{\graph}(\tilde{X} \mid Y) - H_{\graph}(\tilde{X} \mid X, Y) \\
&= H_{\graph}(X \mid \tilde{X}=0,Y) + H_{\graph}(\tilde{X} \mid Y) \\
&= H_{\graph+1}(X \mid Y) + H_{\graph}(\tilde{X} \mid Y) \, .
\end{align*}
The second equality follows since 
$H_{\graph}(\tilde{X} \mid X, Y)=0$ and by using the channel symmetry.
The third step is a consequence of the definition 
of $\graph+1$.  Since $\tilde{X}$ is a bit, its entropy is  between $0$ and $1$
and therefore
\begin{align}
| H_{\graph}(X \mid Y) - H_{\graph+1}(X \mid Y)|\le 1\, .
\label{eq:OneParityCheckVariation}
\end{align}

Recall that $\graph$ and $\graph'$ differ in at most ($\rdegree+1$) 
parity-checks, where $\rdegree$ is upper bounded by $\drmax$,
the maximal check-node degree.
Equation (\ref{eq:OneParityCheckVariation}) implies 
$|H_{\graph}(X \mid Y)-H_{\graph'}(X \mid Y)|\le (\rdegree+1)$
and, therefore, Theorem~\ref{theo:concentrationml}.
\subsection{Concentration of the Derivative of the Conditional Entropy}
It is convenient to introduce the per-bit conditional 
entropy $h_n(\cp) \defas \frac{1}{n}H_{\graph}(X|Y)$ and its expected
value $\hb_n(\cp) \defas \frac{1}{n}\expectation H_{\graph}(X|Y)$ when $\graph$
is a random code drawn uniformly from the $\ldpc(\ledge,\redge,\n)$ ensemble. 

Since the channel family $\{\text{BMS}(\cp)\}_{\cp \in I}$ 
is smooth and ordered by physical degradation, $h_n(\cp)$ is 
differentiable convex function of $\cp\in I$. Therefore
\begin{align}
\frac{1}{\Delta}[h_n(\cp)-h_n(\cp-\Delta)]\le h'_n(\cp)\le
\frac{1}{\Delta}[h_n(\cp+\Delta)-h_n(\cp)]\, ,\label{eq:ConvexDerivativeBound}
\end{align}
for any $\Delta>0$ such that $[\epsilon-\Delta,\epsilon+\Delta]\in I$.
Because of Theorem \ref{theo:concentrationml}, we also have
\begin{align*}
\frac{1}{\Delta}[\hb_n(\cp)-\hb_n(\cp-\Delta)-2\tilde{\xi}]& \le  h'_n(\cp)\le\\
&\le \frac{1}{\Delta}[\hb_n(\cp+\Delta)-\hb_n(\cp)+2\tilde{\xi}]\, ,
\end{align*}
with probability greater than $1-Ae^{-nB\tilde{\xi}^2}$ (it follows from the proof
in the previous subsection that $A$ and $B$ can be chosen uniformly in 
$\epsilon$). By averaging (\ref{eq:ConvexDerivativeBound})
over the code $\graph$, and subtracting it from the last equation, we get
\begin{align*}
|h'_n(\cp)-\hb'_n(\cp)|\le \frac{1}{\Delta}
[\hb_n(\cp\! +\!\Delta)-2\hb_n(\cp)+\hb_n(\cp\!-\!\Delta)+2\tilde{\xi}]\, ,
\end{align*}
which, using the convexity of $\hb_n(\epsilon)$, and fixing 
$\Delta = \tilde{\xi}^{1/2}$, implies
\begin{align*}
|h'_n(\cp)-\hb'_n(\cp)|\le [\hb_n'(\cp\! +\!\tilde{\xi}^{1/2})-
\hb_n'(\cp\!-\!\tilde{\xi}^{1/2})]+2\tilde{\xi}^{1/2}\, .
\end{align*}
The functions $\hb_n$ are differentiable and convex and (by hypothesis)
they converge to $\hb(\cp) = \xh^\MAP(\cp)=
\lim_{\n\to\infty}\frac{1}{n}\expectation H_n$ 
which is differentiable in $J$. It is a standard result in convex 
analysis (see \cite{Roc70}) that the derivatives $\hb'_n$ converge to $\hb'$ 
uniformly in $J$. Therefore, there exists a sequence $\delta_n\to 0$,
such that 
\begin{align*}
|h'_n(\cp)-\hb'_n(\cp)|\le [\hb'(\cp\! +\!\tilde{\xi}^{1/2})-
\hb'(\cp\!-\!\tilde{\xi}^{1/2})]+\delta_n+2\tilde{\xi}^{1/2}\, .
\end{align*}
with probability greater than $1-Ae^{-nB\tilde{\xi}^2}$. In order to complete 
the proof, it is sufficient to let $\tilde{\xi}_*(\xi)$ be the largest 
value of $\tilde{\xi}$,  such that  $[\hb'(\cp\! +\!\tilde{\xi}^{1/2})-
\hb'(\cp\!-\!\tilde{\xi}^{1/2})]+2\tilde{\xi}^{1/2}<\xi/2$.
Then the thesis holds with $\alpha_{\xi} = B\tilde{\xi}^2_*(\xi)/2$.
In particular, if $\hb(\cp)$ is twice differentiable with respect to
$\cp\in J$, then $[\hb'(\cp\! +\!\tilde{\xi}^{1/2})-
\hb'(\cp\!-\!\tilde{\xi}^{1/2})]\le \tilde{A} \tilde{\xi}^{1/2}$,
and $\tilde{\xi}_*(\xi)\ge \tilde{A}'\xi^2$.
%

%
%
\section{Proofs of Lemmas in the Counting Argument}
\label{ProofCounting}

%
%
\subsection{Proof of Lemma \ref{ConcentrationDegreeLemma}}
\label{ProofCountingDegree}

Let $\graph(t)$ denote the residual graph after $t$ iterations 
of the message passing decoder, and 
$\nddp_{\graph(t)}=(\lnode_{\graph(t)},\Gamma_{\graph(t)})$ be the 
corresponding degree distribution pair. Moreover, denote
by $\nddp_t = (\lnode_t,\Gamma_t)$ the typical degree
distribution pair of $\graph(t)$. Explicitly
\begin{align*}
\lnode_t(\zl) & \defas \lnode(\zl \xl_t)\, ,\\
\Gamma_{t}(\zl) &\defas \Gamma(1-\xr_t+\zl\xr_t)-\Gamma(1-\xr_t)-
\zl\xr_t\Gamma'(1-\xr_t)\, ,
\end{align*}
where $\xl_{t},\xr_t$ denote the typical fractions of erased messages 
after $t$ iterations of the decoder. These are obtained by solving
the density evolution equations $\xl_{t+1} = \cp\ledge(\xr_t)$,
$\xr_{t+1}= 1-\redge(1-\xl_t)$ with initial condition $\xl_0=\xr_0=1$.

Notice that 
\begin{align*}
d(\nddp_{\cp},\nddp_{\graph{\cp}}) &\le d(\nddp_{\cp},\nddp_{t})
+d(\nddp_{t},\nddp_{\graph(t)})+d(\nddp_{\graph(t)},\nddp_{\graph(\cp)})\, .
\end{align*}
We claim that
\begin{eqnarray}
\lim_{t\to\infty} d(\nddp_{\graph(t)},\nddp_{\graph(\cp)}) & = &0\, ,
\label{FirstClaim}\\
\lim_{n\to\infty} \expectation [d(\nddp_{t},\nddp_{\graph(t)})] & = & 0\, ,
\label{SecondClaim}\\
\lim_{t\to\infty}\lim_{n\to\infty} \expectation[ d(\nddp_{\cp},\nddp_{t})]
& = & 0\, .\label{ThirdClaim}
\end{eqnarray}
Before proving those claims, let us show that they imply the thesis.
It follows from the triangular inequality above that
$\lim_{t\to\infty}\lim_{n\to\infty} \expectation \, 
d(\nddp_{\cp},\nddp_{\graph(\cp)})= 0$. But 
$d(\nddp_{\cp},\nddp_{\graph(\cp)})$ does not depend upon $t$, therefore
\begin{align*}
\lim_{n\to\infty} \expectation[ d(\nddp_{\cp},\nddp_{\graph(\cp)})] &= 0\, .
\end{align*}
This in turns imply the thesis via Markov inequality.

We must now prove the inequalities (\ref{FirstClaim}) to (\ref{ThirdClaim}).
The first one is a trivial consequence of the convergence of 
DE to its fixed point: $\lim_{t\to\infty}\xl_t=\xl$,
$\lim_{t\to\infty}\xr_t=\xr$, together with the continuity of
the expressions (\ref{LambdaRes}), (\ref{GammaRes}) with $\xl,\xr$.
Eq. (\ref{SecondClaim}) follows from the general concentration 
analysis in \cite{RiU01}.

In order to prove (\ref{ThirdClaim}), consider  a variable node $i$
in the residual graph and imagine changing the received symbol at $i$,
and update all the messages consequently. Consider the edges whose distance
from $i$ is larger than $t$, and denote by $W^{(t)}_i$ the number of 
messages on such edges that change of value after the received symbol at $i$
has been changed. It is clear that 
\begin{align}
\expectation[ d(\nddp_{\cp},\nddp_{t})] &\le \expectation[ W^{(t)}_i] ,
\end{align}
The limit $\lim_{n\to\infty} \expectation[W^{(t)}_i]$ can be computed 
through a branching process analysis. The calculation is very similar
to the one in \cite{RiU00encoding} and we do not reproduce it here.
The final result is that, as long as $\cp\ledge'(\xr)\redge'(1-\xl)<1$,
there exist two positive constants $A$, $b$ with $b<1$ such that
$\expectation [W^{(t)}_i ]\le A\, b^t$. The proof is finished by noticing
that the condition $\cp\ledge'(\xr)\redge'(1-\xl)<1$ is satisfied whenever 
$\cp$ is a continuity point of $\xl(\cp)$.
%
%
\subsection{Proof of Lemma \ref{PsiProp}}

Notice that the function $u\mapsto v(u)$ defined in (\ref{xtoy})
enjoys the property $v(1/u) = 1/v(u)$ for any $u>0$.
Assume {\em ab absurdum} that $\Psi_{\nddp}$ does not achieves 
its maximum in the interval $[0,1]$. Therefore, there exist $u>1$ such that 
$\Psi_{\nddp}(u')<\Psi_{\nddp}(u)$ for any $u'\in [0,1]$.
We will show that $\Psi_{\nddp}(1/u)\ge \Psi_{\nddp}(u)$ thus 
reaching a contradiction.
In fact, some algebra shows that
\begin{align*}
\Psi_{\nddp}(1/u) &=  -\lnode'(1)\log_2\left[\frac{(1+uv)}{(1+u)(1+v)}\right]
\\
&~~ +\sum_\ldegree\lnode_\ldegree\log_2\left[
\frac{1+u^\ldegree}{2(1+u)^\ldegree}\right]\\
&~~ +\frac{\lnode'(1)}{\Gamma'(1)}\sum_{\rdegree}\Gamma_\rdegree
\log_2\left[1+\left(\frac{v-1}{v+1}\right)^\rdegree\right].
\end{align*}
The claim follows from $0<\frac{v-1}{v+1}<1$ together with the monotonicity
of the logarithm.

In order to prove the second claim, i.e., the regularity of
$\Psi_{\nddp}$ with respect to the \ddp write  $\Psi_{\nddp}^{(1)}(u)
+\Psi_{\nddp}^{(2)}(u)+\Psi_{\nddp}^{(3)}(u)$ with $\Psi^{(1,2,3)}_{\nddp}$
the three summands in (\ref{Psidef}). The estimate (\ref{PsiLip})
can be proved for each of the three terms separately. 
Here, we limit ourselves to consider $ \Psi_{\nddp}^{(1)}(u)$, the derivation 
being nearly identical for the two other summands.
Start by noticing that, for any $u\in [0,1]$ and any \ddp, we have
\begin{eqnarray*}
\frac{1}{2}\le \sum_{\ldegree}\frac{\ledge_{\ldegree}}{1+u^{\ldegree}}\le 1\, ,
\;\;\;\;\;\;\;\;\;\;\;
\sum_{\ldegree}\frac{\ledge_{\ldegree}u^{\ldegree-1}}{1+u^{\ldegree}}\le 1\,.
\end{eqnarray*}
Now fix two \ddp $\nddp$ and $\tilde{\nddp}$. Let $v(u)$ and $\tilde{v}(u)$
the corresponding functions defined as in (\ref{xtoy}). 
Notice that
\begin{align*}
\left|\sum_{\ldegree}\frac{\ledge_{\ldegree}-\ledge_{\ldegree}}
{1+u^{\ldegree}}\right| 
&=\left|\sum_{\ldegree}\left(\frac{1}{1+u^{\ldegree}}-\frac{1}{2}\right)
(\ledge_{\ldegree}-\ledge_{\ldegree})\right|\\
&\le  \frac{\ldegree_{\rm max}}{2}(1-u) \sum_{\ldegree}| \ledge_{\ldegree}-\ledge_{\ldegree}|\\
&\le \frac{1}{2}\ldegree_{\rm max}^2(1-u)\,d(\nddp,\tilde{\nddp})
\end{align*}
Using these inequalities, some calculus shows that
\begin{align*}
1\ge v(u), \tilde{v}(u) & \ge  1-2\, \ldegree_{\rm max}(1-u)\, ,\\
|v(u)-\tilde{v}(u)| & \le  3\, \ldegree^2_{\rm max}\,\, (1-u)\, 
d(\nddp, \tilde{\nddp})\, .
\end{align*}
Next notice that, if we set $f(u,v) \defas  \log_2\left[\frac{2(1+uv)}{(1+u)(1+v)}\right]$, then, for any $u,v,\tilde{v}\in [0,1]$, we have
\begin{align*}
\left| f(u,v)\right| & \le \frac{(1-u)(1-v)}{\log 2}\, ,\\
\left|f(u,v)-f(u,\tilde{v})\right|
&\le\frac{(1-u)}{\log 2}\, |v-\tilde{v}|\, .
\end{align*}
Using these observations we obtain
\begin{align*}
%
|\Psi_{\nddp}(u)-\Psi_{\tilde{\nddp}}(u)| 
& \le    \max [f(u,v),f(u,\tilde{v})]
\,|\lnode'(1)-\tilde{\lnode}'(1)| \\
&~~ + \max[\lnode'(1),\tilde{\lnode}'(1)]
\,|f(u,v)-f(u,\tilde{v})| \\
&\le \frac{2\ldegree_{\rm max}}{\log 2}(1-u)^2|\lnode'(1)-\tilde{\lnode}'(1)|\\
&~~ +\frac{\ldegree_{\rm max}}{\log 2}(1-u)|v-\tilde{v}| \\
& \le  A_1\, (1-u)^2\, d(\nddp,\tilde{\nddp})\, , 
\end{align*}
which confirms our thesis with constant $A_1 =  (2\, \ldegree^2_{\rm max}+3
\ldegree^3_{\rm max})/\log 2$. The variations of $\Psi^{(2)}_{\nddp}$
and $\Psi^{(3)}_{\nddp}$ are bounded analogously.
%
%

\section{Area and BP \exitentropy} \label{app:areass}

\subsection{Two Useful Tricks}
\label{sec:tricks}
We give here two lemmas which contain the two computational tricks which 
are used all along this paper. Lemma~\ref{lemma:ibp1} and 
Lemma~\ref{lemma:ibp2} will be again used in the next subsection of the
appendix. 
Observe that the function $\xl \mapsto \xh \defas \lnode(\xr(\xl))$ is composed by 
 two functions $\xr$ and $\lnode$ which are strictly increasing over $[0,1]$. Therefore, 
 the inverse function $\xl(\xh)$ exists and 
 $\xh \mapsto \xl(\xh)\defas \xr^{-1} \circ \lnode^{-1}(\xh)$ is a continuous and strictly 
 increasing bijection from $[0,1]$ to $[0,1]$. The values  
 $\ih(\xl) \defas \frac{\xl}{\ledge(\xr(\xl))}$ can then equivalently 
 be described by $\ih(\xh) \defas \frac{ \xr^{-1} \circ \lnode^{-1} }{\ledge\circ\lnode^{-1}}(\xh)$.
 \blemma \label{lemma:ibp1} Given a  \ddp $(\ledge,\redge)$ and any couple $(\xl_a,\xl_b)\in[0,1]^2$. 
 With the notations  
 $\xh_a=\xh(\xl_a)\defas \lnode\circ\xr(\xl_a)$ and $\xh_b=\xh(\xl_b)$, we can 
 then write
 $$
 \int_{\xh_a}^{\xh_b} \ih(\xh)\text{d}\xh = \frac{1}{\int \ledge}\left(\xl_b \xr(\xl_b ) - \xl_a \xr(\xl_a ) -\int_{\xl_a}^{\xl_b}\xr(\xl)\text{d}\xl  \right).
 $$
 \elemma
 \bproof This is a simple integration by parts once it has been observed  
 $\ih(\xl)\cdot\frac{\text{d}\xh(\xl)}{\text{d}\xl}=\frac{\xl}{\ledge\circ\xr (\xl)}\cdot\frac{(\ledge\circ \xr)(\xl)\cdot\xr'(\xl)}{\int \ledge}=\frac{\xl \xr'(\xl)}{\int\ledge}$.
 \eproof
 \blemma \label{lemma:ibp2} Given a  \ddp $(\ledge,\redge)$ and any interval $(\xl_a\xl_b)\subseteq[0,1]$, $\xl^\BP\leq \xl_a$ over which $\ih(\xr)\defas\frac{\xl}{\ledge\circ\xr(\xl)}$ is increasing. Then, the function $\xh^\BP(\ih)$ is continuous over $(\ih_a, \ih_b)$, where $\ih_a\defas\ih(\xl_a)$ and  $\ih_b\defas\ih(\xl_b)$, and
 \begin{multline*}
 \int_{\ih_a}^{\ih_b} \xh^\BP(\ih)\text{d}\ih = \frac{1}{\int \ledge}\Big( \ih_b \int_0^{\xr(\xl_b)}\ledge(\xr)\text{d}\xr - \ih_a \int_0^{\xr(\xl_a)}\ledge(\xr)\text{d}\xr\\ -\xl_b \xr(\xl_b ) + \xl_a \xr(\xl_a ) +\int_{\xl_a}^{\xl_b}\xr(\xl)\text{d}\xl \Big).
 \end{multline*}
 \elemma
 \bproof
 This is proved by, first, integrating by parts and, second, using Lemma~\ref{lemma:ibp1}. 
 \eproof

\subsection{Area under the BP \exitentropy Curve}

 \btheo[Area Theorem for BP Decoding]
\label{theo:bpintegrationfact}
 Given a \ddp  $(\ledge,\redge)$  and the asymptotic BP \exitentropy entropy as defined in Corollary~\ref{cor:asymptoticbpexit}, 
then 
$$
\drate+\frac{1}{\int \ledge}\sum_{i=1}^JD_i = \int_0^1 \xh^\BP(\ih)\text{d}\ih,
$$ 
 where $D_i=A_i-B_i-C_i$ with $A_i\defas\lxl^{i} \xr(\lxl^{i})- \uxl^{i-1} \xr(\uxl^{i-1})$, $B_i\defas\ih^i \int_{\xr(\uxl^{i-1})}^{\xr(\lxl^i)}  \ledge(\xr)\text{d}\xr$, and $C_i=\int_{\uxl^{i-1}}^{\lxl^i}  \xr(\xl)\text{d}\xl 
$.
 \etheo

\bproof
 Using Corollary~\ref{cor:asymptoticbpexit}, we can derive (\ref{eq:bpareatheo}) as shown above
\begin{figure*}[hbt]
\normalsize
\begin{align}
 &~ \notag  \int_0^1 \xh^\BP(\ih)\text{d}\ih \\
\notag &= \int_0^{\ih^\BP} \xh^\BP(\ih)\text{d}\ih + \sum_{i=1}^J\int_{\ih^i}^{\ih^{i+1}} \xh^\BP(\ih)\text{d}\ih \\ \notag 
& \overset{(a)}{=} 0 +  \frac{1}{\int \ledge}\sum_{i=1}^J \left(\left[\ih(\xl)\int_0^{\xh(\xl)}\ledge(\xr)\text{d}\xr\right]_{\lxl^i}^{\uxl^i} - \Big[ \xl \xr(\xl) \Big]_{\lxl^i}^{\uxl^i} + \int_{\lxl^i}^{\uxl^i} \xr(\xl)\text{d}\xl \right)\\
\notag &= \frac{ 
\left(\int_0^1 \ledge(\xr)\text{d}\xr - \sum_{i=1}^J \left[\ih(\xl)\int_0^{\xh(\xl)}\ledge(\xr)\text{d}\xr \right]_{\uxl^{i-1}}^{\lxl^i} \right) 
- \left( 1 - \sum_{i=1}^J \Big[   \xl \xr(\xl)   \Big]_{\uxl^{i-1}}^{\lxl^i} \right)
 + \left( \int_0^1 \xr(\xl)\text{d}\xl -  \sum_{i=1}^J  \int_{\uxl^{i-1}}^{\lxl^i} \xr(\xl)\text{d}\xl  \right)  
}{\int \ledge}\\
\label{eq:bpareatheo} & \overset{(b)}{=}\frac{\int \ledge -1 + \int \xr }{\ledge}+\frac{1}{\int \ledge}\sum_{i=1}^J 
\left(
\Big[\xl \xr(\xl)   \Big]_{\uxl^{i-1}}^{\lxl^i} 
- \ih^i \int_{\xr(\uxl^{i-1})}^{\xr(\lxl^i)}  \ledge(\xr)\text{d}\xr 
- \int_{\uxl^{i-1}}^{\lxl^i}  \xr(\xl)\text{d}\xl 
\right)
 \end{align}
\hrulefill
\vspace*{4pt}
\end{figure*}
 where $(a)$ comes from Lemma~\ref{lemma:ibp2} and $(b)$ uses the fact that $\ih^i =\ih(\uxl^{i-1})=\ih(\lxl^i)$.
\eproof 

 First, observe that Theorem~\ref{theo:bpintegrationfact} quantifies the average 
sub-optimality of BP decoding compared to  MAP decoding. The area under the BP 
\exitentropy curve is trivially larger or equal than the design rate since the 
$D_i$'s are non-negative. Moreover, it seems to indicate that there performance 
loss occurs at each phase transition.
 
Second, Theorem~\ref{theo:bpintegrationfact} has a pleasing geometric interpretation 
which goes back to the asymptotic analysis and which is explained in 
appendix~\ref{app:gapinterpretation}.

\section{Dynamic Interpretation of the Average Gap between MAP and BP Decoding}
\label{app:gapinterpretation}
It is now well-known that the determination of capacity-achieving sequences on the 
erasure channel reduces to 
a curve-fitting problem, see, e.g., \cite{Sho00}, \cite{AKtB04}. This was the motivation 
for the Area Theorem and - so far - its unique application. Let us recall this view. 
For the purpose of illustration, and without essential loss of generality, 
we focus on the case of (G)LDPC ensembles. 

\begin{figure}[htp]
\centering
\setlength{\unitlength}{1bp}
\begin{picture}(230,70)
\put(0,20){\includegraphics[scale=0.3125]{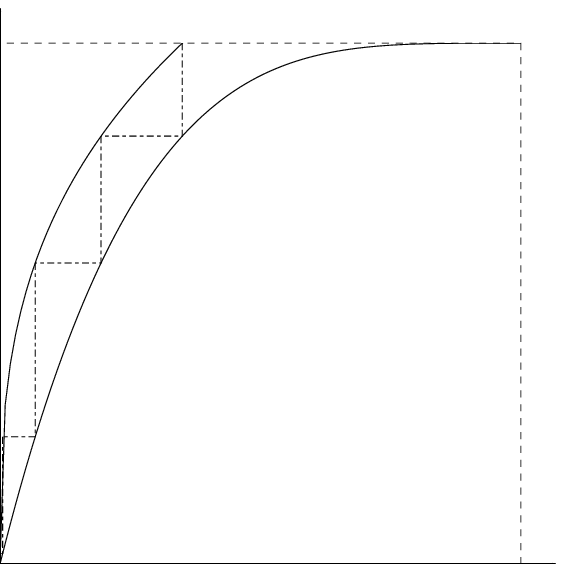}} 
\put(60,20){\includegraphics[scale=0.3125]{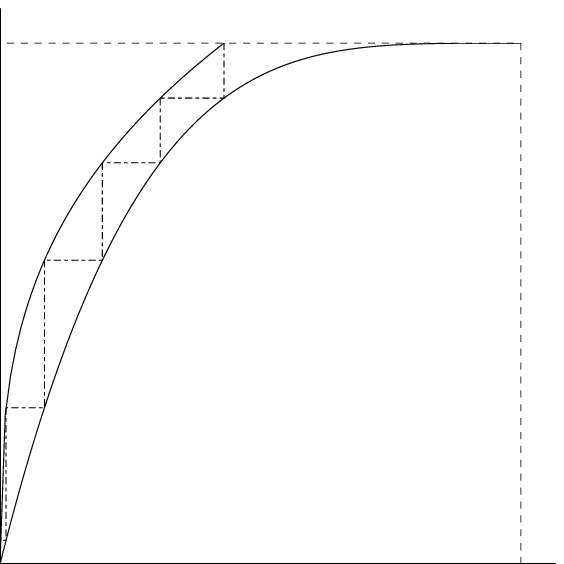}}
\put(120,20){\includegraphics[scale=0.3125]{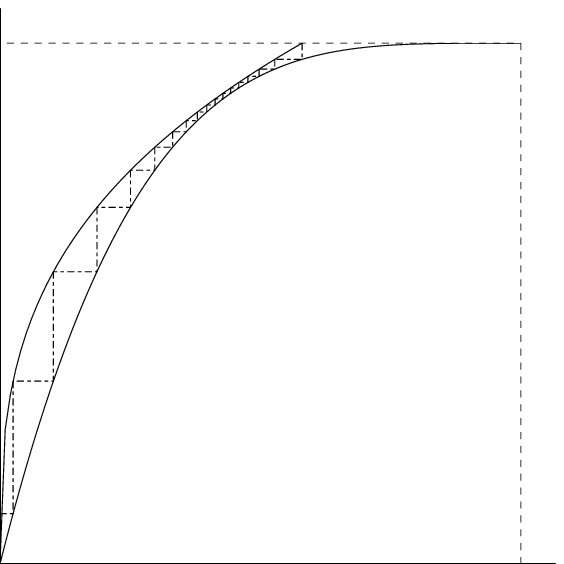}}
\put(180,20){\includegraphics[scale=0.3125]{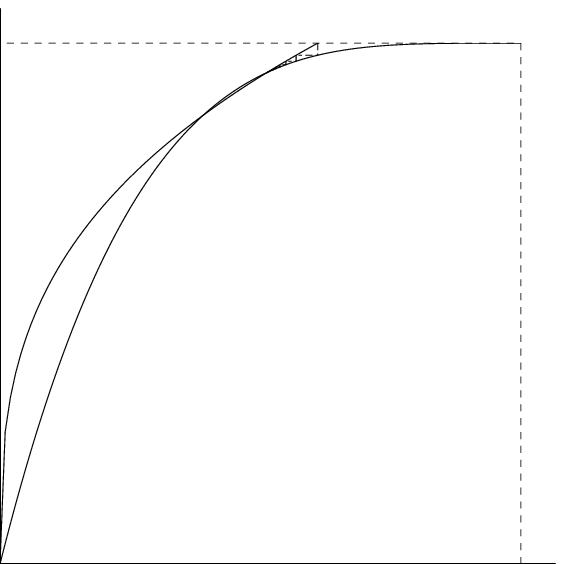}}
\put(-4,15){\makebox(0,0){\footnotesize{$0$}}}
\put(-4,67){\makebox(0,0){\footnotesize{$1$}}}
\put(47,15){\makebox(0,0){\footnotesize{$1$}}}
\put(25,5){\makebox(0,0){\footnotesize{$\ih=0.35<\ih^\BP$}}}
\put(60,0)
{
\put(-4,15){\makebox(0,0){\footnotesize{$0$}}}
\put(-4,67){\makebox(0,0){\footnotesize{$1$}}}
\put(47,15){\makebox(0,0){\footnotesize{$1$}}}
\put(25,5){\makebox(0,0){\footnotesize{$\ih=0.43<\ih^\BP$}}}
}
\put(120,0)
{
\put(-4,15){\makebox(0,0){\footnotesize{$0$}}}
\put(-4,67){\makebox(0,0){\footnotesize{$1$}}}
\put(47,15){\makebox(0,0){\footnotesize{$1$}}}
\put(25,5){\makebox(0,0){\footnotesize{$\ih=0.58<\ih^\BP$}}}
}
\put(180,0)
{
\put(-4,15){\makebox(0,0){\footnotesize{$0$}}}
\put(-4,67){\makebox(0,0){\footnotesize{$1$}}}
\put(47,15){\makebox(0,0){\footnotesize{$1$}}}
\put(25,5){\makebox(0,0){\footnotesize{$\ih=0.61<\ih^\BP$}}}
}
\end{picture}
\caption{Iterative decoding trajectory for the ensemble $\ldpc(\n,\xl^3,\xl^4)$ (in the limit when $\n\to\infty$): 
increasing values of the channel parameter $\ih$.}
\label{fig:iterativetrajectory}
\end{figure}

\subsection{\exitentropy Chart}
Fig.~\ref{fig:iterativetrajectory} summarizes the DE analysis of the BP decoding 
by showing the convergence of the recursive sequence formed the edge entropy $\{\xl_\iter\}_\iter$ 
(i.e., the edge erasure probability).   
 Such a representation (which emphasizes two component 
\exitentropy functions, one associated to the left nodes and one associated to the right nodes) is 
called \exitentropy chart in \cite{teB01}. This representation is 
(asymptotically) 
exact for the binary erasure channel (since it is DE) whereas it is only 
approximate in the general case.

\begin{figure}[htp]

\vspace{10bp}
\centering
\setlength{\unitlength}{0.5bp}
\begin{picture}(160,160)
\put(0,0){\includegraphics[scale=0.5]{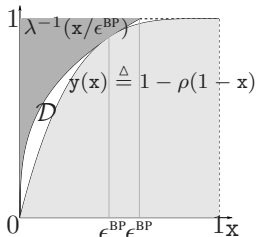}}
\put(-5,-5){\makebox(0,0){$0$}}
\put(150,-5){\makebox(0,0){$1$}}
\put(-5,150){\makebox(0,0){$1$}}
\put(160, -10){\makebox(0,0){$\xl$}}
\put(70, -10){\makebox(0,0){$\ih^\BP$}}
\put(90, -10){\makebox(0,0){$\ih^\BP$}}
\put(20, 80){\makebox(0,0){${\cal{D}}$}}
\put(40, 143){\makebox(0,0){\footnotesize{ ${\ledge}^{-1}(\xl/\ih^\BP)$}}}
\put(105, 105){\makebox(0,0){\footnotesize{ $\xr(\xl)\defas 1-\redge(1-\xl)$}}}
\end{picture}
\caption{ Additive gap to capacity for the \ddp $(x^3,x^4)$. } 
\label{fig:areatheoremflatness}
\vspace{5bp}
\end{figure}

Fig. \ref{fig:areatheoremflatness} represents the \exitentropy chart when transmission 
takes place at the BP threshold $\ih=\ih^\BP$. The  \exitentropy functions are here 
the ones associated to the component of the LDPC ensemble. The function on 
the left is associated to repetition codes on the left while the one on the
right is associated to  parity-check codes. At  channel parameter $\ih=\ih^\BP$, 
the two \exitentropy curves are tangent in $(\xl^\BP,\xr^\BP)$ and
the \exitentropy chart offers also  a graphical representation of the limiting gap to capacity of
the LDPC ensemble. 
The additive gap $C(\ih^\BP)-\drate$ 
to the Shannon threshold is indeed represented by the entire white area ${\cal{D}}$ 
such that 
$$
C(\ih^\BP)-\drate=\ih^\Sh-\ih^\BP=\frac{{{\cal D }}}{\int\ledge},
$$
where $\frac{1}{\int \ledge}=\lnode'(1)$ is the average left degree. In words, the 
area ${\cal{D}}$ is the area between the left \exitentropy curve $\xl\mapsto \ledge^{-1}(\xl/\ih^\BP)$ 
(at the BP threshold) and the right \exitentropy curve $\xl\mapsto1-\redge(1-\xl)$ which is bounded 
away by the unit square. This statement is presented, e.g., in \cite{AKtB04}. 
We will now refine this statement by applying the Area Theorem to 
the \exitentropy curve of the LDPC ensemble  previous statement (i.e., using 
the  basic 
principle of our method). We will see that, in short, the 
area ${\cal{D}}$ can be itself divided into two parts where the subarea below $\xl^\BP$ 
represents the average gap between MAP and BP decoding. The determination of 
LDPC codes for which BP decoding is MAP reduces then again to a 
curve-fitting problem below $\xl^\BP$.

\subsection{Geometric Interpretation at the Component Level}\label{sec:geometryMAP-BP}

\begin{figure}[htp]
\centering
\setlength{\unitlength}{1bp}
\begin{picture}(240,140)
\put(5,28)
{
\put(5,5){\includegraphics[width=100bp]{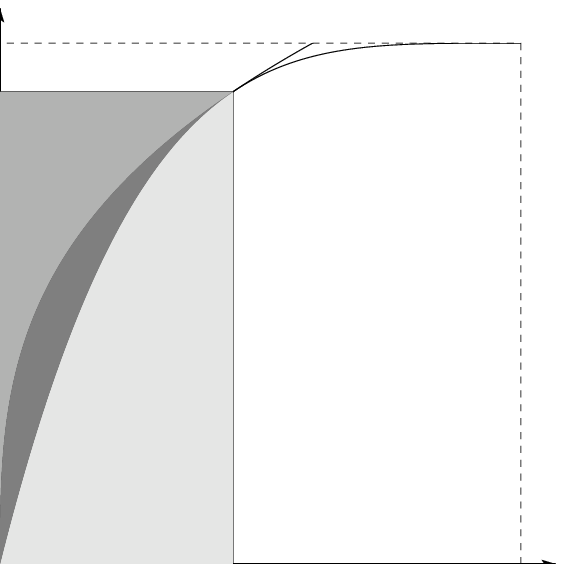}}
\put(135,5){\includegraphics[width=100bp]{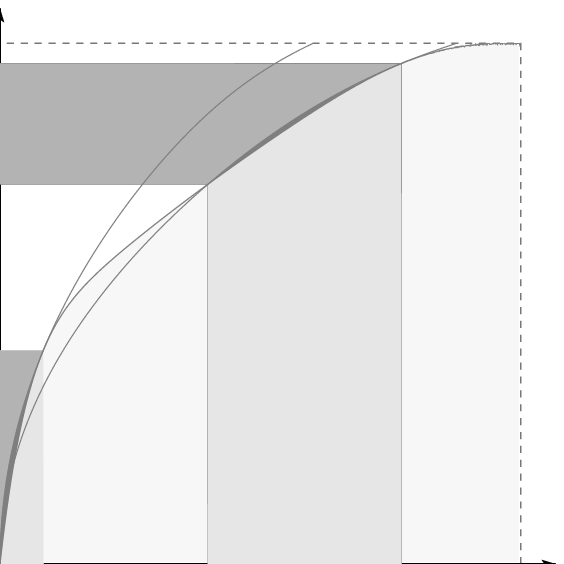}}
\put(0,0){\makebox(0,0){\small{$0$}}}
\put(100,0){\makebox(0,0){\small{$1$}}}
\put(0,100){\makebox(0,0){\small{$1$}}}
\put(130,0){\makebox(0,0){\small{$0$}}}
\put(230,0){\makebox(0,0){\small{$1$}}}
\put(130,100){\makebox(0,0){\small{$1$}}}
\put(7,-6){\makebox(0,0){\small{$\lxl^0$}}} 
\put(66,108){\makebox(0,0){\small{$\ih^\BP$}}} 
\put(18,80){\makebox(0,0){\small{$B_1$}}}
\put(18,55){\makebox(0,0){\small{$D_1$}}}
\put(38,40){\makebox(0,0){\small{$C_1$}}}
\put(48,-6){\makebox(0,0){\small{$\lxl^1=\xl^\BP$}}}
\put(37,100){\makebox(0,0){\small{$\ledge^{-1}(\frac{\xl}{\ih^\BP})$}}}
\put(80,80){\makebox(0,0){\small{$\xr(\xl)\defas 1-\redge(1-\xl)$}}}
\put(148,-5){\makebox(0,0){\small{$\xl^\BP$}}}
\put(178,-5){\makebox(0,0){\small{$\uxl^1$}}}
\put(208,-5){\makebox(0,0){\small{$\lxl^2$}}}
\put(195,108){\makebox(0,0){\small{$\ih^\BP$}}} 
\put(220,108){\makebox(0,0){\small{$\ih^2$}}} 
\put(230,90){\makebox(0,0){\small{$\xr(\xl)$}}} 
\put(138,38){\makebox(0,0){\small{$B_1$}}}
\put(142,09){\makebox(0,0){\small{$C_1$}}}
\put(140,25){\makebox(0,0){\small{$D_1$}}}
\put(158,80){\makebox(0,0){\small{$B_2$}}}
\put(188,80){\makebox(0,0){\small{$D_2$}}}
\put(188,40){\makebox(0,0){\small{$C_2$}}}

}
\put(60,7){\makebox(0,0){{\small{(a)}}}}
\put(190,7){\makebox(0,0){{\small{(b)}}}}
\end{picture}
\caption{
Graphical interpretation of Theorem~\ref{theo:bpintegrationfact} at a microscopic (dynamic) level: 
(a) Standard one-jump case: Ensemble $\ldpc(x^3,x^4)$ and transmission at $\ih=\ih^\BP$
(b) Double-jump case: The left distribution is 
$\ledge(x)=0.78x^2 + 0.1x^3 + 0.12x^{14}$ and the $\xr$ represents the 
\exitentropy function of a mixture of component codes composed by 
50\% of $[19,18]$ single parity-check codes, 35\% of $[7,4]$ Hamming codes 
and 15\% of $[15,11]$ Hamming codes in the edge perspective. Transmission is represented 
for two channel parameters.
}
\label{fig:geometricinterpretation}
\end{figure}

Fig.~\ref{fig:geometricinterpretation} shows a geometric 
representation of Theorem~\ref{theo:bpintegrationfact}. 
In (a) one see that the additive gap between BP threshold and Shannon threshold 
is represented by the total area between the component \exitentropy functions.
Further, the part of this area which 
corresponds to the average gap between MAP and BP decoding is $D_1$ as defined in Theorem~\ref{theo:bpintegrationfact}. 

\bibliographystyle{IEEEtran} 

\newcommand{\SortNoop}[1]{}

\end{document}